\documentclass[twocolumn]{svjour3}
\pdfoutput=1

\usepackage[table,dvipsnames]{xcolor}
\usepackage{data-index}
\usepackage{appendix}
\usepackage{cuted}
\usepackage[figuresright]{rotating}
\usepackage{multirow}
\usepackage{caption}
\usepackage{longtable}
\usepackage{booktabs}
\usepackage{tablenote}

\usepackage[biblabel,nomove,nocompress]{cite}

\usepackage{graphicx}
\graphicspath{{figures/},{./}}

\ifxetex
	\usepackage{fontspec}
	\usepackage{newtxmath}
\else
    \usepackage[T1]{fontenc}
    \usepackage[nomath]{lmodern}
\fi

\usepackage{amssymb}

\usepackage{marvosym}

\usepackage[switch,columnwise]{lineno}

\def\draft#1#2{~}

\let\oldtitle\title
\renewcommand{\title}[1]{\def\ztitle{#1}\oldtitle{#1}}

\newcommand{\twol}[2]{\vbox{\vspace{0.2em}\hbox{#1}\vspace{0.1em}\hbox{#2}}}

\begin{document}


	\title{A Systematic Collection of Medical Image Datasets for Deep Learning}

	\author{ Johann Li $^1$ 
		\and Guangming Zhu $^1$ 
		\and Cong Hua $^1$ 
		\and Mingtao Feng $^1$ 
		\and Basheer Bennamoun $^2$ 
		\and Ping Li $^3$ 
		\and Xiaoyuan Lu $^3$ 
		\and Juan Song $^1$ 
		\and Peiyi Shen $^1$ 
		\and Xu Xu $^4$ 
		\and Lin Mei $^4$ 
		\and Liang Zhang $^1$ 
		\and Syed Afaq Ali Shah $^5$ 
		\and Mohammed Bennamoun $^6$ 
	}

	\institute{
		\Letter Corresponding author: Guangming Zhu, Liang Zhang, and Syed Afaq Ali Shah
		\at \email{gmzhu@xidian.edu.cn}
		\at \email{liangzhang@xidian.edu.cn}
		\at \email{Afaq.Shah@murdoch.edu.au} \\ \\
		1\hspace{1em} School of Computer Science and Technology, Xidian University, China\\
		2\hspace{1em} School of Medicine, The University of Notre Dame, Australia \\
		3\hspace{1em} Data and Virtual Research Room, Shanghai Broadband Network Center, China \\
		4\hspace{1em} The Third Research Institute of The Ministry of Public Security, China \\
		5\hspace{1em} Discipline of Information Technology, Media and Communications, Murdoch University, Australia \\
		6\hspace{1em} Department of Computer Science and Software Engineering, The University of Western Australia, Australia
	}



	\date{Received: date / Accepted: date}

	\maketitle


\begin{abstract}
    The astounding success made by artificial intelligence (AI) in healthcare and other fields proves that AI can achieve human-like performance.
    However, success always comes with challenges.
    Deep learning algorithms are data-dependent and require large datasets for training. The lack of data in the medical imaging field creates a bottleneck for the application of deep learning to medical image analysis.
    Medical image acquisition, annotation, and analysis are costly, and their usage is constrained by ethical restrictions. They also require many resources, such as human expertise and funding. That makes it difficult for non-medical researchers to have access to useful and large medical data.
    Thus, as comprehensive as possible, this paper provides a collection of medical image datasets with their associated challenges for deep learning research.
    We have collected information of around three hundred datasets and challenges mainly reported between 2013 and 2020 and categorized them into four categories: head \& neck, chest \& abdomen, pathology \& blood, and ``others''.
    Our paper has three purposes:
    1) to provide a most up to date and complete list that can be used as a universal reference to easily find the datasets for clinical image analysis,
    2) to guide researchers on the methodology to test and evaluate their methods' performance and robustness on relevant datasets,
    3) to provide a ``route'' to relevant algorithms for the relevant medical topics, and challenge leaderboards.
\end{abstract}



\section{Introduction}
\label{sec:intro}

Since the invention of medical imaging technology, the field of medicine had entered a new era.
The beginning of medical imaging started with the adoption of X-Rays. With further technical advancements, many other imaging methods, including 3D computed tomography (CT), magnetic resonance imaging (MRI), nuclear medicine, ultrasound, endoscopy, and optical coherence tomography (OCT), were also exploited.
Directly or indirectly, these imaging modalities have contributed to the diagnosis and treatment of various diseases, and the research related to the human body's structure and intrinsic mechanisms.

Medical images can provide critical insight into the diagnosis and treatment of many diseases.
The human body's different reactions to imaging modalities are used to produce scans of the body. Reflection and transmission are commonly used in medical imaging because the reflection or transmission ratio of different body tissues and substances are different.
Some other methods acquire images by changing the energy transferred to the body, e.\,g., magnetic field changes or the rays radiated from a chemical agent.

Before modern AI was applied in medical image analysis, radiologists and pathologists needed to manually look for the critical ``biomarkers'' in the patient's scans. These ``biomarkers'', such as tumors and nodules, are the basis for the medics to diagnose and devise treatment plans. Such a diagnostic process needs to be performed by medics with extensive medical knowledge and clinical experience. However, problems such as diagnostic bias and the lack of medical resources are prevalent and cannot be avoided.
After the recent breakthroughs in AI (which achieve human-like performance, e.\,g., for image recognition \cite{Simonyan2014,Szegedy2015,He2015}, and can win games such as Go \cite{Silver2016} and real-time strategy games \cite{Vinyals2019}), the development of AI-based automatic medical image analysis algorithms has attracted lots of attention. Recently, the application of AI in medical image analysis has become one of the major research focuses and has attained many significant achievements \cite{Ronneberger2015,Peng2021a,Mehdizadeh2021}.

Many researchers brought their focus to AI-based medical image analysis methods thinking that it might be one of the solutions to the challenges (e.g., medical resource scarcity) and taking advantage of the technological progress \cite{Wang2021b,Shi2020,Mao2020,Liu2019a,Khan2018}.
Traditional medical image analysis focuses on detecting and identifying biomarkers for diagnosis and treatment. AI imitates the medic's diagnosis through classification, segmentation, detection, regression, and other AI tasks in an automated or semi-automated way.

AI has achieved a significant performance for many computer vision tasks. This success is yet to be translated to the medical image analysis domain.
Deep learning (DL), a branch of AI, is a data-dependent method as it needs massive training data. However, when DL is applied to medical image analysis, the paucity of labeled data becomes a major challenge and a bottleneck.

Data scarcity is a common problem when applying DL methods to a specific domain, and this problem becomes more severe in the case of medical image analysis. Researchers, who apply DL methods to medical image analysis research, do not usually have a medical background, commonly computer scientists. They cannot collect data independently because of the lack of access to medical equipment and patients, and they cannot annotate the acquired data either because they lack the relevant medical knowledge.
Furthermore, medical data is owned by institutions who cannot easily make it public due to privacy and ethics restrictions. When researchers evaluate their algorithms on their private data, the results of their research become incomparable.

To address some of these problems, MICCAI, ISBI, AAPM, and other conferences and institutions have launched many DL-related medical image analysis challenges. These aim to design and develop automatic or semi-automatic algorithms and promote medical image analysis research with computer-aided methods. Concurrently, some researchers and institutions also organize projects to collect medical datasets and publish them for research purposes.

Despite all these developments, it is still challenging for novice medical image analysis researchers to find medical data.
This paper addresses this challenge and presents a comprehensive survey of existing medical datasets. The paper also identifies and summarizes medical image analysis challenges.
It also provides a pathway to identify the most relevant datasets for evaluation and the suitable methods they need in the respective challenge leader board.

This paper refers to other research papers with a number between square brackets and refers to the datasets listed in the tables with numbers between parentheses.

The following sections present the details of the key datasets and challenges.
\textbf{Section \ref{sec:brief-summary}} summarizes the datasets and challenges, including the years, body parts, tasks, and other information that is relevant to the dataset development.
\textbf{Section \ref{sec:head}} discusses the datasets and challenges of the head and neck.
\textbf{Section \ref{sec:chest-abdomen}} covers the datasets and challenges related to the chest and abdomen organs.
\textbf{Section \ref{sec:path-blood}} examines the datasets and challenges of pathology and blood related tasks.
\textbf{Section \ref{sec:other}} introduces other datasets and challenges related to bone, skin, phantom, and animals.
We have also created a \href{https://medical-image-datasets-list.wareless.group/}{website} with a git \href{https://github.com/ETVP/medical-image-list}{repo}\footnote{The website with the git repo will be public after the paper is accepted.}, which shows the list of these datasets and their respective challenges.


\begin{figure*}
    \centering
    \includegraphics[width=.95\linewidth]{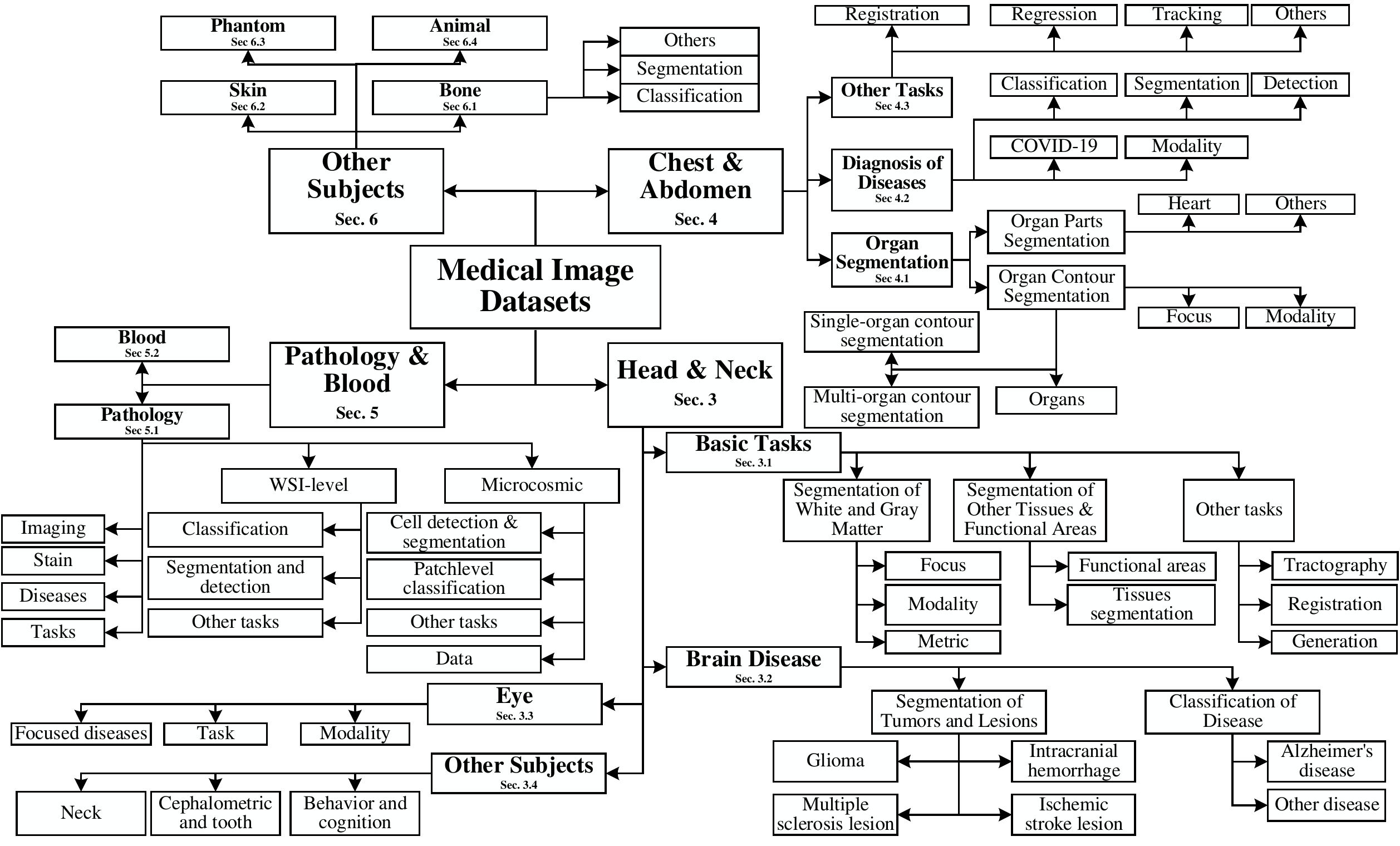}
    \caption{An overall taxonomy to outline the organization of the paper.}
    \label{fig:taxonomy}
\end{figure*}

\begin{figure*}
    \centering
    \makebox[\textwidth][c]{\includegraphics[width=\textwidth]{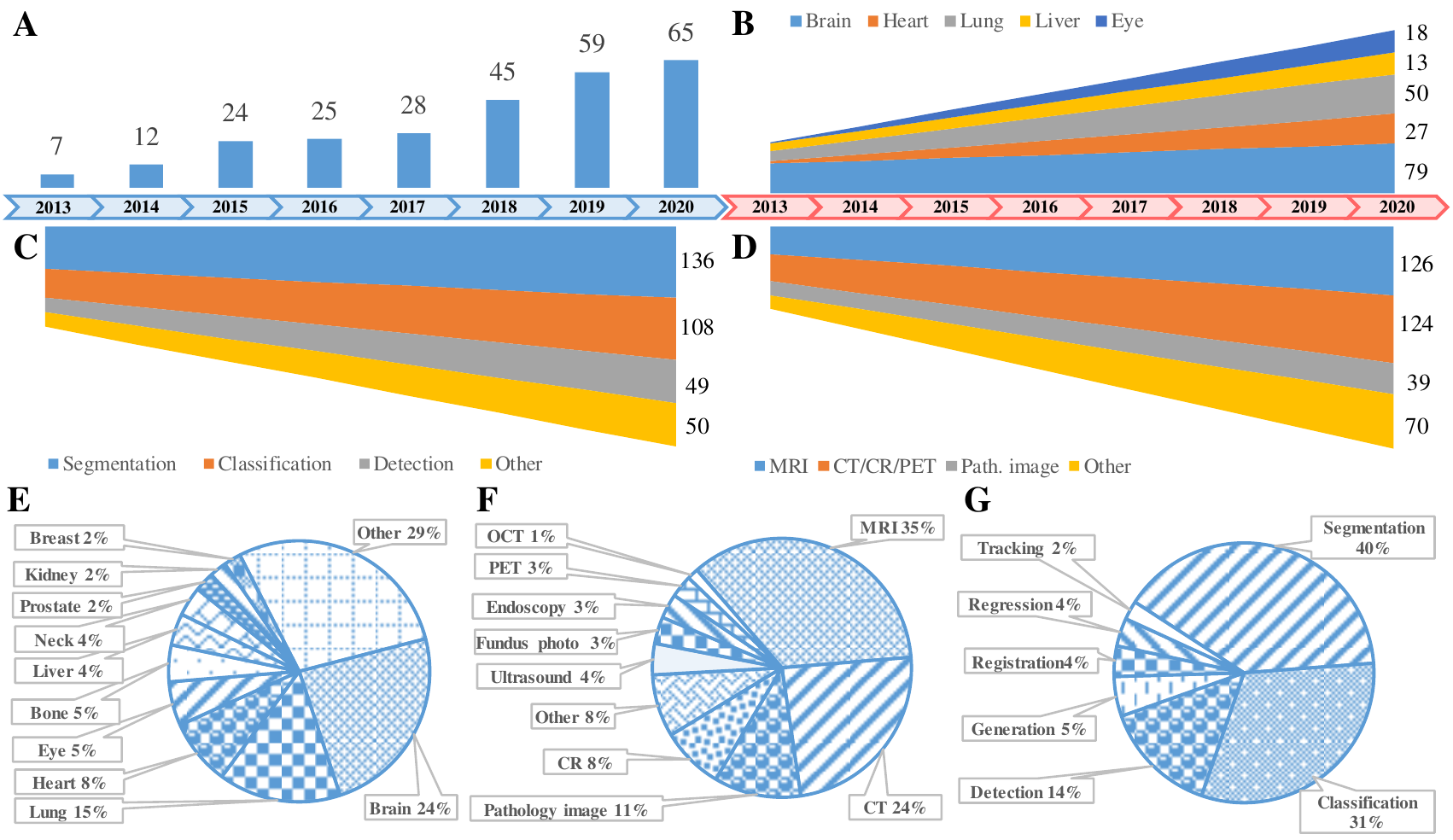}}
    \caption{Summary of medical image datasets and challenges from 2013 to 2020. Figure \ref{fig:brief}\textbf{A} shows the number of datasets and challenges published in each year. Figures \ref{fig:brief} \textbf{B}, \textbf{C}, and \textbf{D} show the year-by-year trends along with the trends in relative numbers for each of the different categories by year. The numbers listed right are the summary count of each category, and the summary counts are not the same with the total numbers, because of \textbf{1)} some of the categories are not shown, and \textbf{2)} a dataset counts two times if it includes two of the categories. Figures \ref{fig:brief} \textbf{E}, \textbf{F}, and \textbf{G} show the most predominant body parts, data modalities, and main tasks with the percentage of their respective dataset.}
    \label{fig:brief}
\end{figure*}

\section{Medical image datasets}
\label{sec:brief-summary}

In this section, we provide an overview of the image datasets and challenges.
Our collection contains over three hundred medical image datasets and challenges organized between 2004 and 2020. This paper focuses mainly on the ones between 2013 and 2020.
\textbf{Subsections} \textbf{\ref{sec:brief-summary:year}}, \textbf{\ref{sec:brief-summary:body}}, \textbf{\ref{sec:brief-summary:modality}}, and \textbf{\ref{sec:brief-summary:task}} provide information about the \textbf{year}, \textbf{body parts}, \textbf{modalities}, and \textbf{tasks}, respectively.
In \textbf{Subsection \ref{sec:brief-summary:st}}, we introduce the sources from where we have collected these datasets and the challenges.
Details about the categorization of these image datasets and challenges into four groups are provided in the subsequent sections. We provide a taxonomy of our paper in Figure \ref{fig:taxonomy} to help the reader navigate through the different sections.

\subsection{Years}
\label{sec:brief-summary:year}

The timeline of these medical image datasets can be split into two, starting from 2013 as the watershed, since Krizhevsky et al.'s excellent success in the ILSVRC competition with their AlexNet \cite{Krizhevsky2017} in 2012. The continuous advancement of deep learning has, to some extent, driven more and more researchers to focus on medical image analysis and indirectly led to an increase in the number of datasets and competitions each year. The number of datasets and challenges before 2013 are irregular according to our statistics.
The main reason is that many datasets developed before 2012 are not aimed at computer-aided diagnosis, for example, ADNI-1 \dRef{data:ADNI1}, although those data could be used for DL. Therefore, we only focus on the datasets and challenges which were released after 2013.

Figure \ref{fig:brief}A shows the statistics of the datasets and challenges per year between 2013 and 2020.
As shown in Figure \ref{fig:brief}A, the number of related datasets and challenges increased year by year because of the progress and success of DL in computer vision and medical image analysis. That led more and more researchers to focus on medical image analysis with DL-based methods and more and more datasets and challenges with different body parts and tasks started to appear.

As shown in Figures \ref{fig:brief}B and \ref{fig:brief}C, there was not only an increase in the number of datasets and challenges but also in their variety with respect to the body parts and types of tasks. The research focus ranges from a simple diagnosis or structural analysis (e.\,g., segmentation and classification) in the early stages to more complex tasks or combinations of tasks that are closer to the clinical needs, including classification, segmentation, detection, regression, generation, tracking, and registration, as time progresses. The focus of these datasets and challenges has also changed from cancer diagnosis to the entire healthcare system. Meanwhile, the organs focused on by researchers also range from the single and simple, but important ones, such as the brain and lungs, to many different other parts of the human body accounting for different sizes, shapes, and other characteristics.

\subsection{Body parts}
\label{sec:brief-summary:body}

With the success of DL, the number of focused body parts has increased, as shown in Figure \ref{fig:brief}B. We also show the most targeted researched body parts in Figure \ref{fig:brief}E, and the top-5 researched organs include the brain, lung, heart, eye, and liver. These organs have been the focus of research because they are the most important parts of the human body.

In the beginning, the main reason which motivated researchers to focus on these organs and parts was that a simple diagnosis and a structural study greatly helped in the diagnosis and treatment of cancer (a major threat to human life). Many datasets focus on brain, lung and other organs, without considering DL, and many challenges focus on simple tasks, such as segmentation and classification.
Subsequently, AI showed to be more competent to tackle complex tasks, and therefore researchers started to focus on several other organs.
For example, eye related diseases, which cause blindness, incited the collection of eye related datasets and the release of challenges. Some other datasets and challenges focus on the small organs, such as the prostate, which are challenging to analyze due to the low resolution of images.

\subsection{Modalities}
\label{sec:brief-summary:modality}

There are several types of medical image modalities.
As shown in Figure \ref{fig:brief}F, the frequently used modalities to acquire medical datasets include Magnetic Resonance Images (MRI), Computed Tomography (CT), Ultrasound (US), Endoscopy, Positron Emission Tomography (PET), Computed Radiography (CR), Electrocardiography, and Optical Coherence Tomography (OCT). We introduce below these main modalities and provide a summary at the end of this subsection.

\paragraph{Radiography:\,}

Radiography is an imaging technique based on the difference of attenuation when X-rays passes through the different organs and tissues of the human body. The primary used modalities include CR and CT. CR is a 2D image, and CT is a volume (3D) image.
Radiography is the most commonly used method to image the human body.
For example, CR is frequently used to diagnose chest related diseases, such as pneumonia, tuberculosis, and COVID-19. Meanwhile, 3D CT plays an important role in the diagnosis and treatment related to cancer and lesions.
The advantages of radiography are \textbf{1)} high resolution of the hard tissues (e.\,g. bones), \textbf{2)} lower cost, and \textbf{3)} compatibility of contrast agents, but the disadvantages are \textbf{1)} X-rays are harmful for human health, \textbf{2)} X-rays are ideal for distinguishing between healthy tissues and tumors without the help of contrast agents, and \textbf{3)} their resolution is limited by the radiation intensity. Moreover, as the main component of the human bone is calcium, CT plays an important role in many bone related diagnoses.

\paragraph{Magnetic resonance:\,}

MR images display the body structure caused by the difference of signal released by the different substances of the imaged organ as the magnetic field is changed. MR has many submodalities, such as T1 and T2. For essential organs and tissues, MR is a commonly used imaging method because it is considered non-invasive, effective, and safe. Due to the principle of MR imaging, MR plays an essential role in the diagnosis of brain, heart, and soft tissues. Because higher resolution MR images can be obtained by increasing the magnetic field strength, MR is also suitable for small organs or tissues.
However, MR images do have disadvantages such as high cost and incompatibility with metal (e.\,g., metallic orthopedic implants).

\paragraph{Nuclear medicine:\,}

Nuclear medicine captures images by the absorption of the targeted tissue of specific chemical components marked by radioactive isotopes.
Tumors and healthy tissues absorb different chemical components, so medics use the specific chemical marked with the radioactive isotope and receive the ray radiated by the chemical.
An example is Positron Emission Computed Tomography, i.\,e., PET, which performs imaging by capturing radiations produced by fluorodeoxyglucose or other similar contrast agents absorbed by the tissue or tumor.
Nuclear medicine is good at imaging regions of interest, such as tumors, but the disadvantage is their high cost and the low-resolution.

\paragraph{Ultrasound:}

Ultrasound operates by acquiring the differences in the absorption and reflection of ultrasound waves when applied to tissues.
It is widely used in imaging the heart and fetus because ultrasound causes no damage to these parts and provides real-time imaging. Nevertheless, the main disadvantage is the noise caused by the reflection of irregular shapes of organs and tissues, and the interference with their imaging.

\paragraph{Eye-related modalities:}

An OCT image is obtained by using low-coherence light to capture 2D and 3D micrometer-resolution images within optical scattering media to diagnose eye-related diseases. The fundus photo is also used for diagnosis purposes. These two methods are non-invasive eye-specific imaging modalities.

\paragraph{Pathology:}

Pathological data is the gold standard in diagnosing diseases. It is taken with microscopy of the stained tissue slides by the camera to show cell-level features. Pathology is used in the cell-level diagnosis for cancer and tumors.

\paragraph{Other modalities:}

Other imaging modalities are usual but specific to certain body parts, such as endoscopy, and provide the medics with various biomarkers to make critical decisions when diagnosing, curing, and researching.

\paragraph*{}
Overall, MR, CT, and other modalities are the most commonly used imaging modalities. MR can provide sharp images without harmful radiations of soft tissues. It is therefore widely used in the imaging of brain, heart and many other small organs. CT is an economical and simple imaging approach, and it is widely used for the diagnosis of cancer, e.\,g., the neck, chest, and abdomen. A pathology image is different from MR and CT, because it is a cell-level imaging method. Pathology is widely used in cancer-related diagnosis.

\subsection{Tasks}
\label{sec:brief-summary:task}

According to our analysis, our collected datasets and challenges have been used for the tasks of classification, prediction, detection, segmentation, location, characterization, up-sampling, tracking, registration, regression, estimation, coding, automatic annotation, and other tasks.
As Figure \ref{fig:brief}G shows, we grouped these tasks into seven categories: classification, segmentation, detection, regression, generation, registration, and tracking. The following subsections briefly describe each task.

\paragraph{Classification:}

Classification is used for qualitative analysis. According to pre-defined specific rules, the classification task aims to group medical images or particular regions of an image into two or more distinct categories.
The classification task can be used alone for medical image analysis or as a subsequent task after other lower level tasks, such as segmentation and detection, in order to analyze the results and further extract features.
There are many ways to express the classification task, such as detection and prediction. The detection tasks (which are also sometimes termed as classification) are different from the ones introduced in the following paragraph, although sometimes the same word is used synonymously.
The typical examples of classification tasks include AD prediction and the attributes classification of pulmonary nodules. AD prediction aims to group MR images in Alzheimer's disease (AD) and normal cognition (NC). The attributes classification of pulmonary nodules aims to analyze the pathology attributes of pulmonary nodules.
Classification performance measures mainly include accuracy, precision, specificity, sensitivity, F-score, ROC, and AUC. All these measures are based on four basic measures: true positive (TP), false positive (FP), true negative (TN), and false negative (FP).

\paragraph{Segmentation:}

The segmentation task can be regarded as a pixel-level or voxel-level classification task, but the difference is that the segmentation task is limited to the context. It aims to split an image into different areas or contour specific regions. The regions can contain tumor, tissue, or other specific targets. The results of the segmentation task consist of areas and boundaries.
Since segmentation can be seen as a pixel-level classification, the average precision (AP) can be used as a metric. Other performance metrics include intersection over union (IoU), Dice index, Jaccard Index, Hausdorff distance, and average surface distance.

\paragraph{Detection:}

The detection task aims to find an object of interest, and it also usually needs to classify such an object (classification task). In this work, we categorize the tasks which aim to determine the location of the object of interest with a bounding box or a point. The detection task is sometimes represented as a localization task. A typical example of detection is pulmonary nodules detection, which aims to find the pulmonary nodules in chest CT images and annotate the nodules with a bounding box. The performance measures used in the detection tasks include mainly the intersection over union (IoU), mean Average Precision (mAP), precision and recall, false positive rate, receiver operating characteristic curve (ROC), and other metrics. For the task to locate an object without the boundary, the Euclidean Distance is the most commonly used measure.

\paragraph{Regression:}

Classification is used for qualitative analysis, while regression is used for quantitative analysis. A typical example is the estimation of the volume of a lesion. For the regression task, the root mean square error, i.\,e., rMSE, mean absolute error, and correlation coefficient are the most commonly used metrics.

\paragraph{Tracking:}

The tracking task aims to locate specific targets, but the tracking is a dynamic process and is therefore different from the detection task. That means the tracking algorithms need to detect or localize targets in different frames. For medical image analysis, the tracking tasks include the tracking of organs and tissues. The tracking is not just of one point, but it can also be of an area, e.\,g., every part of an organ or tissue. An example is the tracking of the lung when the subject breathes.

\paragraph{Generation:}

The image data generation task has many different aims, but for simplicity we categorize all of these aims under the ``generation task'' because they focus on generating image data from other image data. Typical generation tasks include \textbf{1)} to generate a T2-weight image from T1-weight images and \textbf{2)} to generate a pathology image stained with one stain from an image stained by another stain.

\paragraph{Registration:}

The image registration task aims to align one image with another image, i.\,e., to find a transformation (e.\,g., rotation and translation) to align the two images. Registration is a necessary process for computer-aided diagnosis algorithms from multi-modalities. During medical scanning, the movement of the human body cannot be avoided and is a challenge. At the same time, imaging cannot be taken instantly. As a result, images from different viewpoints cannot be aligned directly or when two or more modalities are used. Therefore, researchers rely on registration techniques to solve these alignment problems.

\subsection{Source and Term}
\label{sec:brief-summary:st}

We collected the datasets and challenges mainly from \href{https://www.cancerimagingarchive.net/}{The Cancer Imaging Archive} \cite{Clark2013}, \href{https://grand-challenge.org/}{Grand Challenge}, \href{https://www.kaggle.com/}{Kaggle}, \href{https://openneuro.org/}{OpenNeuro}, \href{https://physionet.org/}{PhysioNet} \cite{Goldberger2000}, and \href{https://competitions.codalab.org/}{Codalab}.

The original records of datasets and challenges that we collected include four to five hundred, and we removed some of them as some datasets are not suitable for DL and AI methods. We then categorized the remaining datasets and challenges into different groups.
Categorizing the datasets and challenges is not easy because all these datasets and challenges are derived from clinical research sources.
Thus we used an asymmetric categorization to group these datasets and challenges into four groups, as shown in Figure \ref{fig:taxonomy}. This means that we did not use the same sub-taxonomy in each category or sub-category.

\textbf{First}, we split the medical datasets and challenges into two groups: body-level and cell-level (\textbf{Section \ref{sec:path-blood}}),
according to the imaged body part. The body-level datasets focus on specific tissues, while the cell-level ones focus on cells.
\textbf{Second}, we grouped the datasets and challenges of the brain, eye, and neck into one group (\textbf{Section \ref{sec:head}}), because these are parts of the head.
\textbf{Third}, we organized the datasets and challenges related to the chest and abdomen into the same group (\textbf{Section \ref{sec:chest-abdomen}}). These datasets and challenges relate to the diagnosis, anatomical segmentation, and treatments.
\textbf{Finally}, for the datasets and challenges that cannot be categorized into the above groups, we grouped them under ``other'' (\textbf{Section \ref{sec:other}}), and these datasets and challenges are related to the skin, bone, phantom, and animals.

The introduction of each group and sub-group includes mainly the type of modality, the task, the disease, and the body part. However, not all the groups of datasets can be introduced in that way. For some groups, we introduce the datasets and challenges according to the domain-specific problems. For example, we categorize the pathology datasets into microcosmic and macrocosmic tasks.


\section{Head and neck related datasets and challenges}
\label{sec:head}

The head and neck are significant parts of the human body because many essential organs, glands, and tissues are located there. Several researchers' image analysis work relate to the head and neck. To make an effective use of computers for research, diagnosis, and treatment, many researchers have released datasets and challenges, for examples: 1) \textbf{the analysis of tissue structure and functions} \dRef{data:iseg2019,data:iseg2017,data:MRBrainS18,data:MRBrainS13} and 2) \textbf{diseases diagnosis} \dRef{data:brats20,data:ISLES2018,data:lgg1p19q}.

Because the brain controls emotions, actions and functions of other organs, the brain's area is significant.
\textbf{First}, we introduce the datasets and challenges related to the analysis of the brain structure, function, imaging, and other basic tasks in Subsection \ref{sec:head:basic}. \textbf{Second}, we introduce the datasets and challenges related to brain disease diagnosis in Subsection \ref{sec:head:disease}.

Moreover, since the eyes are crucial to our vision, the computer-aided diagnosis of eye-related diseases is also an important research focus.
The eye-related datasets and challenges are covered in Subsection \ref{sec:head:eye}.
We introduce other datasets and challenges of the neck and the datasets related to the brain's behavior and cognition in Subsection \ref{sec:head:other}.

\begin{sidewaystable*}
	\centering
	\caption{Summary of datasets and challenges for the basic brain image analysis.}
	\label{tab:data:gwmatter-seg}
	\small
		\begin{tabular}{clccccclccc}
			\toprule 
			\textsc{\textbf{Reference}} & \textsc{\textbf{Dataset/Challenge}} & \textsc{\textbf{Year}} & \multicolumn{4}{l}{\textsc{\textbf{Modalities}}} & \textsc{\textbf{Focus}} & \multicolumn{3}{l}{\textsc{\textbf{Tasks}}} \\ 
			\cline{4-7} \cline{9-11}
			\textsc{\textbf{Index}} & & & T1 & T2 & dMR & OT & & Segmentation & Generation & Registration \\
			\midrule
			\dataindex{data:cerebrum} & \href{https://openneuro.org/datasets/ds002207/versions/1.0.0}{CEREBRuM} \cite{Bontempi2019} & 2019 & \checkmark & & & & WM, GM, CSF & \checkmark & & \\ 
			\hline
			\dataindex{data:iseg2019} & \href{http://iseg2019.web.unc.edu/}{iSeg 2019} \cite{Sun2021} & 2019 & \checkmark & \checkmark & & & WM, GM, CSF & \checkmark & & \\ 
			\hline
			\dataindex{data:iseg2017} & \href{http://iseg2017.web.unc.edu/}{iSeg 2017} \cite{Wang2019} & 2017 & \checkmark & \checkmark & & & WM, GM, CSF & \checkmark & & \\ 
			\hline
			\dataindex{data:MRBrainS18} & \href{https://mrbrains18.isi.uu.nl/}{MRBrainS 18} & 2018 & \checkmark \tnote{tnote:data:gwmatter-seg:1} & \checkmark \tnote{tnote:data:gwmatter-seg:2} & & & WM, GM, CSF & \checkmark & & \\ 
			\hline
			\dataindex{data:NEATBrainS15} & \href{https://www.isi.uu.nl/research/challenges/neatbrains/}{NEATBrainS 15} \cite{NEATBrainS} & 2015 & \checkmark \tnote{tnote:data:gwmatter-seg:1} & \checkmark \tnote{tnote:data:gwmatter-seg:2} & & & WM, GM, CSF & \checkmark & & \\
			\hline
			\dataindex{data:MRBrainS13} & \href{https://mrbrains13.isi.uu.nl/}{MRBrainS 13} \cite{Mendrik2015} & 2013 & \checkmark \tnote{tnote:data:gwmatter-seg:1} & \checkmark \tnote{tnote:data:gwmatter-seg:2} & & & WM, GM, CSF & \checkmark & & \\ 
			\hline
			\dataindex{data:NeoBrainS12} & \href{https://neobrains12.isi.uu.nl/}{Neonatal MRBrainS 12} \cite{Isgum2015} & 2012 & \checkmark & \checkmark & & & WM, GM, CSF & \checkmark & & \\ 
			\hline
			\dataindex{data:WHM-Seg-C} & \href{https://wmh.isi.uu.nl/}{WMH Seg. Chlg.} \cite{Kuijf2019} & 2017 & \checkmark & & & & WM, GM, CSF & \checkmark & & \\ 
			\hline
			\dataindex{data:ENIGMA-cerebellum} & \href{https://my.vanderbilt.edu/enigmacerebellum/}{ENIGMA Cerebellum} & 2017 & & & \checkmark & & Cerebellum & \checkmark & &  \\ 
			\hline
			\dataindex{data:mindboggle} & \href{https://mindboggle.info/data.html}{Mindboggle} \cite{Klein2012} & 2012 & \checkmark & & & & Brain atlases & \checkmark & &  \\ 
			\hline
			\dataindex{data:neuromorphometrics} & \href{http://www.neuromorphometrics.com/?page_id=23}{Labeled Brain Scans} & 2012 & \checkmark & & & & Brain atlases & \checkmark & &  \\ 
			\hline
			\dataindex{data:cause07} & \href{https://cause07.grand-challenge.org/}{CAUSE 07} \cite{VanGinneken2007} & 2007 & \checkmark & & & & Caudate & \checkmark & &  \\ 
			\hline
			\dataindex{data:autoimplant}& \href{https://autoimplant.grand-challenge.org/}{AutoImplant} \cite{Li2020} & 2020 & & & & \checkmark\tnote{tnote:data:gwmatter-seg:3} & Cranioplasty  & & \checkmark &  \\ 
			\hline
			\dataindex{data:accelmr} & \href{https://accelmr.org/}{AccelMR 2020}\cite{AccelMR2020}  & 2020 & \checkmark & \checkmark & & & Non-linear mapping of different resolutions & & \checkmark & \\ 
			\hline
			\dataindex{data:memento} & \href{https://my.vanderbilt.edu/memento/}{MRI WM Reconstruction} \cite{MEMENTO} & 2020 & && \checkmark & & White matter reconstruction & & \checkmark & \\ 
			\hline
			\dataindex{data:ccpd} & \href{https://sites.google.com/view/calgary-campinas-dataset/home/mr-reconstruction-challenge}{Calgary Campinas Brain Dataset} \cite{Souza2018} & 2020 & \checkmark & & & & Brain image reconstruction & & \checkmark & \\ 
			\hline
			\dataindex{data:mc-mrrec} & \href{https://sites.google.com/view/calgary-campinas-dataset/mr-reconstruction-challenge}{MRI Reconstruction Challenge} & 2020 & \checkmark & & & & MR reconstruction & & \checkmark & \\ 
			\hline
			\dataindex{data:fastmri} & \href{https://fastmri.org/}{FastMRI} \cite{Zbontar2018} & 2018 & \checkmark & \checkmark & & & Accelerating magnetic resonance imaging & & \checkmark & \\ 
			\hline
			\dataindex{data:MUSHAC} & \href{https://projects.iq.harvard.edu/cdmri2018/challenge}{MUSHAC 2018} & 2018 & & & \checkmark & & DW MRI registration and enhancement & & \checkmark & \checkmark \\ 
			\hline
			\dataindex{data:HARDI2013} & \href{http://hardi.epfl.ch/static/events/2013_ISBI/}{HARDI 2013} & 2013 & & & \checkmark & & Diffusion MRI reconstruction & & \checkmark &\\
			\hline
			\dataindex{data:HARDI2012} & \href{http://hardi.epfl.ch/static/events/2012_ISBI/}{HARDI 2012} & 2012 & & & \checkmark & & Diffusion MRI reconstruction & & \checkmark &\\
			\hline
			\dataindex{data:curious2019} & \href{https://curious2019.grand-challenge.org/}{CuRIOUS 2019} & 2019 & & & \checkmark & \checkmark\tnote{tnote:data:gwmatter-seg:4} & Image registration & & & \checkmark \\
			\hline
			\dataindex{data:curious2018} & \href{https://curious2018.grand-challenge.org/}{CuRIOUS 2018} & 2018 & & & \checkmark & \checkmark\tnote{tnote:data:gwmatter-seg:4} & Image registration & & & \checkmark \\
            \hline
            \dataindex{data:3dvotem} & \href{https://my.vanderbilt.edu/votem/}{3D VoTEM} & 2018 & \checkmark & & & & Fiber tractography & \checkmark & & \\ 
            \hline
            \dataindex{data:dti2012} & \href{https://projects.iq.harvard.edu/dti_challenge/}{DTI Tractography Challenge 2012} & 2012 & & & \checkmark & & DTI tractography & \checkmark & & \\ 
            \hline
            \dataindex{data:DTIT2011} & \href{http://www.na-mic.org/Wiki/index.php/Events:_DTI_Tractography_Challenge_MICCAI_2011}{DTI Tractography Challenge 2011} & 2011 & & & \checkmark & & DTI tractography & \checkmark & & \\ 
			\bottomrule
		\end{tabular}
		\raggedright
		\begin{tablenotes}{llll}
			\item[tnote:data:gwmatter-seg:1] T1w and T1w-IR &
			\item[tnote:data:gwmatter-seg:2] T2-FLAIR       &
			\item[tnote:data:gwmatter-seg:3] CT             &
			\item[tnote:data:gwmatter-seg:4] Ultrasound
		\end{tablenotes}
\end{sidewaystable*}

\subsection{Structural analysis tasks of the brain}
\label{sec:head:basic}

The basic analysis and processing of the brain medical images are clinically critical for the diagnosis, treatment, and other brain-related analysis tasks. The datasets and challenges we discuss are mainly for the segmentation tasks and center around the brain structure. In contrast, some datasets focus on imaging, including MR imaging acceleration, the non-linear registration of different resolutions, and tissue reconstruction. One of the most popular tasks is the segmentation of white matter (WM), gray matter (G<), and cerebrospinal fluid (CSF), and their respective datasets and challenges are introduced in Subsection \ref{sec:head:basic:wg-seg}. Meanwhile, other tissues and functional areas' segmentation are also the focus of research, and their related datasets and challenges are discussed in Subsection \ref{sec:head:basic:area-seg}. Subsection \ref{sec:head:basic:oth} describes the other basic tasks. Table \ref{tab:data:gwmatter-seg} shows the datasets and challenges of these basic tasks.

\subsubsection{Segmentation of white and gray matter}
\label{sec:head:basic:wg-seg}

The segmentation of WM, GM, and CSF has great significance for brain structure research and computer-aided diagnosis, particularly using AI. Similarly, for AI algorithms, it is also of great significance to understand the human brain's structure. Therefore, MICCAI and others have held many challenges with this research focus, and researchers could design automatic algorithms to segment magnetic resonance images into different parts. We introduce these datasets and challenges with respect to their modalities and tasks.

\paragraph{Modality:} The datasets and challenges which focus on the WM, GM, and CSF segmentation, usually provide MR images. Challenges \dRef{data:iseg2019,data:iseg2017,data:MRBrainS18,data:NEATBrainS15,data:MRBrainS13,data:NeoBrainS12} provide mainly two modalities: T1, T2, while datasets \dRef{data:WHM-Seg-C,data:cerebrum} only provide T1 for the white matter hyperintensities segmentation task. Note that, MR scans are sensitive to the hydrogen atom, and such a feature can effectively help image analysts to distinguish between different tissues and parts of the image. Moreover, due to the color of the tissue imaged by MR, these scans are named as ``white matter'' and ``gray matter''.

\paragraph{Task:} The main focus of these datasets and challenges is the segmentation of WM, GM, and CSF. However, they do not only focus on that.
Challenges \dRef{data:MRBrainS18,data:NEATBrainS15,data:MRBrainS13,data:NeoBrainS12,data:cerebrum} also provide the annotation of other parts of the brain, including basal ganglia, white matter lesions, cerebellum, and infarction.
One of the challenge for segmentation is the presence of a lesion because of the unnatural characterization of lesions. A well-annotated data can help AI to overcome this problem and also achieve more robust results.
Challenges \dRef{data:NEATBrainS15,data:NeoBrainS12} use MR images of the neonatal brain, and consider tissue volumes as an indicator of long-term neurodevelopmental performance \cite{Isgum2015}.

\paragraph{Performance metric:} For the segmentation task, the Dice score is one of the most commonly used metrics, and all these datasets and challenges adopt it as a performance measure. Besides the Dice score, datasets \dRef{data:MRBrainS18,data:MRBrainS13,data:WHM-Seg-C} also use Hausdorff distance and volumetric similarity as metrics; datasets \dRef{data:iseg2019,data:iseg2017} use the average the Hausdorff distance and the average surface distance as one of their metrics; moreover, dataset \dRef{data:WHM-Seg-C} also uses sensitivity and F1-score as metrics for performance evaluation.

\subsubsection{Segmentation of functional areas \& other tissues}
\label{sec:head:basic:area-seg}

The segmentation of functional areas and tissues has also an essential meaning for brain-related research and computer-aided diagnosis. In this subsection, we introduce the datasets and challenges that are related to the segmentation of functional areas and tissues.

\paragraph{Tissues segmentation:}

While, WM, GM, and CSF were introduced in Subsection \ref{sec:head:basic:wg-seg}, the segmentation of other brain tissues is also an active research area. Challenges \dRef{data:MRBrainS18,data:MRBrainS13,data:NeoBrainS12,data:cerebrum} aim to segment brain images into different tissues, including ventricles, cerebellum, brainstem, and basal ganglia. These challenges provide MR images and the voxel-level annotations of the regions of interest with thirty or forty scans.
Because these regions are essential for brain health, researchers need to overcome the challenges related to their size and shape in order to segment them.
Dataset \dRef{data:ENIGMA-cerebellum} focuses on the cerebellum segmentation from the diffusion-weighted image (DWI), while dataset \dRef{data:cause07} focuses on the segmentation of caudate from the brain MR image.

\paragraph{Functional areas:}

The segmentation of the human brain cortex into different functional areas is of great significance in education, clinical research, treatment, and other applications. Datasets \dRef{data:mindboggle,data:neuromorphometrics} provide images and annotations for the design of automatic algorithms to segment the brain cortex into different functional areas. Dataset \dRef{data:mindboggle} uses DTK protocol \cite{Klein2012}, which is modified from DK protocol \cite{Desikan2006}, and the DTK protocol includes 31 labels, details of which are listed in \url{https://mindboggle.readthedocs.io/en/latest/labels.html}. Dataset \dRef{data:neuromorphometrics} is a commercial dataset for research in the segmentation of functional areas of the brain cortex.

\subsubsection{Imaging-related tasks}
\label{sec:head:basic:oth}

In addition to the segmentation tasks of the brain tissues and the functional areas, some of the datasets and challenges also focus on the generation, registration, and tractography.

\paragraph{Generation:}

Datasets and challenges \dRef{data:fastmri,data:accelmr,data:mc-mrrec} aim to accelerate MR imaging or generate high-resolution MR images from low-resolution ones. Usually, high-resolution imaging requires higher cost, while low-resolution imaging is cheaper but affects the analytical judgment and may lead to an incorrect diagnosis. These challenges provide many scans at low-resolution to allow researchers to design algorithms to convert or map low-resolution images onto higher-resolution ones. The datasets and challenges mainly focus on the generation tasks.
Another focus is the cranioplasty \dRef{data:autoimplant} to generate a part of broken skull from CT images of the models of the broken skull.
Other datasets and challenges \dRef{data:HARDI2012,data:HARDI2013,data:MUSHAC,data:memento} focus on the reconstruction of MR images.

\paragraph{Registration:}

The registration between different modalities is another research focus. Challenges \dRef{data:curious2019,data:curious2018} focus on the registration between ultrasound data and MR images of the brain.
Cross-modality registration is difficult because the subject is not absolutely static. Moreover, the MR is a 3D volume imaging modality and hence is different from ultrasound, which is a 2D imaging modality. Thus, these challenges focus on establishing the topological relation between Preoperative MR image and intraoperative ultrasound.
Challenge \dRef{data:MUSHAC} also focuses on the diffusion MR image registration to eliminate differences between different vendors' hardware devices and protocols.

\paragraph{Tractography:}

Tractography is another segmentation task and focuses on the segmentation and imaging of the fiber in the WM. Dataset \dRef{data:3dvotem} aims to segment the fiber bundles from brain images, including phantom, squirrel monkey, and macaque, while challenges \dRef{data:dti2012,data:DTIT2011} focus on the tractography with DTI, another type of MR image.

\subsection{Brain diseases related datasets and challenges}
\label{sec:head:disease}

Besides the structural analysis and image processing tasks, computer-aided diagnosis is also a research focus in healthcare. Medical image analysis plays a critical role in clinical research, diagnosis, and treatment. The datasets and challenges we have included are for two tasks: \textbf{1)} the segmentation of lesions and tumors and \textbf{2)} the classification of diseases. For the segmentation task, the respective datasets and challenges focus on the tumor and lesion segmentation of the human brain, mark the lesion's contour for diagnosis and treatment, and the relevant details are shown in Subsection \ref{sec:head:disease:tumor}. For classification tasks, the datasets and challenges have been used for the development of automatic algorithms to classify or predict diseases from medical images, and these datasets and challenges are presented in Subsection \ref{sec:head:disease:cls}.

\begin{sidewaystable*}
	\centering
	\caption{Summary of datasets and challenges for the brain lesion and tumor segmentation task.}
	\label{tab:data:brain-tumor}
	\small
		\begin{tabular}{clcccclcccc}
			\toprule 
			\textsc{\textbf{Reference}} & \textsc{\textbf{Dataset/Challenge}} & \textsc{\textbf{Year}} & \multicolumn{3}{l}{\textsc{\textbf{Modalities}}} & \textsc{\textbf{Focus}} & \multicolumn{4}{l}{\textsc{\textbf{Lesion/Tumor}}} \\ 
			\cline{4-6} \cline{8-11}
			\textsc{\textbf{Index}} & & & T1 & T2 & OT & & \twol{brain}{tumor} & \twol{stroke}{lesion} & \twol{intracranial}{hemorrhage} & \twol{sclerosis}{lesion} \\
			\midrule
			\dataindex{data:cada-as} & \href{https://cada-as.grand-challenge.org/}{CADA-AS} & 2020 & & & \checkmark\tnote{tnote:data:brain-tumor:1} & Cerebral aneurysm & \checkmark & & & \\ 
			\hline
			\dataindex{data:cada-rre} & \href{https://cada-rre.grand-challenge.org/}{CADA-RRE} & 2020 & & & \checkmark\tnote{tnote:data:brain-tumor:1} & Cerebral aneurysm & \checkmark & & & \\ 
			\hline
			\dataindex{data:cada} & \href{https://cada-rre.grand-challenge.org/}{CADA} & 2020 & & & \checkmark\tnote{tnote:data:brain-tumor:1} &  Cerebral aneurysm& \checkmark & & & \\ 
			\hline
			\dataindex{data:brats20} & \href{https://www.med.upenn.edu/cbica/brats2020/}{BraTS 2020} \cite{Menze2015,Bakas2017,Bakas2018} & 2020 & \checkmark\tnote{tnote:data:brain-tumor:2.1} & \checkmark\tnote{tnote:data:brain-tumor:3.1} & & Mutli-modalities & \checkmark & & & \\ 
			\hline
			\dataindex{data:brats19} & \href{https://www.med.upenn.edu/cbica/brats-2019/}{BraTS 2019} \cite{Menze2015,Bakas2017,Bakas2018} & 2019 & \checkmark\tnote{tnote:data:brain-tumor:2.1} & \checkmark\tnote{tnote:data:brain-tumor:3.1} & & Mutli-modalities & \checkmark & & & \\ 
			\hline
			\dataindex{data:brats18} & \href{https://www.med.upenn.edu/sbia/brats2018.html}{BraTS 2018} \cite{Bakas2018} & 2018 & \checkmark\tnote{tnote:data:brain-tumor:2.1} & \checkmark\tnote{tnote:data:brain-tumor:3.1} & & Mutli-modalities & \checkmark & & & \\ 
			\hline
			\dataindex{data:brats17} & \href{https://www.med.upenn.edu/sbia/brats2017.html}{BraTS 2017} \cite{Bakas2017a} & 2017 & \checkmark\tnote{tnote:data:brain-tumor:2.1} & \checkmark\tnote{tnote:data:brain-tumor:3.1} & & Mutli-modalities & \checkmark & & & \\ 
			\hline
			\dataindex{data:brats16} & \href{https://sites.google.com/site/braintumorsegmentation/home/brats_2016}{BraTS 2016} \cite{Menze2015} & 2016 & \checkmark\tnote{tnote:data:brain-tumor:2.2} & \checkmark\tnote{tnote:data:brain-tumor:3.2} & & Mutli-modalities & \checkmark & & & \\ 
			\hline
			\dataindex{data:brats15} & \href{https://sites.google.com/site/braintumorsegmentation/home/brats2015}{BraTS 2015} \cite{Menze2015} & 2015 & \checkmark\tnote{tnote:data:brain-tumor:2.2} & \checkmark\tnote{tnote:data:brain-tumor:3.2} & & Mutli-modalities & \checkmark & & & \\ 
			\hline
			\dataindex{data:brats14} & \href{https://sites.google.com/site/miccaibrats2014/}{BraTS 2014} \cite{Menze2015} & 2014 & \checkmark\tnote{tnote:data:brain-tumor:2.2} & \checkmark\tnote{tnote:data:brain-tumor:3.2} & & Mutli-modalities & \checkmark & & & \\ 
			\hline
			\dataindex{data:brats13} & \href{https://www.smir.ch/BRATS/Start2013}{BraTS 2013} \cite{Menze2015} & 2013 & \checkmark\tnote{tnote:data:brain-tumor:2.2} & \checkmark\tnote{tnote:data:brain-tumor:3.2} & & Mutli-modalities & \checkmark & & & \\ 
			\hline
			\dataindex{data:brats12} & \href{http://www2.imm.dtu.dk/projects/BRATS2012/}{BraTS 2012} \cite{Menze2015} & 2012 & \checkmark\tnote{tnote:data:brain-tumor:2.2} & \checkmark\tnote{tnote:data:brain-tumor:3.2} & & Mutli-modalities & \checkmark & & & \\ 
			\hline
			\dataindex{data:ISLES2018} & \href{http://www.isles-challenge.org/}{ISLES 2018} & 2018 & & & \checkmark\tnote{tnote:data:brain-tumor:4} & Ischemic stroke & & \checkmark & & \\ 
			\hline
			\dataindex{data:ISLES2017} & \href{http://www.isles-challenge.org/ISLES2017/}{ISLES 2017} & 2017 &  & \checkmark & \checkmark\tnote{tnote:data:brain-tumor:5} & Ischemic stroke & & \checkmark & & \\ 
			\hline
			\dataindex{data:ISLES2016} & \href{http://www.isles-challenge.org/ISLES2016/}{ISLES 2016} & 2016 &  & & \checkmark\tnote{tnote:data:brain-tumor:6} & Ischemic stroke & & \checkmark & & \\ 
			\hline
			\dataindex{data:ISLES2015} & \href{http://www.isles-challenge.org/ISLES2015/}{ISLES 2015} \cite{Maier2017} & 2015 & \checkmark & \checkmark & \checkmark\tnote{tnote:data:brain-tumor:7} & Ischemic stroke & & \checkmark & & \\ 
			\hline
			\dataindex{data:ct-ich} & \href{https://physionet.org/content/ct-ich/1.3.1/}{CT-ICH} \cite{Hssayeni2020} & 2020 & & & \checkmark\tnote{tnote:data:brain-tumor:8} & Intracranial hemorrhage & & & \checkmark & \\ 
			\hline
			\dataindex{data:RSNA-IHD} & \href{https://www.kaggle.com/c/rsna-intracranial-hemorrhage-detection/overview}{RSNA Intracranial Hemorrhage Detection} & 2019 & & & \checkmark\tnote{tnote:data:brain-tumor:8} & Intracranial hemorrhage & & & \checkmark & \\
			\hline
			\dataindex{data:HeadCT–hemorrhage} & \href{https://www.kaggle.com/felipekitamura/head-ct-hemorrhage}{Head CT - Hemorrhage} & 2019 & & & \checkmark\tnote{tnote:data:brain-tumor:8} & Intracranial hemorrhage & & & \checkmark & \\
			\hline
			\dataindex{data:schmainda2018} & \href{https://wiki.cancerimagingarchive.net/display/Public/Brain-Tumor-Progression}{Brain Tumor Progression} \cite{Schmainda2018} & 2018 & \checkmark & \checkmark & \checkmark\tnote{tnote:data:brain-tumor:7} & Brain tumor & \checkmark & & & \\ 
			\hline
			\dataindex{data:lgg1p19q} & \href{https://wiki.cancerimagingarchive.net/display/Public/LGG-1p19qDeletion}{LGG-1p19qDeletion} \cite{Akkus2017, Erickson2017} & 2017 & \checkmark & \checkmark & & Low-grade gliomas & \checkmark & & & \\ 
			\hline
			\dataindex{data:atlas} & \href{http://fcon_1000.projects.nitrc.org/indi/retro/atlas.html}{ATLAS} \cite{Liew2017} & 2017 & \checkmark & & & Anatomical segmentation & & \checkmark & & \\
			\hline
			\dataindex{data:MSSEG} & \href{https://portal.fli-iam.irisa.fr/msseg-challenge/overview}{MSSEG Challenge} \cite{Commowick2018,Commowick2018a} & 2016 & \checkmark & \checkmark & \checkmark\tnote{tnote:data:brain-tumor:9} & Multiple sclerosis & & & & \checkmark \\ 
			\hline
			\dataindex{data:LMSLSC2015} & \href{http://iacl.ece.jhu.edu/index.php/MSChallenge}{MSChallenge 2015} \cite{Carass2017,Carass2017a} & 2015 & \checkmark & \checkmark & \checkmark\tnote{tnote:data:brain-tumor:10} & Longitudinal multiple sclerosis & & & & \checkmark \\ 
			\hline
			\dataindex{data:MSlesion2008} & \href{http://www.ia.unc.edu/MSseg/}{MSSeg 2008} & 2008 & \checkmark & \checkmark & \checkmark \tnote{tnote:data:brain-tumor:7} & Multiple sclerosis & & & & \checkmark \\ 
			\bottomrule
		\end{tabular}
		\raggedright
		\begin{tablenotes}{llll}
				\multicolumn{2}{l}{\item[tnote:data:brain-tumor:2.1] For BraTS 17 to 20, T1 modality includes T1 image and T1Gd image.} &
				\multicolumn{2}{l}{\item[tnote:data:brain-tumor:3.1] For BraTS 17 to 20, T2 modality includes T2 image and T2-FLAIR image.} \\
				\multicolumn{2}{l}{\item[tnote:data:brain-tumor:2.2] For BraTS 12 to 16, T1 modality includes T1 image and T1c image.} &
				\multicolumn{2}{l}{\item[tnote:data:brain-tumor:3.2] For BraTS 12 to 16, T2 modality includes T2 image and T2w-FLAIR image.} \\
				\item[tnote:data:brain-tumor:1] MR Angiography             &
				\item[tnote:data:brain-tumor:4] DWI and Perfusion MR image &
				\item[tnote:data:brain-tumor:5] T2w and FLAIR              &
				\item[tnote:data:brain-tumor:6] ADC and Perfusion MR image \\
				\item[tnote:data:brain-tumor:7] FLAIR                      &
				\item[tnote:data:brain-tumor:8] CT                         &
				\item[tnote:data:brain-tumor:9] DP/T2 and FLAIR            &
				\item[tnote:data:brain-tumor:10] PDw and FLAIR
		\end{tablenotes}
\end{sidewaystable*}

\subsubsection{Datasets for segmentation of tumors and lesions}
\label{sec:head:disease:tumor}

Tumors and lesions in the brain affect human's healthy life and safety, and image analysis is an effective way to diagnose the relevant diseases. In this subsection, related datasets and challenges are introduced, and they are reported in Table \ref{tab:data:brain-tumor}.

\paragraph{Glioma datasets and challenges:}

Gliomas are one of the most common brain malignancies for adults. Therefore, many challenges and datasets focus on the segmentation of glioma for its diagnosis and treatment.
BraTS challenge series \dRef{data:brats13,data:brats12,data:brats14,data:brats15,data:brats16,data:brats17,data:brats18,data:brats19,data:brats20} have been occurring since 2012 to segment the glioma. The challenges of such a segmentation task are caused by the heterogeneous appearance and shape of gliomas.
The heterogeneity of glioma reflects its shape, modalities, and many different histological sub-regions, such as the peritumoral edema, the necrotic core, enhancing, and the non-enhancing tumor core. Therefore, these series of challenges provide multi-modal MR scans to help researchers design and train algorithms to segment tumors and their sub-regions.
The tasks of this challenge series include \textbf{1)} low- and high-grade glioma segmentation \dRef{data:brats12,data:brats13}, \textbf{2)} survival prediction from pre-operative images \dRef{data:brats17,data:brats18}, and \textbf{3)} the quantification of segmentation uncertainty \dRef{data:brats19,data:brats20}.
Besides the BraTS challenge series, dataset \dRef{data:lgg1p19q} is another one for the segmentation of low-grade glioma and provides T1-weight and T2-weight MR images with biopsy-proven gene status of each subject by fluorescence in-situ hybridization, a.\,k.\,a. FISH \cite{Scheie2006}. Dataset \dRef{data:schmainda2018} focuses on the processing of brain tumor and aims to design and evaluate DL-based automatic algorithms for glioblastoma segmentation and further research.

\paragraph{Ischemic stroke lesion  datasets and challenges:}

Similar to tumor segmentation, brain lesion segmentation also focuses on detecting brain abnormalities. However, the difference is that lesion segmentation deals with damaged tissues. Challenges \dRef{data:ISLES2015,data:ISLES2016,data:ISLES2017,data:ISLES2018,data:atlas} focus on stroke lesion segmentation because stroke is also life-threatening and can disable the surviving patients. Stroke is often associated with high socioeconomic costs and disabilities.
Automatic analysis algorithms help to diagnose and treat stroke, since its manifestation is triggered by local thrombosis, hemodynamic factors, or embolic causes.
In MR images, the infarct core can be identified with diffusion MR images, while the penumbra (which can be treated) can be characterized by perfusion MR images.
The challenge ISLES 2015 \dRef{data:ISLES2015} focuses on sub-acute ischemic stroke lesion segmentation and acute stroke outcome/penumbra estimation and provides 50 and 60 multi-modalities MR scans of data for training and validation, respectively, for two subtasks, i.\,e., sub-acute ischemic stroke lesion segmentation and acute stroke outcome/penumbra estimation. The subsequent year's challenge, ISLES 2016 \dRef{data:ISLES2016}, focuses on the segmentation of lesions and the prediction of the degree of disability. This challenge provides about 70 scans, including clinical parameters and MR modalities, such as DWI, ADC, and perfusion maps. The challenge ISLES 2017 \dRef{data:ISLES2017} focuses on the segmentation with acute MR images, and ISLES 2018 \dRef{data:ISLES2018}, focuses on the segmentation task based on acute CT perfusion data. Moreover, dataset \dRef{data:atlas} focuses on the segmentation of the brain after stroke for further treatments.

\paragraph{Intracranial hemorrhage related datasets:}

Intracranial hemorrhage is another type of medical condition that affects our health. Dataset and challenge \dRef{data:ct-ich,data:RSNA-IHD} focus on the detection and segmentation of intracranial hemorrhage to help medics locate the hemorrhage regions and decide on a treatment plan. Dataset \dRef{data:HeadCT–hemorrhage} also provides data for the classification of normal or hemorrhage CT images.

\paragraph{Multiple sclerosis lesion related datasets:}

Multiple sclerosis lesion is another kind of lesion in the brain which is not life-threatening and deadly but can cause disabilities. Datasets and challenges \dRef{data:LMSLSC2015,data:MSSEG,data:MSlesion2008} are about the multiple sclerosis lesion segmentation with multi-modalities MR data (T1w, T2w, FLAIR, etc.).

\begin{table*}
	\centering
	\caption{Summary of datasets and challenges for brain disease classification tasks. }
	\label{tab:head:dis}
	\small
		\begin{tabular}{clccccccccccm{12em}}
			\toprule
			\textsc{\textbf{Reference}} & \textsc{\textbf{Dataset/Challenge}} & \textsc{\textbf{Year}} & \multicolumn{5}{l}{\textsc{\textbf{Modalities}}} & & \multicolumn{3}{l}{\textsc{\textbf{Diseases}}} & \textsc{\textbf{Category}} \\ 
			\cline{4-8} \cline{10-12}
			\textsc{\textbf{Index}} & & & T1 & T2 & DWI & PT & OT & & AD & PD & OT &  \\
			\midrule
			\dataindex{data:ADNI1} & \href{http://adni.loni.usc.edu/}{ADNI-1} \cite{Mueller2005} & 2004 & \checkmark & \checkmark & \checkmark & \checkmark & & & \checkmark & & & NC, MCI, AD \\ 
			\hline
			\dataindex{data:ADNIGO} & \href{http://adni.loni.usc.edu/}{ADNI-GO} \cite{Mueller2005} & 2009 & \checkmark & \checkmark & \checkmark & \checkmark & & & \checkmark & & & NC, MCI, AD \\ 
			\hline
			\dataindex{data:ADNI2} & \href{http://adni.loni.usc.edu/}{ADNI-2} \cite{Aisen2015} & 2011 &  \checkmark & \checkmark & \checkmark & \checkmark & & & \checkmark & & & NC, EMCI, LMCI, AD \\ 
			\hline
			\dataindex{data:ADNI3} & \href{http://adni.loni.usc.edu/}{ADNI-3} \cite{Weiner2017} & 2016 & \checkmark & \checkmark & \checkmark & \checkmark & & & \checkmark & & & NC, EMCI, LMCI, AD \\ 
			\hline
			\dataindex{data:OASIS-1} & \href{http://www.oasis-brains.org/}{OASIS 1} \cite{Marcus2007} & 2007 & \checkmark & & & & & & \checkmark & &  & NC, AD \\ 
			\hline
			\dataindex{data:OASIS-2} & \href{http://www.oasis-brains.org/}{OASIS 2} \cite{Marcus2010} & 2009 & \checkmark & & & & & & \checkmark & &  & NC, AD \\ 
			\hline
			\dataindex{data:OASIS-3} & \href{http://www.oasis-brains.org/}{OASIS 3} \cite{LaMontagne2019} & 2019 & \checkmark & \checkmark & & \checkmark & \checkmark \tnote{tnote:head:dis:6} & & \checkmark & &  & NC, AD \\ 
			\hline
			\dataindex{data:MRIHS} & \href{https://www.kaggle.com/sabermalek/mrihs}{MRIHS} \cite{Varmazyar2020,Malekzadeh2019} & 2019 & \checkmark & & & & & & \checkmark & &  & NC, AD \\ 
			\hline
			\dataindex{data:TADPOLE} & \href{https://tadpole.grand-challenge.org/}{TADPOLE} \cite{Marinescu2020} & 2017 & \checkmark & \checkmark & \checkmark & \checkmark & \checkmark\tnote{tnote:head:dis:1} & & \checkmark & &  & NC, MCI, AD \\ 
			\hline
			\dataindex{data:MRIandAlzheimersKaggle} & \href{https://www.kaggle.com/jboysen/mri-and-alzheimers}{MRI and Alzheimers} & 2017 & \checkmark & & & & & & \checkmark & &  & NC, AD \\ 
			\hline
			\dataindex{data:CADDementia} & \href{https://caddementia.grand-challenge.org/}{CADDementia} \cite{Bron2015} & 2014 & \checkmark & & & & & & \checkmark & &  & NC, MCI, AD \\ 
			\hline
			\dataindex{data:ANT-Day2019} & \href{https://openneuro.org/datasets/ds001907/versions/2.0.3}{ANT} \cite{Boord2017,Madhyastha2015} & 2019 & \checkmark & & & & \checkmark\tnote{tnote:head:dis:2} & & & \checkmark &  & NC, PD \\ 
			\hline
			\dataindex{data:Tessa2018} & \href{https://openneuro.org/datasets/ds001354/versions/1.0.0}{PD De Novo} \cite{Tessa2018,Tessa2019} & 2018 & \checkmark & & & & \checkmark\tnote{tnote:head:dis:3} & & & \checkmark & & NC, PD \\ 
			\hline
			\dataindex{data:Mascalchi2018} & \href{https://openneuro.org/datasets/ds001378/versions/00003}{SCA2 DTI} \cite{Mascalchi2018a,Mascalchi2018} & 2018 & \checkmark & & \checkmark & & & & & & \checkmark\tnote{tnote:head:dis:4} & NC, SCA2 \\ 
			\hline
			\dataindex{data:mtop} & \href{https://tbichallenge.wordpress.com/}{MTOP} & 2016 & \checkmark & & \checkmark & & & & & & \checkmark\tnote{tnote:head:dis:5} & Healthy, category I or category II \\ 
			\bottomrule
		\end{tabular}
		\raggedright
		\begin{tablenotes}{llllll}
			\item[tnote:head:dis:1] CSF &
			\item[tnote:head:dis:2] Events and bold &
			\item[tnote:head:dis:3] Bold &
			\item[tnote:head:dis:4] Spinocerebellar ataxia type II &
			\item[tnote:head:dis:5] Mild traumatic brain injury &
			\item[tnote:head:dis:6] MRI FLAIR
		\end{tablenotes}
\end{table*}

\subsubsection{Classification of brain disease}
\label{sec:head:disease:cls}

Except for the tumor and lesion segmentation, brain disease classification also plays an essential role in healthcare. Brain related diseases have a severe effect on patients' health and their lives, e.\,g., Alzheimer's disease (AD) \cite{Fan2021,Khagi2019,Bae2020a,Tang2019} and Parkinson's disease (PD). Therefore, effective diagnosis and early intervention can effectively reduce the health damage to patients, the effect on the social times of families, and the economical impact on society.
In this section, we first introduce the datasets and challenges of AD \dRef{data:ADNI1,data:ADNIGO,data:ADNI2,data:ADNI3,data:CADDementia,data:OASIS-1}, and then we introduce other diseases \dRef{data:ANT-Day2019,data:Tessa2018,data:Mascalchi2018}. Table \ref{tab:head:dis} shows the relevant challenges and datasets.

\paragraph{Alzheimer's disease:}

AD affects a person's behavior, cognition, memory, and daily life activities. Such a progressive neurodegenerative disorder affects the normal daily life of patients because suffering from such a disease makes patients not know who they are and what they should do which then progresses to the point until they forget everything they know. The disease takes an unbearable toll on the patient and leads to a high cost to their loved ones and to the society. For example, according to \cite{2020ADFact}, AD became the sixth deadly cause in the U.S. in 2018 and costs more than two to three hundred billion U.S. dollars.

Therefore, researchers are doing everything they could to explore the causes of AD and its treatments. Diagnosis based on medical images has become a research focus because early diagnosis and intervention have significance on the progress of this disease. Hence many researchers work on the classification, i.\,e., prediction of AD using brain images.
The datasets mainly include ``Alzheimer's Disease Neuroimaging Initiative (ADNI)'' and ``Open Access Series of Imaging Studies (OASIS)''.

The ADNI is a series of projects that aim to develop clinical, imaging, genetic, and biochemical biomarkers for the early detection and tracking of AD. It includes four stages: ADNI-1 \dRef{data:ADNI1}, ADNI-GO \dRef{data:ADNIGO}, ADNI-2 \dRef{data:ADNI2}, and ADNI-3 \dRef{data:ADNI3}. These projects provide image data of the brain for researchers, and the modalities of images include MR (T1 and T2) and PET (FDG, PIB, Florbetapir, and AV-1451). These four stages consists of 1400 subjects. The subjects can be categorized into normal cognition (NC), mild cognitive impairment (MCI), and AD, where MCI can be split into early mild cognitive impairment (EMCI), later mild cognitive impairment (LMCI).

The OASIS is a series of projects aiming to provide neuroimaging data of the brain, which researchers can freely access. OASIS released three datasets, which are named OASIS-1, OASIS-2, and OASIS-3. All these three datasets are related to AD, but these datasets are also used in functional areas segmentation and other tasks. The OASIS-1 \dRef{data:OASIS-1} contains 418 subjects aged from 18 to 96, and for the subjects older than 60, there are 100 subjects diagnosed with AD. The dataset includes 434 MR sessions. The OASIS-2 \dRef{data:OASIS-2} contains 150 subjects, aged between 60 to 96, and each subject includes three or four MR sessions (T1). About 72 subjects were diagnosed as normal, while 51 subjects were diagnosed with AD. Besides, there are 14 subjects who were diagnosed as normal but were characterized as AD at a later visit. The OASIS-3 \dRef{data:OASIS-3} includes more than 1000 subjects, more than 2000 MR sessions (T1w, T2w, FLAIR, etc.), and more than 1500 PET sessions (PIB, AV45, and FDG). The dataset includes 609 normal subjects and 489 AD subjects.

Moreover, there are many other challenges based on ADNI and OASIS or independence datasets. Challenge \dRef{data:TADPOLE} is based on ADNI and aims at the prediction of  the longitudinal evolution. Dataset \dRef{data:MRIandAlzheimersKaggle} is based on OASIS and it is released on Kaggle for the classification of AD. Challenge \dRef{data:CADDementia} is an independent AD-related challenge to classify subjects into NC, MCI, and AD.

\paragraph{Other diseases:}

Similar to AD, other brain diseases are also important from the diagnosis and treatment perspective. However, the number of datasets and challenges of these diseases is not as large as AD. A few datasets focus on  Parkinson’s disease (PD) and spinocerebellar ataxia type II (SCA2). Datasets \dRef{data:ANT-Day2019,data:Tessa2018} provide images of PD with MR images and classification labels. Dataset \dRef{data:Mascalchi2018} provides images and classification labels of spinocerebellar ataxia-II, i.e., SCA2. Dataset \dRef{data:mtop} provides images and annotations for the diagnosis of mild traumatic brain injury.

\begin{sidewaystable*}
	\centering
	\caption{Summary of datasets and challenges of eye-disease related tasks.}
	\label{tab:head:eye}
	\small
		\begin{tabular}{clcccccccccccccc}
			\toprule 
			\textsc{\textbf{Reference}} & \textsc{\textbf{Dataset/Challenge}} & \textsc{\textbf{Year}} & \multicolumn{3}{l}{\textsc{\textbf{Modalities}}} & & \multicolumn{4}{l}{\textsc{\textbf{Diseases}}\tnote{tnote:head:eye:2}} & & \multicolumn{4}{l}{\textsc{\textbf{Tasks}}} \\ 
			\cline{4-6} \cline{8-11} \cline{13-16}
			\textsc{\textbf{Index}} & & & OCT & FP & OT & & AMD & DR & G & OT & & Classification & Segmentation & Dectection & Other \\
			\midrule
			\dataindex{data:riadd} & \href{https://riadd.grand-challenge.org/}{RIADD} \cite{Quellec2020} & 2020 & & \checkmark & & & \checkmark & \checkmark & & \checkmark \tnote{tnote:head:eye:3} & & \checkmark & & & \\ 
			\hline
			\dataindex{data:DeepDRiD:2} & \href{https://isbi.deepdr.org/}{The 2nd Deep DRiD} & 2020 & & \checkmark & & & & \checkmark & & & & \checkmark & & & \\ 
			\hline
			\dataindex{data:REFUGE2} & \href{https://refuge.grand-challenge.org/Home2020/}{REFUGE 2} \cite{Orlando2020} & 2020 & & \checkmark & & & & & \checkmark & & & \checkmark & \checkmark & \checkmark & \\ 
			\hline
			\dataindex{data:REFUGE} & \href{https://refuge.grand-challenge.org/REFUGE2018/}{REFUGE} \cite{Orlando2020} & 2018 & & \checkmark & & & & & \checkmark & & & \checkmark & \checkmark & \checkmark & \\ 
			\hline
			\dataindex{data:AGE} & \href{https://age.grand-challenge.org/}{AGE} \cite{Fu2020} & 2019 & \checkmark & & & & & & & \checkmark\tnote{tnote:head:eye:7} & & \checkmark & & \checkmark & \\ 
			\hline
			\dataindex{data:DRIVE} & \href{https://drive.grand-challenge.org/}{DRIVE} & 2019 & & \checkmark & & & & & & \checkmark \tnote{tnote:head:eye:6} & & & \checkmark & & \\ 
			\hline
			\dataindex{data:ODIR2019} & \href{https://odir2019.grand-challenge.org/}{ODIR-2019} & 2019 & & \checkmark & & & \checkmark & & \checkmark & \checkmark \tnote{tnote:head:eye:8} & & \checkmark & & & \\ 
			\hline
			\dataindex{data:PALM} & \href{https://palm.grand-challenge.org/}{PALM} & 2019 & & \checkmark & & & & & & \checkmark \tnote{tnote:head:eye:9} & & \checkmark & \checkmark & \checkmark & \\ 
			\hline
			\dataindex{data:APTOS2019} & \href{http://kaggle.com/c/aptos2019-blindness-detection/}{APTOS 2019} & 2019 & & \checkmark & & & & \checkmark & & & & \checkmark & & & \\ 
			\hline
			\dataindex{data:ADAM} & \href{https://amd.grand-challenge.org/}{ADAM} \cite{GrandChallengeADAM} & 2018 & & \checkmark & & & \checkmark & & & & & \checkmark & & \checkmark & \\ 
			\hline
			\dataindex{data:IDRiD} & \href{https://idrid.grand-challenge.org/}{IDRiD} \cite{GrandChallengeIDRID,Porwal2020,Porwal2018} & 2018 & & \checkmark & & & & \checkmark & & \checkmark\tnote{tnote:head:eye:10} & & \checkmark & \checkmark & \checkmark & \\ 
			\hline
			\dataindex{data:Retinal} & \href{https://www.kaggle.com/paultimothymooney/kermany2018}{Retinal OCT Images} \cite{DanielKermanyKangZhang2018,Kermany2018} & 2018 & \checkmark & & & & & \checkmark & & \checkmark\tnote{tnote:head:eye:11} & & \checkmark & & & \\ 
			\hline
			\dataindex{data:ROCC} & \href{https://rocc.grand-challenge.org/}{ROCC} & 2017 & \checkmark & & & & & \checkmark & & & & \checkmark & & & \\ 
			\hline
			\dataindex{data:CATARACTS} & \href{https://cataracts.grand-challenge.org/}{CATARACTS} \cite{AlHajj2019,Grammatikopoulou2019} & 2017 & & & \checkmark\tnote{tnote:head:eye:1} & & & & & \checkmark \tnote{tnote:head:eye:4} & & & & & \checkmark\tnote{tnote:head:eye:12} \\ 
			\hline
			\dataindex{data:RETOUCH} & \href{https://retouch.grand-challenge.org/}{RETOUCH} \cite{Bogunovic2019} & 2017 & \checkmark & & & & & & & \checkmark \tnote{tnote:head:eye:5} & & & \checkmark & & \\ 
			\hline
			\dataindex{data:DRD} & \href{https://www.kaggle.com/c/diabetic-retinopathy-detection}{Diabetic Retinopathy Detection} \cite{KaggleDiabeticRetinopathyDetection} & 2015 & & \checkmark & & & & \checkmark & & & & \checkmark & & & \\ 
			\hline
			\dataindex{data:DME} & \href{https://www.kaggle.com/paultimothymooney/chiu-2015}{Seg OCT (DME)} \cite{Chiu2015} & 2015 & \checkmark & & & & & & & \checkmark\tnote{tnote:head:eye:10} & & & \checkmark & & \\ 
			\hline
			\dataindex{data:ROC} & \href{http://webeye.ophth.uiowa.edu/ROC/}{ROC} \cite{Niemeijer2010} & 2009 & & \checkmark & & & & \checkmark & & & & & & \checkmark & \\ 
			\bottomrule
		\end{tabular}
		\raggedright
		\begin{tablenotes}{llll}
			\multicolumn{4}{l}{\item[tnote:head:eye:2] AMD: age-related macular degeneration; DR: diabetic retinopathy; G: glaucoma.} \\ 
			\multicolumn{4}{l}{\item[tnote:head:eye:3] All the diseases of this dataset are listed on the official website. See \url{https://riadd.grand-challenge.org/Data/}.} \\
			\item[tnote:head:eye:1] Video &
			\item[tnote:head:eye:4] Surgery tools detection &
			\item[tnote:head:eye:5] Fluid segmentation &
			\item[tnote:head:eye:6] Vessel extraction \\
			\item[tnote:head:eye:7] Closure glaucoma &
			\item[tnote:head:eye:8] Diabetes, cataract, hypertension, and myopia. &
			\item[tnote:head:eye:9] Pathologic myopia &
			\item[tnote:head:eye:10] Diabetic macular edema \\
			\item[tnote:head:eye:11] Macular degeneration &
			\item[tnote:head:eye:12] Tool annotation
		\end{tablenotes}
\end{sidewaystable*}

\subsection{Eye related datasets and challenges}
\label{sec:head:eye}

As the human's imaging sensor, the eyes' health is essential for human beings, and eye diseases may lead to blindness.
We introduce the relevant challenges and datasets in this subsection and list them in Table \ref{tab:head:eye}.

\paragraph{Datasets according to the modality:}

With regards to the eye-related datasets and challenges, the main used modalities are the fundus photo \dRef{data:REFUGE,data:ODIR2019,data:PALM,data:APTOS2019,data:ADAM,data:IDRiD,data:DRD,data:ROC,data:DRIVE} and OCT \dRef{data:Retinal,data:ROCC,data:AGE,data:RETOUCH,data:DME}.
The fundus photo can help medics evaluate the eye's health and locate the retinal lesions because the fundus photo clearly shows the important parts of the eye, such as the blood vessels and the optic disc.
OCT is a new imaging approach that is safe for the eye and shows the retinal tissues' in details. However, it has also disadvantages -- it is not suitable for diagnosing microangioma and the planning of retinal laser for the photocoagulation treatment.

\paragraph{Datasets according to the analysis task:}

These datasets and challenges can be used for four tasks.

\textbf{1)} \textbf{Classification} tasks focus on classifying whether the subject has specific diseases or judging whether the subject is abnormal. Datasets and challenges \dRef{data:REFUGE2,data:REFUGE,data:AGE,data:PALM,data:APTOS2019,data:ADAM,data:DRD} focus on predicting a single disease, while others \dRef{data:ODIR2019,data:IDRiD,data:Retinal,data:ROCC} focus on diagnosing multiple diseases.

\textbf{2)} \textbf{Segmentation} is another task, which provides more information compared to classification. Datasets and challenges \dRef{data:REFUGE,data:PALM,data:IDRiD,data:DRIVE,data:RETOUCH,data:DME} focus on the segmentation of the tissues and lesions for further diagnosis and disease analysis.

\textbf{3)} Datasets and challenges \dRef{data:REFUGE,data:PALM,data:ADAM,data:IDRiD,data:ROC,data:AGE} focus on the \textbf{detection} of lesions or other landmarks. These tasks help medics locate key targets, such as areas and tissues, for effective diagnosis or provide feature details for other automated algorithms.

\textbf{4)} Unlike other tasks, the last one focuses on the \textbf{annotation} of the tools used for eye-related surgery \dRef{data:CATARACTS}.

\paragraph{Datasets according to focused eye diseases:}

Researchers mainly focus on these diseases:
\begin{itemize}
	\item Diabetes retinopathy \dRef{data:ODIR2019,data:APTOS2019,data:IDRiD,data:Retinal,data:ROCC,data:DRD,data:ROC}
	\item (Age-related) macular degeneration \dRef{data:ODIR2019,data:ADAM,data:Retinal}
	\item Pathologic myopia \dRef{data:ODIR2019,data:PALM}
	\item Diabetic macular edema \dRef{data:IDRiD,data:DME}
	\item Glaucoma \dRef{data:REFUGE,data:ODIR2019}
	\item Cataract \dRef{data:ODIR2019}
	\item Closure glaucoma \dRef{data:AGE}.
	\item Hypertension \dRef{data:ODIR2019}
\end{itemize}
Besides these diseases, dataset \dRef{data:CATARACTS} aims at the annotation of images.

\subsection{Datasets and challenges of other Subjects}
\label{sec:head:other}

\begin{table*}
	\centering
	\caption{Summary of datasets and challenges of head and neck related diseases.}
	\label{tab:head:oth}
	\small
		\begin{tabular}{clcccclcc}
			\toprule 
			\textsc{\textbf{Reference}} & \textsc{\textbf{Dataset/Challenge}} & \textsc{\textbf{Year}} & \multicolumn{3}{l}{\textsc{\textbf{Modalities}}} & \textsc{\textbf{Focus}} & \multicolumn{2}{l}{\textsc{\textbf{Tasks}}} \\ 
			\cline{4-6} \cline{8-9}
			\textsc{\textbf{Index}} & & & CT & PT & OT & & Seg. & Other \\
			\midrule
			\dataindex{data:HECKTOR} & \href{https://www.aicrowd.com/challenges/hecktor}{MICCAI 2020: HECKTOR} & 2020 & \checkmark & \checkmark & & Head and neck primary tumors & \checkmark &  \\ 
			\hline
			\dataindex{data:TN-SCUI2020} & \href{https://tn-scui2020.grand-challenge.org/}{TN-SCUI 2020} \cite{M2019} & 2020 & & & \checkmark\tnote{tnote:head:oth:1} & Thyroid gland nodules diagnosis & & \checkmark \tnote{tnote:head:oth:2} \\ 
			\hline
			\dataindex{data:Head-Neck-Radiomics-HN1} & \href{https://wiki.cancerimagingarchive.net/display/Public/Head-Neck-Radiomics-HN1}{Head Neck Radiomics HN1} \cite{Aerts2014,TCIAHead-Neck-Radiomics-HN1} & 2019 & \checkmark & & & Head and neck squamous cell carcinoma & \checkmark & \\ 
			\hline
			\dataindex{data:rtmac2019} & \href{https://wiki.cancerimagingarchive.net/display/Public/AAPM+RT-MAC+Grand+Challenge+2019}{AAPM RT-MAC 2019} \cite{Cardenas2019} & 2019 & & & \checkmark\tnote{tnote:head:oth:3} & Soft tissue and tumor & \checkmark & \\ 
			\hline
			\dataindex{data:vallieres2017} & \href{https://wiki.cancerimagingarchive.net/display/Public/Head-Neck-PET-CT}{Head Neck PET-CT} \cite{Vallieres2017, Vallieres2017a} & 2017 & \checkmark & \checkmark & & Tumor & & \checkmark\tnote{tnote:head:oth:4} \\ 
			\hline
			\dataindex{data:Head-and-Neck-Auto-Seg} & \href{http://www.imagenglab.com/wiki/mediawiki/index.php?title=2015_MICCAI_Challenge}{Head \& Neck AutoSeg Challenge} \cite{Raudaschl2017} & 2015 & \checkmark & & & Tumor & \checkmark &  \\ 
			\hline
			\dataindex{data:ultrasound-nerve-segmentation} & \href{https://www.kaggle.com/c/ultrasound-nerve-segmentation}{Ultrasound Nerve Segmentation} & 2016 & & & \checkmark\tnote{tnote:head:oth:1} & Nerve & \checkmark & \\
			\hline
			\dataindex{data:GCDXIA:C1} & \href{http://www-o.ntust.edu.tw/~cweiwang/ISBI2015/challenge1/index.html}{Dental X-Ray Analysis 2} & 2015 & & & \checkmark\tnote{tnote:head:oth:5} & Cephalometric & & \checkmark\tnote{tnote:head:oth:6} \\ 
			\hline
			\dataindex{data:GCDXIA:C2} & \href{http://www-o.ntust.edu.tw/~cweiwang/ISBI2015/challenge2/}{Dental X-Ray Analysis 1} & 2015 & & & \checkmark\tnote{tnote:head:oth:5} & Caries & \checkmark & \\ 
			\hline
			\dataindex{data:HeadNeckSeg2010} & \href{https://www.sciencedirect.com/science/article/pii/S0360301614040577}{Head \& Neck AutoSeg 2010} \cite{Yang2014} & 2010 & & & \checkmark\tnote{tnote:head:oth:7} & Parotid gland & \checkmark & \\ 
			\hline
			\dataindex{data:HeadNeckSeg2009} & \href{https://www.midasjournal.org/browse/publication/703}{Head \& Neck AutoSeg 2009} & 2009 & \checkmark & & & Multi-organs and tissues & \checkmark & \\ 
			\hline
			\dataindex{data:CLS2009} & \href{http://cls2009.bigr.nl/}{CLS 2009} \cite{Hameeteman2011} & 2009 & \checkmark\tnote{tnote:head:oth:8} & & & Carotid bifurcation & \checkmark & \\ 
			\bottomrule
		\end{tabular}
		\raggedright
		\begin{tablenotes}{llllll}
			\item[tnote:head:oth:1] Ultrasound &
			\item[tnote:head:oth:2] Detection &
			\item[tnote:head:oth:3] MR T2-weighted image &
			\item[tnote:head:oth:4] Classification &
			\item[tnote:head:oth:5] Computed Radiography &
			\item[tnote:head:oth:6] Localization \\
			\item[tnote:head:oth:7] MR &
			\item[tnote:head:oth:8] CT Angiography
		\end{tablenotes}
\end{table*}

Besides the brain's structural analysis, the image processing, and the computer-aided diagnosis tasks, another important research focus is the human neck because it holds many essential glands and organs. This subsection discusses the datasets and challenges of the neck and teeth, covered in Subsection \ref{sec:head:other:neck} and Subsection \ref{sec:head:other:tooth}, respectively. Moreover, many researchers are working on the analysis of behavior and cognition with DL-based methods. We discuss the details in Subsection \ref{sec:head:other:behavior}.

\subsubsection{Neck related datasets}
\label{sec:head:other:neck}

The neck is also essential for our health. The neck holds many glands and organs, and when these become abnormal, effective diagnosis and segmentation play an essential role in their treatments. The related image datasets and challenges are listed in Table \ref{tab:head:oth}.

Datasets and challenges \dRef{data:Head-Neck-Radiomics-HN1,data:Head-and-Neck-Auto-Seg,data:HeadNeckSeg2010,data:HeadNeckSeg2009,data:HECKTOR,data:rtmac2019} focus on the segmentation of glands and the lesions and tumors in relevant glands. Dataset \dRef{data:vallieres2017} focuses on the binary classification tumor vs. normal. Challenge \dRef{data:TN-SCUI2020} aims at the task of thyroid gland nodules detection with ultrasound images and videos. Challenge \dRef{data:ultrasound-nerve-segmentation} focuses on the nerves segmentation in the neck, while challenge \dRef{data:CLS2009} focuses on evaluating carotid bifurcation.

\subsubsection{Cephalometric and teeth related datasets}
\label{sec:head:other:tooth}

Challenges \dRef{data:GCDXIA:C1,data:GCDXIA:C2} focus on the diagnosis of dental X-Ray images. The main tasks of these two challenges include landmark localization and caries segmentation. Challenge \dRef{data:GCDXIA:C1} provides around 400 cephalometric X-ray images with the annotation of landmarks by two experienced dentists. Challenge \dRef{data:GCDXIA:C2} provides about 120 bitewing images with experts' annotations of different parts of the teeth.

\begin{table*}
	\centering
	\caption{Summary of datasets and challenges that are used for behavioral and perception related tasks.}
	\label{tab:head:beh}
	\small
		\begin{tabular}{cm{14em}ccccccm{11em}c}
			\toprule 
			\textsc{\textbf{Reference}} & \textsc{\textbf{Dataset/Challenge}} & \textsc{\textbf{Year}} & \multicolumn{5}{l}{\textsc{\textbf{Modalities}}} & \textsc{\textbf{Tasks}} & \textsc{\textbf{Stimulation}} \\ 
			\cline{4-8}
			\textsc{\textbf{Index}} & & & T1 & T2 & Bold & Events & OT & & \\
			\midrule
			\dataindex{data:Zhang2020} & \href{https://openneuro.org/datasets/ds002596/versions/1.0.1}{Cognitive control of sensory pain encoding in the pregenual anterior cingulate cortex.} \cite{Zhang2020} & 2020 & \checkmark & & \checkmark & & & Sensory pain encoding & pain \\ 
			\hline
			\dataindex{data:Ikutani2020} & \href{https://openneuro.org/datasets/ds002411/versions/1.1.0}{fMRI dataset on program comprehension and expertise} \cite{Ikutani2020,Ikutani2020a} & 2020 & \checkmark & & \checkmark & \checkmark & & Different between expert and novices in cortical representations of source code & brain action \\ 
			\hline
			\dataindex{data:VanRullen2018} & \href{https://openneuro.org/datasets/ds001761/versions/2.0.0}{Reconstructing Faces from fMRI Patterns using Deep Generative Neural Networks.} \cite{VanRullen2018} & 2019 & \checkmark & & & & & Face Reconstruction & vision \\ 
			\hline
			\dataindex{data:Resting State-TMS} & \href{https://openneuro.org/datasets/ds001832/versions/1.0.1}{Resting State - TMS} \cite{Alkhasli2019,Alkhasli2019a} & 2019 & \checkmark & & \checkmark & & &Effect of iTBS on fronto-striatal network and ROI segmentation & iTBS \\
			\hline
			\dataindex{data:Shen2019} & \href{https://openneuro.org/datasets/ds001506/versions/1.3.1}{Deep Image Reconstruction} \cite{Shen2019,GuohuaShen2020} & 2018 & \checkmark & \checkmark & \checkmark & \checkmark & & Image reconstruction from human brain activity & vision \\ 
			\hline
			\dataindex{data:BOLD5000} & \href{https://openneuro.org/datasets/ds001499/versions/1.3.0}{BOLD5000} \cite{Chang2019} & 2018 & \checkmark & \checkmark & \checkmark & & \checkmark \tnote{tnote:head:beh:1} & Brain reaction to vision & vision \\ 
			\hline
			\dataindex{data:Lepping2016} & \href{https://openneuro.org/datasets/ds000171/versions/00001}{Neural Processing of Emotional Musical and Nonmusical Stimuli in Depression} \cite{Lepping2016,Lepping2016a,Dugre2019} & 2018 & \checkmark & & \checkmark & & & Brain reaction to audition & audio \\ 
			\hline
			\dataindex{data:Miyawaki2008} & \href{https://openneuro.org/datasets/ds000255/versions/00002}{Visual image reconstruction} \cite{Miyawaki2008} & 2018 & \checkmark & \checkmark & \checkmark & & & Image reconstruction from human brain activity & vision \\ 
			\hline
			\dataindex{data:Horikawa2017} & \href{https://openneuro.org/datasets/ds001246/versions/1.2.1}{Generic Object Decoding} \cite{Horikawa2017} & 2018 & \checkmark & \checkmark & \checkmark & \checkmark & & Image reconstruction from human brain activity & vision \\ 
			\hline
			\dataindex{data:Carlin2017} & \href{https://openneuro.org/datasets/ds000232/versions/00001}{Adjudicating between face-coding models with individual-face fMRI responses} \cite{Carlin2017} & 2018 & \checkmark & & & & & Decoding face from brain activity & vision \\ 
			\hline
			\dataindex{data:Koenders2016} & \href{https://openneuro.org/datasets/ds000174/versions/1.0.1}{T1w structural MRI study of cannabis users at baseline and 3 years follow up} \cite{Koenders2016} & 2018 & \checkmark & & & & & Impact of cannabis on brain & cannabis \\ 
			\bottomrule
		\end{tabular}
		\raggedright
		\begin{tablenotes}{l}
			\item[tnote:head:beh:1] DWI, Field map
		\end{tablenotes}
\end{table*}

\subsubsection{Behavior and cognition datasets}
\label{sec:head:other:behavior}

To understand what we see, hear, smell, and feel, our brain draws on neurons in our brain to compute and analyze the stimulations and understand what, where, why, and when questions and scenarios are. Many researchers now use Artificial Neural Networks as a research method to analyze the relationship between brain activities and stimulation. They use functional MR images to scan our brain activity, analyze the hemodynamic feedback, and identify the area of the neurons which react. Therefore, the analysis of the reactions of the brain in response to a specific stimulation is an important research focus. Researchers use DL to detect or decode the stimulation of subjects to work out the brain's functionality. The related datasets are listed in Table \ref{tab:head:beh}.

Some datasets \dRef{data:Ikutani2020,data:Lepping2016,data:Koenders2016} focus on classifying the stimulations or the subject's attribution based on the subject's functional MR images. Dataset \dRef{data:Ikutani2020} aims to identify whether the subject is a beginner or an expert in programming via the reaction of their brain to source codes. Dataset \dRef{data:Lepping2016} focuses on diagnosing subjects with depression vs. subjects with no-depression using audio stimulations and analyzing the subjects' brain activity. Dataset \dRef{data:Koenders2016} works on the influence of cannabis on the brain.

Datasets \dRef{data:Zhang2020,data:Shen2019,data:BOLD5000,data:Miyawaki2008,data:Horikawa2017,data:Carlin2017,data:VanRullen2018} focus on the encoding of the stimulations, i.\,e., brain activities' decoding. Datasets \dRef{data:Shen2019,data:Miyawaki2008,data:Horikawa2017} aim to rebuild what subjects have seen using DL-based methods from their brain activities using functional MR images. On the other hand, datasets \dRef{data:Carlin2017,data:VanRullen2018} work on the encoding of faces that subjects have seen from functional MR images with similar modalities.


\section{Chest and abdomen related datasets and challenges}
\label{sec:chest-abdomen}

There are many vital organs in the chest and abdomen. For example, the heart is responsible for the blood supply; the lungs are responsible for breathing; the kidneys are responsible for the production of urine to eliminate toxins from the body.
Therefore, the medical image analysis of organs in the chest and abdomen is an important research focus.
Most of the tasks are computer-aided diagnosis with classification, detection, and segmentation of lesions being the most targeted tasks.

Many datasets and challenges aim to segment one or more organs in the chest and abdomen for diagnosis or treatment planning. Subsection \ref{sec:chest-abdomen:organ-seg} discusses the datasets and challenges for segmentation. Subsection \ref{sec:chest-abdomen:diagnosis} introduces the datasets and challenges which focus on the diagnosis of organs in the chest and abdomen. While, Subsection \ref{sec:chest-abdomen:other} describes the datasets and challenges of the chest and abdomen that are not categorized above, including regression, tracking, registration, and other tasks related to the chest and abdomen organs.

\subsection{Datasets for chest \& abdomen organ segmentation}
\label{sec:chest-abdomen:organ-seg}

This subsection covers the datasets and challenges of the chest and abdomen organs that are used for anatomic segmentation tasks.
The anatomic segmentation tasks include the organ contour segmentation (Subsection \ref{sec:chest-abdomen:organ-seg:contour}) and organ segmentation (Subsection \ref{sec:chest-abdomen:organ-seg:seg}).
The contour segmentation is different from organ segmentation--the former aims to separate an organ from the backgroup or mark the boundaries between multiple organs and the background. The latter aims to segment the organ into different parts at the anatomical level. Table \ref{tab:ca:os} presents the datasets and challenges that are used for the segmentation of the chest and abdomen organs.

\begin{table*}
	\centering
	\caption{Summary of datasets and challenges for the chest and abdomen organ segmentation tasks.}
	\label{tab:ca:os}
	\small
		\begin{tabular}{cm{13em}cccccccccccc}
			\toprule
			\textsc{\textbf{Reference}}& \textsc{\textbf{Dataset/Challenge}} & \textsc{\textbf{Year}} & \multicolumn{4}{l}{\textsc{\textbf{Modalities}}} & & \multicolumn{6}{l}{\textsc{\textbf{Organs}}} \\ 
			\cline{4-7} \cline{9-14}
			\textsc{\textbf{Index}} & & & MR & CT & CR & OT & & Liver & Lung & Kidney & Prostate & Heart & Other \\
			\midrule
			\dataindex{data:mnms} & \href{https://www.ub.edu/mnms/}{MNMS Challenge} \cite{Campello2020} & 2020 & \checkmark & & & & & & & & & \checkmark & \\ 
			\hline
			\dataindex{data:ASOCA} & \href{https://asoca.grand-challenge.org/}{Automated Segmentation of Coronary Arteries} & 2020 & & \checkmark\tnote{tnote:ca:os:1} & & & & & & & & & \checkmark\tnote{tnote:ca:os:2} \\ 
			\hline
			\dataindex{data:c4kc-kits} & \href{https://kits19.grand-challenge.org/}{C4KC-KiTS} \cite{Heller2019,TCIA-C4KC-KiTS} & 2019 & & \checkmark & & & & & & \checkmark & & & \\ 
			\hline
			\dataindex{data:ms-cmrseg2019} & \href{http://www.sdspeople.fudan.edu.cn/zhuangxiahai/0/mscmrseg19/}{MS-CMRSeg 2019} \cite{Zhuang2019,Zhuang2016} & 2019 & \checkmark & & & & & & & & & \checkmark & \\ 
			\hline
			\dataindex{data:CAMUS} & \href{https://www.creatis.insa-lyon.fr/Challenge/camus/}{CAMUS} \cite{Leclerc2019} & 2019 & & & & \checkmark\tnote{tnote:ca:os:3} & & & & & & \checkmark & \\ 
			\hline
			\dataindex{data:CT-ORG} & \href{https://wiki.cancerimagingarchive.net/display/Public/CT-ORG\%3A+CT+volumes+with+multiple+organ+segmentations}{CT-ORG} \cite{TCIA-CT-ORG,TCIA-NCI-ISBI-2013,Bilic2019} & 2019 & & \checkmark & & & & \checkmark & \checkmark & \checkmark & & & \checkmark\tnote{tnote:ca:os:4} \\ 
			\hline
			\dataindex{data:CHAOS} & \href{https://chaos.grand-challenge.org/}{CHAOS} \cite{CHAOSdata2019,Kavur2020,Kavur2020a} & 2019 & \checkmark & & & & & \checkmark & & \checkmark & & & \checkmark\tnote{tnote:ca:os:5} \\ 
			\hline
			\dataindex{data:segthor} & \href{https://competitions.codalab.org/competitions/21145}{SegTHOR} \cite{Trullo2019} & 2019 & & \checkmark & & & & & & & & \checkmark & \checkmark\tnote{tnote:ca:os:6} \\ 
			\hline
			\dataindex{data:msd} & \href{http://medicaldecathlon.com/}{Medical Segmentation Decathlon} \cite{Simpson2019} & 2019 & \checkmark & \checkmark & & & & \checkmark & \checkmark & & & \checkmark & \checkmark\tnote{tnote:ca:os:7} \\ 
			\hline
			\dataindex{data:PAVES} & \href{https://paves.grand-challenge.org/}{PAVES} & 2018 & & \checkmark & & & & & & & & & \checkmark\tnote{tnote:ca:os:8} \\
			\hline
			\dataindex{data:shcxr-lung-mask} & \href{https://www.kaggle.com/yoctoman/shcxr-lung-mask}{SHCXR Lung Mask} \cite{Jaeger2014,Candemir2014,Stirenko2018} & 2018 & & & \checkmark & & & & \checkmark & & & & \\ 
			\hline
			\dataindex{data:atriaseg2018} & \href{http://atriaseg2018.cardiacatlas.org/}{AtriaSeg 2018} \cite{Xiong2021} & 2018 & \checkmark & & & & & & & & & \checkmark & \\ 
			\hline
			\dataindex{data:TCIA-Lung-CT-Segmentation-Challenge} & \href{https://wiki.cancerimagingarchive.net/display/Public/Lung+CT+Segmentation+Challenge+2017}{Lung CT Segmentation Challenge 2017} \cite{Yang2018,TCIA-Lung-CT-Segmentation-Challenge} & 2017 & & \checkmark & & \checkmark\tnote{tnote:ca:os:9} & & & \checkmark & & & \checkmark & \\ 
			\hline
			\dataindex{data:aapm-tasc} & \href{http://aapmchallenges.cloudapp.net/competitions/3}{AAPM Thoracic AutoSeg} & 2017 & & \checkmark & & & & & \checkmark & & & \checkmark & \\ 
			\hline
			\dataindex{data:LiTS} & \href{https://competitions.codalab.org/competitions/17094}{LiTS} \cite{Bilic2019} & 2017 & & \checkmark & & & & \checkmark & & & & & \\ 
			\hline
			\dataindex{data:Pancreas-CT} & \href{https://wiki.cancerimagingarchive.net/display/Public/Pancreas-CT}{Pancreas CT} \cite{Roth2015,TCIA-Pancreas-CT} & 2016 & & \checkmark & & & & & & & & & \checkmark\tnote{tnote:ca:os:10} \\ 
			\hline
			\dataindex{data:Breast-MRI-NACT-Pilot} & \href{https://wiki.cancerimagingarchive.net/display/Public/Breast-MRI-NACT-Pilot}{Breast MRI NACT Pilot} \cite{TCIA-Breast-MRI-NACT-Pilot} & 2016 & \checkmark & & & & & & & & & & \checkmark\tnote{tnote:ca:os:11} \\ 
			\hline
			\dataindex{data:HVSMR2016} & \href{http://segchd.csail.mit.edu/index.html}{HVSMR 2016} \cite{Pace2015,Trullo2019} & 2016 & \checkmark & & & & & & & & & \checkmark & \\ 
			\hline
			\dataindex{data:PROSTATE-DIAGNOSIS} & \href{https://wiki.cancerimagingarchive.net/display/Public/PROSTATE-DIAGNOSIS}{Prostate Diagnosis} \cite{TCIA-PROSTATE-DIAGNOSIS} & 2015 & \checkmark & & & & & & & & \checkmark & & \\ 
			\hline
			\dataindex{data:synapse-malbcv} & \href{https://www.synapse.org/\#!Synapse:syn3193805/wiki/}{Multi-Atlas Labeling Beyond the Cranial Vault} & 2015 & & \checkmark & & & & \checkmark & & \checkmark & & & \checkmark\tnote{tnote:ca:os:12} \\ 
			\hline
			\dataindex{data:anatomy3} & \href{http://www.visceral.eu/closed-benchmarks/anatomy3/}{Anatomy3} \cite{Jimenez-Del-Toro2016} & 2015 & \checkmark & \checkmark & & & & \checkmark & \checkmark & \checkmark & & & \checkmark\tnote{tnote:ca:os:13} \\ 
			\hline
			\dataindex{data:CT+Lymph+Nodes} & \href{https://wiki.cancerimagingarchive.net/display/Public/CT+Lymph+Nodes}{CT Lymph Nodes} \cite{Seff2015,Seff2014,Roth2014,TCIA-CT-Lymph-Nodes} & 2015 & & \checkmark & & & & & & & & & \checkmark\tnote{tnote:ca:os:14} \\ 
			\hline
			\dataindex{data:CETUS} & \href{http://www.creatis.insa-lyon.fr/Challenge/CETUS/index.html}{CETUS} & 2014 & & & & \checkmark\tnote{tnote:ca:os:3} & & & & & & \checkmark & \\ 
			\hline
			\dataindex{data:benchmark-1b} & \href{http://www.visceral.eu/benchmark-1b-isbi/}{VISCERAL Benchmark 2} \cite{Spanier2014} & 2014 & \checkmark & \checkmark & & & & \checkmark & \checkmark & \checkmark & & & \checkmark\tnote{tnote:ca:os:13} \\
			\hline
			\dataindex{data:left-atrium-segmentation-challenge} & \href{http://www.cardiacatlas.org/challenges/left-atrium-segmentation-challenge/}{Left Atrium Segmentation Challenge} \cite{Tobon-Gomez2015} & 2013 & \checkmark & \checkmark & & & & & & & & \checkmark & \\ 
			\hline
			\dataindex{data:NCI-ISBI-2013} & \href{https://wiki.cancerimagingarchive.net/display/DOI/NCI-ISBI+2013+Challenge\%3A+Automated+Segmentation+of+Prostate+Structures}{NCI-ISBI 2013} \cite{TCIA-NCI-ISBI-2013} & 2013 & \checkmark & & & & & & & & \checkmark & & \\ 
			\hline
			\dataindex{data:left-atrium-fibrosis-and-scar-segmentation-challenge} & \href{http://www.cardiacatlas.org/challenges/left-atrium-fibrosis-and-scar-segmentation-challenge/}{Left Atrium Fibrosis and Scar Segmentation Challenge} & 2012 & \checkmark & & & & & & & & & \checkmark\tnote{tnote:ca:os:15} & \\
			\hline
			\dataindex{data:vessel12} & \href{https://vessel12.grand-challenge.org/}{VESSEL12} & 2012 & & \checkmark & & & & & \checkmark & & & & \\
			\hline
			\dataindex{data:} & \href{https://crass.grand-challenge.org/}{CRASS12} & 2012 & & & \checkmark & & & & & & & & \checkmark\tnote{tnote:ca:os:16} \\ 
			\hline
			\dataindex{data:PROMISE12} & \href{https://promise12.grand-challenge.org/}{PROMISE12} \cite{Litjens2014} & 2012 & \checkmark & & & & & & & & \checkmark & & \\ 
			\hline
			\dataindex{data:Prostate-3T} & \href{https://wiki.cancerimagingarchive.net/display/Public/Prostate-3T}{Prostate-3T} \cite{TCIA-Prostate-3T} & 2012 & \checkmark & & & & & & & & \checkmark & & \\ 
			\hline
			\dataindex{data:LOLA11} & \href{https://lola11.grand-challenge.org/}{LOLA11} & 2011 & & \checkmark & & & & & \checkmark & & & & \\ 
			\hline
			\dataindex{data:lvseg} & \href{http://www.cardiacatlas.org/challenges/lv-segmentation-challenge/}{Left Ventricular Segmentation Challenge} & 2011 & \checkmark & & & & & & & & & \checkmark & \\
			\hline
			\dataindex{data:IVUS11} & \href{https://www.cvc.uab.es/IVUSchallenge2011/}{IVUS11} \cite{Balocco2014} & 2011 & \checkmark & & & & & & & & & & \checkmark\tnote{tnote:ca:os:17} \\ 
			\hline
			\dataindex{data:Promise09} & \href{http://wiki.na-mic.org/Wiki/index.php/2009_prostate_segmentation_challenge_MICCAI}{Promise09} & 2009 & \checkmark & & & & & & & & \checkmark & & \\ 
			\hline
			\dataindex{data:exact09} & \href{http://image.diku.dk/exact/}{EXACT09} \cite{Lo2012} & 2009 & & \checkmark & & & & & \checkmark & & & & \\ 
			\hline
			\dataindex{data:sliver07} & \href{https://sliver07.grand-challenge.org/}{SLIVER07} \cite{Heimann2009} & 2007 & & \checkmark & & & & \checkmark & & & & & \\ 
			\bottomrule
		\end{tabular}
		\raggedright
		\begin{tablenotes}{lllll}
			\item[tnote:ca:os:1] Coronary CT angiography &
			\item[tnote:ca:os:2] Coronary arteries &
			\item[tnote:ca:os:3] Ultrasound &
			\item[tnote:ca:os:4] Bladder &
			\item[tnote:ca:os:5] Spleen \\
			\item[tnote:ca:os:6] Aorta, trachea, and esophagus &
			\multicolumn{3}{l}{\item[tnote:ca:os:7] Pancreas, colon, and other non-chest or -abdomen organs.} &
			\item[tnote:ca:os:8] Vein \\
			\item[tnote:ca:os:9] Radiotherapy Structure Set &
			\item[tnote:ca:os:10] Pancreas &
			\item[tnote:ca:os:11] Breast \\
			\item[tnote:ca:os:14] Lymph &
			\multicolumn{2}{l}{\item[tnote:ca:os:15] Left atrium fibrosis and scar} &
			\item[tnote:ca:os:16] Chest structure &
			\item[tnote:ca:os:17] Vessel \\
			\multicolumn{5}{l}{\item[tnote:ca:os:12] Adrenalglands, aorta, esophagus, gall bladder, pancreas, splenic/portal veins, spleen, stomach, vena cava} \\
			\multicolumn{5}{p{0.8\linewidth}}{ \item[tnote:ca:os:13] Spleen, urinary bladder, rectus abdominis muscle, 1st lumbar vertebra, pancreas, psoas major, muscle, gall bladder, sternum, aorta, trachea, and adrenal gland}
		\end{tablenotes}
\end{table*}

\subsubsection{Datasets of chest and abdomen organs}
\label{sec:chest-abdomen:organ-seg:contour}

Organ contour segmentation is a necessary information for the preplanning of surgery and diagnosis. A well-segmented contour of the organs provides a precise mask, which helps to produce accurate segmentation results for the diagnosis, treatment, and operation. This subsection introduces datasets and challenges for the contour segmentation of a single organ and of multiple organs.

\paragraph{Chest \& abdomen datasets according to the organ:}

The datasets and challenges that we have covered are shown here. The following organs and parts are involved in:

\begin{itemize}
	\item Liver \dRef{data:LiTS,data:sliver07,data:CT-ORG,data:CHAOS,data:msd,data:synapse-malbcv,data:anatomy3}
	\item Lung \dRef{data:shcxr-lung-mask,data:LOLA11,data:exact09,data:CT-ORG,data:msd,data:TCIA-Lung-CT-Segmentation-Challenge,data:anatomy3}
	\item Kidney \dRef{data:c4kc-kits,data:CT-ORG,data:CHAOS,data:synapse-malbcv,data:anatomy3}
	\item Prostate \dRef{data:Prostate-3T,data:Promise09,data:PROMISE12,data:PROSTATE-DIAGNOSIS,data:msd}
	\item Heart \dRef{data:ms-cmrseg2019,data:segthor,data:msd,data:TCIA-Lung-CT-Segmentation-Challenge}
	\item Pancreas \dRef{data:Pancreas-CT,data:msd,data:synapse-malbcv,data:anatomy3}
	\item Aorta \dRef{data:segthor,data:synapse-malbcv,data:anatomy3}
	\item Esophagus \dRef{data:segthor,data:TCIA-Lung-CT-Segmentation-Challenge,data:synapse-malbcv}
	\item Spleen \dRef{data:CHAOS,data:synapse-malbcv,data:anatomy3}
	\item Adrenal glands \dRef{data:synapse-malbcv,data:anatomy3}
	\item Bladder \dRef{data:CT-ORG,data:anatomy3}
	\item Gall bladder \dRef{data:synapse-malbcv,data:anatomy3}
	\item Trachea \dRef{data:segthor,data:anatomy3}
	\item Colon \dRef{data:msd}
	\item Breast \dRef{data:Breast-MRI-NACT-Pilot}
	\item Lymph \dRef{data:CT+Lymph+Nodes}
	\item Spinal cord \dRef{data:TCIA-Lung-CT-Segmentation-Challenge}
	\item Stomach \dRef{data:synapse-malbcv}
\end{itemize}

Generally, these datasets and challenges focus on the larger organs, such as the liver and the lungs, with the aim to diagnose tumors and lesions, and where contour segmentation is a pre-processing step.
However, it is challenging to segment smaller organs with low-resolution images, particularly for radiotherapy,
because an incorrect contour segmentation of these small organs can lead to severe consequences (e.\,g., organ damage).
Small organs' incorrect contour can lead to their damage during radiotherapy.

\paragraph{Chest \& abdomen datasets according to modality:}

The most commonly used image modalities for chest and abdomen organs segmentation are MR and CT.
As Table 1 shows, many datasets and challenges use MR images. MR images have higher resolution under certain conditions and have better resolution for soft body tissues and organs, such as the heart and prostate.
Meanwhile, CT is the most widely used modality for organ segmentation and other tasks and diagnosis that are related to chest and abdomen, such as the lung and liver, according to our research, because of its convenience, effectiveness, and low cost.

\paragraph{Chest \& abdomen datasets according to focus:}

The purpose of these datasets and challenges can be categorized into three groups: further analysis, benchmark, and radiotherapy.
Most datasets and challenges which provide annotated organs' contours are provided with the objective to focus on further analysis and treatments. One of the challenges of segmentation is to achieve a robust segmentation of the whole organ and separate it from the background, without omitting the lesions and tumors, and thus, some test benchmarks \dRef{data:anatomy3,data:msd} are provided for researchers to evaluate their algorithms. Another challenge, which is  addressed by datasets and challenges \dRef{data:TCIA-Lung-CT-Segmentation-Challenge,data:segthor} is the imbalance between different organs because of their sizes and shapes, and such an imbalance makes it challenging to segment small organs and provide valuable information for analysis and treatment.

\paragraph{Single chest \& abdomen organ contour segmentation:}

The single organ's contour segmentation tasks usually focus on segmenting a region for subsequent tasks \dRef{data:c4kc-kits,data:Pancreas-CT,data:Breast-MRI-NACT-Pilot,data:CT+Lymph+Nodes,data:Prostate-3T,data:sliver07,data:LiTS} or with an anatomical purpose \dRef{data:shcxr-lung-mask,data:PROMISE12,data:PROSTATE-DIAGNOSIS,data:Promise09,data:NCI-ISBI-2013,data:exact09} for research. The difficulty of the former task is that the lesions and tumors may affect the segmentation by separating the organ from the background, while the latter's difficulty is to perform more precise segmentation.

\paragraph{Chest \& abdomen multi-organs contours segmentation:}

The chest and abdomen multiple organs contour segmentation focuses on splitting the organs from each other. Some of these datasets and challenges \dRef{data:CT-ORG,data:CHAOS,data:msd} focus on the segmentation of multiple organs, including the relatively larger organs, which are easier to segment, and the relatively smaller organs, which can be more challenging to segment compared to the larger ones, especially when the model is handling the larger and smaller organs at the same time. Similarly, some of these datasets and challenges \dRef{data:TCIA-Lung-CT-Segmentation-Challenge,data:segthor} focus on the ``organ at risk'' which means that these organs are healthy but might be at risk because of radiation therapy. Dataset \dRef{data:synapse-malbcv} focuses on multi-atlas-based methods, which are widely used in brain-related research. Dataset \dRef{data:anatomy3} aims to provide a benchmark for the segmentation algorithms.

\subsubsection{Chest \& abdomen organ parts segmentation}
\label{sec:chest-abdomen:organ-seg:seg}

Different from contour segmentation of the chest and abdomen organs, the organ segmentation aims to segment the organ into different parts. Just as the hand has five fingers, organs are made up of multiple parts, and a typical example is the Couinaud liver segmentation method.
This subsection introduces the datasets and challenges for organ segmentation. These datasets and challenges are listed in Table \ref{tab:ca:os}.

\paragraph{Heart realted datasets and challenges:}

Most of these datasets and challenges \dRef{data:CAMUS,data:atriaseg2018,data:HVSMR2016,data:left-atrium-segmentation-challenge,data:CETUS,data:lvseg,data:left-atrium-fibrosis-and-scar-segmentation-challenge} are related to the heart segmentation. The most frequently used modalities are MR and ultrasound, and the aim is to segment the heart into the left atrium, chambers, valves, and other parts.
Though MR and ultrasound can effectively image the different tissues of the heart, the heartbeat results in blurred images, which makes the segmentation task more difficult, while for ultrasound, the dynamic nature of ultrasound images is another challenge for the segmentation algorithm.

\paragraph{Others chest \& abdomen body parts:}

Challenge \dRef{data:LOLA11} provides 55 CT scans and focuses on the segmentation of the lung with the labeling of its different parts: outside the lungs, the left lung, the upper lobe of the left lung, the lower lobe of the left lung, the upper lobe of the right lung, the middle lobe of the right lung, and the lower lobe of the right lung. The biggest challenge is the effect of the lung lesions and diseases, such as tuberculosis and pulmonary emphysema, on the performance of the segmentation.
Moreover, challenges \dRef{data:vessel12,data:IVUS11} focus on the segmentation of the lung vessels.

\begin{sidewaystable*}
	\centering
	\caption{Summary of datasets and challenges for chest and abdomen organs-related tasks \textbf{I}. }
	\label{tab:ca:t:1}
	\small
		\begin{tabular}{cm{18em}ccccccm{12em}ccc}
			\toprule 
			\textsc{\textbf{Reference}} & \textsc{\textbf{Dataset/Challenge}} & \textsc{\textbf{Year}} & \multicolumn{5}{l}{\textsc{\textbf{Modalities}}} & \textsc{\textbf{Focus}} & \multicolumn{3}{l}{\textsc{\textbf{Tasks}}} \\ 
			\cline{4-8} \cline{10-12}
			\textsc{\textbf{Index}} & & & CT & MR & PT & CR & OT & & Classification & Segmentation & Detection \\
			\midrule
			\dataindex{data:COVID-CT} & \href{https://covid-ct.grand-challenge.org/}{CT Diagnosis of COVID-19} \cite{Zhao2020} & 2020 & \checkmark & & & & & COVID-19 & \checkmark & & \\ 
			\hline
			\dataindex{data:covid19-eu} & \href{https://www.covid19challenge.eu/}{Covid19 Challenge.eu} & 2020 & \checkmark & & & & & COVID-19 & \checkmark & & \\ 
			\hline
			\dataindex{data:objCXR} & \href{https://jfhealthcare.github.io/object-CXR/}{Object CXR} & 2020 & \checkmark & & & & & COVID-19 & \checkmark & & \checkmark \\ 
			\hline
			\dataindex{data:CORD19} & \href{https://www.semanticscholar.org/cord19/}{CORD-19} \cite{Wang2020b} & 2020 & \checkmark & & & & & COVID-19 & \checkmark & & \\ 
			\hline
			\dataindex{data:covid-chestxray-dataset} & \href{https://github.com/ieee8023/covid-chestxray-dataset}{Covid Chest X-Ray Dataset} \cite{Maguolo2020,Tartaglione2020} & 2020 & & & & \checkmark & & COVID-19 & \checkmark & & \\ 
			\hline
			\dataindex{data:Covid19PocusUltrasound} & \href{https://github.com/jannisborn/covid19_pocus_ultrasound}{Detection of COVID-19 from Ultrasound} \cite{Born2020,Born2020a} & 2020 & & & & & \checkmark\tnote{tnote:ca:t:1:1} & COVID-19 & \checkmark & & \\ 
			\hline
			\dataindex{data:COVID-Net} & \href{https://github.com/lindawangg/COVID-Net/}{COVID-Net} \cite{Wang2020c} & 2020 & & & & \checkmark & & COVID-19 & \checkmark & & \\ 
			\hline
			\dataindex{data:covid-19-sirm} & \href{https://www.sirm.org/category/senza-categoria/covid-19/}{COVID-19} & 2020 & \checkmark & & & & & COVID-19 & \checkmark & & \\ 
			\hline
			\dataindex{data:covid19-radiography-database} & \href{https://www.kaggle.com/tawsifurrahman/covid19-radiography-database}{COVID-19 Radiography Database} \cite{Chowdhury2020} & 2020 & & & & \checkmark & & COVID-19 & \checkmark & & \\ 
			\hline
			\dataindex{data:COVID-19-chest-xray} & \href{https://www.kaggle.com/bachrr/covid-chest-xray}{COVID-19 Chest X-Ray} & 2020 & & & & \checkmark & & COVID-19 & \checkmark & & \\ 
			\hline
			\dataindex{data:COVID-19-seg} & \href{http://medicalsegmentation.com/covid19/}{COVID-19 CT Segmentation Dataset} & 2020 & \checkmark & & & & & COVID-19 & & \checkmark & \\ 
			\hline
			\dataindex{data:COVID-19-SEG} & \href{https://covid-segmentation.grand-challenge.org/}{COVID-19 Lung CT Lesion Segmentation Challenge 2020} & 2020 & \checkmark & & & & & COVID-19 & & \checkmark & \\ 
			\hline
			\dataindex{data:CT-COVID-19} & \href{https://wiki.cancerimagingarchive.net/display/Public/CT+Images+in+COVID-19}{CT Images in COVID-19} \cite{Harmon2020,An2020} & 2020 & \checkmark & & & & & COVID-19 & \checkmark & & \\ 
			\hline
			\dataindex{data:COVID-19-AR} & \href{https://wiki.cancerimagingarchive.net/pages/viewpage.action?pageId=70226443}{COVID-19 AR} \cite{Desai2020,Desai2020a} & 2020 & \checkmark & & & \checkmark & & COVID-19 & \checkmark & & \\ 
			\hline
			\dataindex{data:bimcv-covid19} & \href{https://bimcv.cipf.es/bimcv-projects/bimcv-covid19/}{BIMCV-COVID19} & 2020 & \checkmark & & & \checkmark & & COVID-19 & \checkmark & & \\ 
			\hline
			\dataindex{data:BCS-DBT} & \href{https://wiki.cancerimagingarchive.net/pages/viewpage.action?pageId=64685580}{BCS-DBT} \cite{Buda2020,Buda2020a} & 2020 & & & & & \checkmark\tnote{tnote:ca:t:1:2} & Breast cancer & \checkmark & & \checkmark \\ 
			\hline
			\dataindex{data:Lung-PET-CT-Dx} & \href{https://wiki.cancerimagingarchive.net/pages/viewpage.action?pageId=70224216}{Lung-PET-CT-Dx} \cite{TCIA-NNC2-0461} & 2020 & \checkmark & & \checkmark & & & Lung cancer & \checkmark & & \\ 
			\hline
			\dataindex{data:lndb} & \href{https://lndb.grand-challenge.org/}{LNDb Challenge} & 2020 & \checkmark & & & & & Pulmonary nodule & \checkmark & & \checkmark \\ 
			\hline
			\dataindex{data:A-AFMA-detection} & \href{https://a-afma-detection.grand-challenge.org/}{A-AFMA-Detection} & 2020 & & & & & \checkmark\tnote{tnote:ca:t:1:1} & Amniotic fluid detection & & & \checkmark \\ 
			\hline
			\dataindex{data:c4kc-kits} & \href{https://kits19.grand-challenge.org/}{C4KC-KiTS} \cite{Heller2019,TCIA-C4KC-KiTS} & 2019 & \checkmark & & & & & Kidney tumor & & \checkmark & \\ 
			\hline
			\dataindex{data:SIIM-ACR} & \href{https://www.kaggle.com/c/siim-acr-pneumothorax-segmentation}{SIIM-ACR Pneumothorax Segmentation} & 2019 & & & & \checkmark & & Pneumothorax & \checkmark & \checkmark & \\ 
			\hline
			\dataindex{data:CTVIE19} & \href{http://aapmchallenges.cloudapp.net/competitions/35}{CT ventilation imaging evaluation 2019} & 2019 & \checkmark\tnote{tnote:ca:t:1:3} & \checkmark & & & & Ventilation imaging & & \checkmark & \\ 
			\hline
			\dataindex{data:chexpert} & \href{https://stanfordmlgroup.github.io/competitions/chexpert/}{CheXpert} \cite{Irvin2019} & 2019 & & & & \checkmark & & Chest & \checkmark & \checkmark & \\ 
			\hline
			\dataindex{data:NSCLC-RI} & \href{https://wiki.cancerimagingarchive.net/display/Public/NSCLC-Radiomics-Interobserver1}{NSCLC-Radiomics-Interobserver1} \cite{Aerts2014,Wee2019} & 2019 & \checkmark & & & & \checkmark\tnote{tnote:ca:t:1:4} & Non-small cell lung cancer & & \checkmark & \\ 
			\hline
			\dataindex{data:StructSeg2019} & \href{https://structseg2019.grand-challenge.org/}{StructSeg2019} & 2019 & \checkmark & & & & & Lung cancer \& organs-at-risk & \checkmark & & \\ 
			\hline
			\dataindex{data:MIMIC-CXR} & \href{https://physionet.org/content/mimic-cxr/2.0.0/}{MIMIC-CXR} \cite{Johnson2019,MIMIC-CXR} & 2019 & & & & \checkmark & \checkmark\tnote{tnote:ca:t:1:5} & Chest image analysis & \checkmark & & \\ 
			\hline
			\dataindex{data:MIMIC-CXR-JPG} & \href{https://physionet.org/content/mimic-cxr-jpg/2.0.0/}{MIMIC-CXR-JPG} & 2019 & & & & \checkmark & & Chest X-Ray & \checkmark & \checkmark & \\ %
			\hline
			\bottomrule
		\end{tabular}
		\raggedright
		\begin{tablenotes}{lllll}
			\item[tnote:ca:t:1:1] Ultrasound &
			\item[tnote:ca:t:1:2] Digital breast tomosynthesis &
			\item[tnote:ca:t:1:3] 4D-CT &
			\item[tnote:ca:t:1:4] Radiotherapy Structure Set &
			\item[tnote:ca:t:1:5] Electronic health record and report
		\end{tablenotes}
\end{sidewaystable*}

\begin{sidewaystable*}
	\centering
	\caption{Summary of datasets and challenges for chest and abdomen organs-related tasks \textbf{II}. }
	\label{tab:ca:t:2}
	\small
		\begin{tabular}{cm{18em}ccccccm{12em}ccc}
			\toprule 
			\textsc{\textbf{Reference}} & \textsc{\textbf{Dataset/Challenge}} & \textsc{\textbf{Year}} & \multicolumn{5}{l}{\textsc{\textbf{Modalities}}} & \textsc{\textbf{Focus}} & \multicolumn{3}{l}{\textsc{\textbf{Tasks}}} \\ 
			\cline{4-8} \cline{10-12}
			\textsc{\textbf{Index}} & & & CT & MR & PT & CR & OT & & Classification & Segmentation & Detection \\
			\midrule
			\dataindex{data:PadChest} & \href{http://bimcv.cipf.es/bimcv-projects/padchest/}{PadChest} & 2019 & & & & \checkmark & & Chest X-Ray & \checkmark & & \\ %
			\hline
			\dataindex{data:rsna-pneumonia-detection-challenge} & \href{https://www.kaggle.com/c/rsna-pneumonia-detection-challenge}{RSNA Pneumonia Detection Challenge} & 2018 & & & & \checkmark & & Pneumonia & & & \checkmark \\ 
			\hline
			\dataindex{data:ImageCLEF2018-tuberculosis} & \href{https://www.imageclef.org/2018/tuberculosis}{ImageCLEF 2018 - Tuberculosis} & 2018 & \checkmark & & & & & Tuberculosis & & & \checkmark \\ 
			\hline
			\dataindex{data:Lung+Fused-CT-Pathology} & \href{https://wiki.cancerimagingarchive.net/display/Public/Lung+Fused-CT-Pathology}{Lung Fused-CT-Pathology} \cite{Rusu2017,Madabhushi2018} & 2018 & \checkmark & & & & \checkmark\tnote{tnote:ca:t:2:6} & Pulmonary nodule & & & \checkmark \\ 
			\hline
			\dataindex{data:ACAD} & \href{https://www.creatis.insa-lyon.fr/Challenge/acdc/index.html}{ACAD} \cite{Bernard2018} & 2017 & & \checkmark & & & & Cardiac diseases & \checkmark & & \\ 
			\hline
			\dataindex{data:NSCLC+Radiogenomics} & \href{https://wiki.cancerimagingarchive.net/display/Public/NSCLC+Radiogenomics}{NSCLC Radiogenomics} \cite{Gevaert2012,Irvin2019} & 2017 & \checkmark & & \checkmark & & & Non-small cell lung cancer & & \checkmark & \\ 
			\hline
			\dataindex{data:ProstateX} & \href{https://prostatex.grand-challenge.org/}{ProstateX} & 2018 & & \checkmark\tnote{tnote:ca:t:2:7} & & & & Prostate lesion & \checkmark & & \\ 
			\hline
			\dataindex{data:PCXR} & \href{https://www.kaggle.com/kmader/pulmonary-chest-xray-abnormalities}{Pulmonary Chest X-Ray Abnormalities} & 2018 & & & & \checkmark & & Tuberculosis & \checkmark & & \\ 
			\hline
			\dataindex{data:chest-xray-pneumonia} & \href{https://www.kaggle.com/paultimothymooney/chest-xray-pneumonia}{Chest X-Ray Images (Pneumonia)} \cite{DanielKermanyKangZhang2018,Kermany2018} & 2018 & & & & \checkmark & & Pneumonia & \checkmark & & \\ 
			\hline
			\dataindex{data:LiTS} & \href{https://competitions.codalab.org/competitions/17094}{LiTS} \cite{Bilic2019} & 2017 & \checkmark & & & & & Liver tumor & & \checkmark & \\ 
			\hline
			\dataindex{data:data-science-bowl-2017} & \href{https://www.kaggle.com/c/data-science-bowl-2017}{Data Science Bowl 2017} & 2017 & \checkmark & & & & & Lung cancer & & & \checkmark \\ 
			\hline
			\dataindex{data:ACRIN-FLT-Breast} & \href{https://wiki.cancerimagingarchive.net/pages/viewpage.action?pageId=30671268}{ACRIN-FLT-Breast (ACRIN 6688)} \cite{Kostakoglu2015,Kinahan2017} & 2017 & \checkmark & & \checkmark & & & Breast cancer & \checkmark & & \\ 
			\hline
			\dataindex{data:intel-mobileodt-cervical-cancer-screening} & \href{https://www.kaggle.com/c/intel-mobileodt-cervical-cancer-screening}{Cervical Cancer Screening} & 2017 & & & & & \checkmark\tnote{tnote:ca:t:2:8} & Cervical cancer & & & \checkmark \\ 
			\hline
			\dataindex{data:nih-chest-xrays} & \href{https://www.kaggle.com/nih-chest-xrays/data}{NIH Chest X-Rays} \cite{Wang2017a} & 2017 & & & & \checkmark & & Lung disease & \checkmark & & \checkmark \\ 
			\hline
			\dataindex{data:SPIE-AAPM-NCI+PROSTATEx+Challenges} & \href{https://wiki.cancerimagingarchive.net/display/Public/SPIE-AAPM-NCI+PROSTATEx+Challenges}{SPIE-AAPM-NCI PROSTATEx Challenges} \cite{Litjens2014b,Litjens2017a} & 2016 & & \checkmark\tnote{tnote:ca:t:2:7} & & & & Prostate lesion & \checkmark & & \\ 
			\hline
			\dataindex{data:ImageCLEFmed2016} & \href{https://www.imageclef.org/2016/medical}{ImageCLEFmed: The Medical Task 2016} & 2016 & & & & & \checkmark & Report-combined medical image classification & \checkmark & & \\ 
			\hline
			\dataindex{data:luna16} & \href{https://luna16.grand-challenge.org/}{LUNA16} \cite{Setio2017} & 2016 & \checkmark & & & & & Pulmonary nodule & \checkmark & & \checkmark \\ %
			\hline
			\dataindex{data:Digital-Mammography-DREAM} & \href{https://www.synapse.org/\#!Synapse:syn4224222/wiki/401743}{The Digital Mammography DREAM Challenge} & 2016 & & & & \checkmark & \checkmark\tnote{tnote:ca:t:2:9} & Breast cancer & \checkmark & & \checkmark \\ 
			\hline
			\dataindex{data:LowDoseCT} & \href{https://www.aapm.org/GrandChallenge/LowDoseCT/}{Low Dose CT Challenge} \cite{Kachelrie2006,Flohr2005} & 2016 & \checkmark & & & & & Low dose CT \& Liver lesion & & & \checkmark \\ 
			\hline
			\dataindex{data:RIDER-Lung-CT} & \href{https://wiki.cancerimagingarchive.net/display/Public/RIDER+Lung+CT}{RIDER Lung CT} \cite{Zhao2009,Zhao2015} & 2015 & \checkmark & & & & & Non–small cell lung cancer & & & \checkmark \\ 
			\hline
			\dataindex{data:phantom-fda} & \href{https://wiki.cancerimagingarchive.net/display/Public/Phantom+FDA}{Phantom FDA} \cite{Gavrielides2015,Gavrielides2010} & 2015 & \checkmark & & & & & Phantom \& pulmonary nodule & & & \\ 
			\hline
			\dataindex{data:BREAST-DIAGNOSIS} & \href{https://wiki.cancerimagingarchive.net/display/Public/BREAST-DIAGNOSIS}{BREAST-DIAGNOSIS} \cite{Bloch2015} & 2015 & & \checkmark & & & & Breast & & & \\ 
			\hline
			\dataindex{data:Soft-tissue-Sarcoma} & \href{https://wiki.cancerimagingarchive.net/display/Public/Soft-tissue-Sarcoma}{Soft-tissue-Sarcoma} \cite{Vallieres2015} & 2015 & \checkmark & \checkmark & \checkmark & & & Soft tissue sarcoma & \checkmark & & \\ 
			\hline
			\dataindex{data:SPIE-AAPM-lung} & \href{https://wiki.canc-erimagingarchive.net/display/Public/SPIE-AAPM+Lung+CT+Challenge}{SPIE-AAPM Lung CT Challenge} \cite{Armato2016,Armato2015,ArmatoIIIS.G.HadjiiskiL.TourassiG.D.DrukkerK.GigerM.L.LiF.2015} & 2014 & \checkmark & & & & & Pulmonary nodule & & & \checkmark \\ 
			\hline
			\dataindex{data:NSCLC-Radiomics} & \href{https://wiki.cancerimagingarchive.net/display/Public/NSCLC-Radiomics}{NSCLC-Radiomics} \cite{Aerts2014} & 2014 & \checkmark & & & & & Non-small cell lung cancer & \checkmark & & \\ %
			\bottomrule
		\end{tabular}
		\raggedright
		\begin{tablenotes}{llll}
			\item[tnote:ca:t:2:6] Pathology image &
			\item[tnote:ca:t:2:7] T2-weighted, Proton density weighted, Dynamic contrast-enhanced, DE &
			\item[tnote:ca:t:2:8] Endoscopy &
			\item[tnote:ca:t:2:9] Mammography
		\end{tablenotes}
\end{sidewaystable*}

\begin{sidewaystable*}
	\centering
	\caption{Summary of datasets and challenges for chest and abdomen organs-related Tasks \textbf{III}. }
	\label{tab:ca:t:3}
	\small
		\begin{tabular}{cm{18em}ccccccm{12em}ccc}
			\toprule 
			\textsc{\textbf{Reference}} & \textsc{\textbf{Dataset/Challenge}} & \textsc{\textbf{Year}} & \multicolumn{5}{l}{\textsc{\textbf{Modalities}}} & \textsc{\textbf{Focus}} & \multicolumn{3}{l}{\textsc{\textbf{Tasks}}} \\ 
			\cline{4-8} \cline{10-12}
			\textsc{\textbf{Index}} & & & CT & MR & PT & CR & OT & & Classification & Segmentation & Detection \\
			\midrule
			\dataindex{data:NCI-ISBI-2013} & \href{https://wiki.cancerimagingarchive.net/display/DOI/NCI-ISBI+2013+Challenge\%3A+Automated+Segmentation+of+Prostate+Structures}{Automated Segmentation of Prostate Structures} \cite{TCIA-NCI-ISBI-2013} & 2013 & & \checkmark & & & & Prostate & & \checkmark & \\ %
			\hline
			\dataindex{data:NSCLC-R-ISS} & \href{https://wiki.cancerimagingarchive.net/display/DOI/NSCLC+Radiogenomics\%3A+Initial+Stanford+Study+of+26+Cases}{NSCLC Radiogenomics: Initial Stanford Study of 26 Cases} \cite{Gevaert2012,Napel2014} & 2013 & \checkmark & & \checkmark & & & Non-small cell lung cancer & & & \checkmark \\ 
			\hline
			\dataindex{data:ventricular-infarct-segmentation} & \href{http://www.cardiacatlas.org/challenges/ventricular-infarct-segmentation/}{Ventricular Infarct Segmentation} & 2012 & & \checkmark & & & & Left ventricular myocardial infarction segmentation & & \checkmark & \\ %
			\hline
			\dataindex{data:LIDC-IDRI} & \href{https://wiki.cancerimagingarchive.net/display/Public/LIDC-IDRI}{LIDC-IDRI} \cite{Armato2011,III2015} & 2011 & \checkmark & & & \checkmark & & Lung cancer \& pulmonary nodule & & & \checkmark \\ 
			\hline
			\dataindex{data:CT+COLONOGRAPHY} & \href{https://wiki.cancerimagingarchive.net/display/Public/CT+COLONOGRAPHY}{CT COLONOGRAPHY} \cite{Smith2015,Johnson2008} & 2011 & \checkmark & & & & & Polyp \& colonography & & & \checkmark \\ 
			\hline
			\dataindex{data:RIDER-Collections} & \href{https://wiki.cancerimagingarchive.net/display/Public/RIDER+Collections}{RIDER Collections} & 2011 & \checkmark & \checkmark & \checkmark & & & Cancer & & & \checkmark \\ 
			\hline
			\dataindex{data:ANODE09} & \href{https://anode09.grand-challenge.org/}{Automatic Nodule Detection 2009} \cite{VanGinneken2010} & 2009 & \checkmark & & & & & Pulmonary nodule & & & \checkmark \\ 
			\hline
			\dataindex{data:volcan09} & \href{http://www.via.cornell.edu/challenge/}{VOLCANO 09} & 2009 & \checkmark & & & & & Pulmonary nodule & & \checkmark & \\ 
			\hline
			\dataindex{data:endo2021} & \href{https://endocv2021.grand-challenge.org/}{EndoCV 2021} \cite{Ali2020,Ali2020a} & 2020 & & & & & \checkmark\tnote{tnote:ca:t:3:8} & Colon (polyp, cancer), oesophagus (Barrett’s, dysplasia and cancer), and stomach & & \checkmark & \checkmark \\ 
			\hline
			\dataindex{data:ead2020} & \href{https://ead2020.grand-challenge.org/}{EAD 2020} & 2020 & & & & & \checkmark\tnote{tnote:ca:t:3:8} & Artefact region & & \checkmark & \checkmark \\ 
			\hline
			\dataindex{data:edd2020} & \href{https://edd2020.grand-challenge.org/}{EDD 2020} & 2020 & & & & & \checkmark\tnote{tnote:ca:t:3:8} & Colon (polyp, cancer), oesophagus (Barrett’s, dysplasia and cancer), and stomach & & \checkmark & \checkmark \\ 
			\hline
			\dataindex{data:SARAS-ESAD:2020} & \href{https://saras-esad.grand-challenge.org/}{SARAS-ESAD} \cite{Bawa2020} & 2020 & & & & & \checkmark\tnote{tnote:ca:t:3:8} & Surgeon action detection & \checkmark & & \checkmark \\ 
			\hline
			\dataindex{data:ead2019} & \href{https://ead2019.grand-challenge.org/}{EAD 2019} \cite{Ali2019} & 2019 & & & & & \checkmark\tnote{tnote:ca:t:3:8} & Artefact region & & \checkmark & \checkmark \\ 
			\hline
			\dataindex{data:aidasub-chromogastro} & \href{https://aidasub-chromogastro.grand-challenge.org/}{AIDA-E Subchallenge 1} & 2016 & & & & & \checkmark\tnote{tnote:ca:t:3:8} & Mucosa damage in celiac disease & \checkmark & & \\ 
			\hline
			\dataindex{data:aidasub-cleceliachy} & \href{https://aidasub-cleceliachy.grand-challenge.org/}{AIDA-E Subchallenge 2} & 2016 & & & & & \checkmark\tnote{tnote:ca:t:3:8} & Mucosa in Barrett's esophagus & \checkmark & & \\ 
			\hline
			\dataindex{data:aidasub-clebarrett} & \href{https://aidasub-clebarrett.grand-challenge.org/}{AIDA-E Subchallenge 3} & 2016 & & & & & \checkmark\tnote{tnote:ca:t:3:8} & Mucosa in gastric chromoendoscopy & \checkmark & & \\ 
			\bottomrule
		\end{tabular}
		\raggedright
		\begin{tablenotes}{l}
			\item[tnote:ca:t:3:8] Endoscopy
		\end{tablenotes}
\end{sidewaystable*}

\subsection{Datasets for diagnosis of chest \& abdomen diseases}
\label{sec:chest-abdomen:diagnosis}

Diseases of organs in the chest and abdomen have a significant impact on human health.
Therefore, many researchers work on this problem by analyzing medical images. Several researchers have designed automatic or semi-automatic algorithms for the classification, segmentation, detection, and characterization tasks to help medics diagnose these diseases. In this subsection, we describe the datasets and challenges related to the diagnosis of diseases of the chest and abdomen that are reported in Tables \ref{tab:ca:t:1}, \ref{tab:ca:t:2}, and \ref{tab:ca:t:3}, respectively.

\paragraph{Chest \& abdomen datasets according to modality:}

According to the datasets and challenges collected, CT is the most commonly used imaging modality for the chest \& abdomen, because of its suitable imaging quality and ability to clearly display tissues and lesions. Some datasets and challenges also provide CT images using contrast agents for clearer images.
Besides CT imaging, there are other modalities, including MR, X-Ray digital radiographs, PET, endoscopy, etc.
The MR images are used in breast-related diagnosis, cardiac-related tasks, soft tissue sarcoma detection, and ventilation imaging.
Because of the organs' size and the CT's resolution, which is limited by the imaging exposure time and radiation dose, MR is a more suitable imaging modality for small or specific organs. The PET is always used with other modalities, such as CT and MRI.
The contrast agent's density is related to the metabolism, which means the density of radiation from contrast agent will be high in the tumor, so PET is always used for tumor related tasks. Endoscopy images are used for medical inspection of the stomach, intestines, and others.

\paragraph{Chest \& abdomen datasets according to classification of diseases:}

The classification of diseases intends to determine whether a subject is healthy or not. It is sometimes called ``detection'' or ``prediction'', and the prediction is different from the detection task presented below.

The main focus of these datasets is to judge whether there is any cancer, lesion, or tumor, such as soft tissue sarcoma \dRef{data:Soft-tissue-Sarcoma}, prostate lesion \dRef{data:SPIE-AAPM-NCI+PROSTATEx+Challenges,data:ProstateX}, lung cancer \dRef{data:Lung-PET-CT-Dx}, and breast cancer \dRef{data:BCS-DBT}.
Classification is an effective task for diagnosis, particularly computer-aided tasks. A quick and early diagnosis can allow effective interventions to increase the probability of the patient recovery before the condition worsens.

Another focus is the classification of diseases. These diseases include mainly pneumothorax \dRef{data:SIIM-ACR}, cardiac diseases \dRef{data:ACAD}, tuberculosis \dRef{data:PCXR}, pneumonia \dRef{data:chest-xray-pneumonia}, and COVID-19, which are discussed at the end of this subsection. The endoscopy related challenges provide data with the aim to classify RGB images and videos to classify patient into ``normal'' vs. ``abnormal''. Dataset \dRef{data:MIMIC-CXR} focuses on the classification based on the diagnostic records. These datasets and challenges provide data for researchers to design AI-based algorithms to diagnose common diseases.

\paragraph{Chest \& abdomen datasets for attribute classification:}

The characterization task of the tumor and lesion is also called attribute classification, which focuses on the subsequent characterization analysis of the tumors and lesions following the detection and segmentation tasks using automatic analysis algorithms.
A typical example is the attributes classification of pulmonary and lung cancer \dRef{data:lndb,data:bimcv-covid19,data:StructSeg2019,data:RIDER-Lung-CT,data:SPIE-AAPM-lung,data:luna16}.
The datasets and challenges usually provide CT scans with the annotation of different attributes, such as lesion type, spiculation, lesion localization, margin, lobulation, calcification, cavity, etc. Each attribute includes two or more categories.
Another focus is the characterization of the breast related lesions and tumors \dRef{data:Digital-Mammography-DREAM,data:BREAST-DIAGNOSIS}.

\paragraph{Chest \& abdomen datasets for detection:}

In most research and clinical situations, classification is not enough. The medics and researchers usually focus on the reason for such a disease, and the localization of the lesion or tumor. Further treatment evaluations, plan and interpretability are the specific focus for medics and DL researchers. Thus, detection and segmentation are the tasks which are receiving a lot of attention at present. The detection task aims to find a region of interest and localize its position. The regions of interest usually include:
\begin{itemize}
	\item Lung cancer and tumor \dRef{data:ImageCLEF2018-tuberculosis,data:data-science-bowl-2017,data:RIDER-Lung-CT,data:NSCLC-R-ISS,data:LIDC-IDRI}
	\item Pulmonary nodule \dRef{data:lndb,data:Lung+Fused-CT-Pathology,data:SPIE-AAPM-lung,data:LIDC-IDRI,data:ANODE09}
	\item Celiac-related damage \dRef{data:ead2019,data:ead2020,data:endo2021,data:edd2020}
	\item Other lung lesions \dRef{data:rsna-pneumonia-detection-challenge,data:nih-chest-xrays}
	\item Polyp \dRef{data:CT+COLONOGRAPHY,data:edd2020}
	\item Cervical cancer \dRef{data:intel-mobileodt-cervical-cancer-screening}
	\item Liver cancer \dRef{data:LowDoseCT}
	\item Breast cancer \dRef{data:Digital-Mammography-DREAM}
	\item Action and artefact of surgeon\dRef{data:SARAS-ESAD:2020,data:ead2019}
\end{itemize}

\paragraph{Chest \& abdomen datasets for segmentation:}

Segmentation is a refinement of the detection task because it provides information about the location and the pixel-level labels.
Pixel-level annotations can help researchers design pixel-level algorithms for accurate and effective quantification, volume calculations, and other analysis and diagnosis of tumors and lesions at the pixel-level (e.\,g., monitoring of tumors size).
According to the datasets and challenges we have collected, most of them aim at the segmentation of the tumor and lesion from CT of:

\begin{itemize}
	\item Lung cancer \dRef{data:NSCLC-RI,data:NSCLC+Radiogenomics}
	\item Kidney tumor \dRef{data:c4kc-kits}
	\item Pulmonary nodule \dRef{data:volcan09}
	\item Pneumothorax \dRef{data:SIIM-ACR}
	\item Liver tumor \dRef{data:LiTS}
	\item Polyp \dRef{data:edd2020}
\end{itemize}

Furthermore, challenges \dRef{data:ead2019,data:ead2020} focus on the segmentation of artifacts (e.\,g., polyps) in endoscopic images.

\paragraph{COVID-19:}

In 2020, COVID-19 became a research focus because it caused more than 100 million infections and two million deaths. Different datasets and challenges focus on this devastating disease and provide data to help researchers develop deep learning models to detect COVID-19 via various medical image modalities.

In the view of modalities, most of these datasets and challenges use either CT or CR images, and some provide both modalities. One exception is dataset \dRef{data:Covid19PocusUltrasound}, which uses ultrasound images. These datasets provide image annotations labeled by radiologists.

Most of these datasets and challenges are related to classification tasks.
Datasets \dRef{data:COVID-CT,data:CT-COVID-19,data:bimcv-covid19,data:covid19-eu,data:objCXR} directly focus on diagnosing COVID-19 from normal subjects.
In contrast, datasets and challenges \dRef{data:covid-chestxray-dataset,data:Covid19PocusUltrasound,data:COVID-Net,data:covid19-radiography-database,data:COVID-19-chest-xray} focus on diagnosing COVID-19 from a few other similar diseases, which can also lead to lung opacity or other symptoms, such as Middle East Respiratory Syndrome (MERS), Severe Acute Respiratory Syndrome (SARS), and Acute Respiratory Distress Syndrome.
Moreover, other datasets and challenges \dRef{data:CORD19,data:covid-19-sirm} focus on the diagnosis task, with natural language processing, genomics, or clinical methods.

Similarly, some other datasets \dRef{data:COVID-19-seg,data:COVID-19-SEG,data:objCXR} focus on the segmentation or detection of COVID-19 related lesions, such as ground-glass opacity, air-containing space, and pleural effusion.

\begin{sidewaystable*}
	\centering
	\caption{Summary of datasets and challenges of other medical applications in chest and abdomen.}
	\label{tab:ca:ot}
	\small
		\begin{tabular}{cm{16em}cccccm{12em}ccccc}
			\toprule 
			\textsc{\textbf{Reference}} & \textsc{\textbf{Dataset/Challenge}} & \textsc{\textbf{Year}} & \multicolumn{4}{l}{\textsc{\textbf{Modalities}}} & \textsc{\textbf{Focus}} & \multicolumn{5}{l}{\textsc{\textbf{Tasks}}} \\ 
			\cline{4-7} \cline{9-12}
			\textsc{\textbf{Index}} & & & CT & MR & US & OT & & Regression & Registration & Tracking & Detection & Other \\
			\midrule
			\dataindex{data:A-AFMA-localization} & \href{https://a-afma-localization.grand-challenge.org/}{A-AFMA-Localization} & 2020 & & & \checkmark & & Amniotic fluid localization & & & & \checkmark & \\ 
			\hline
			\dataindex{data:surgvisdom} & \href{https://surgvisdom.grand-challenge.org/}{SurgVisDom 2020} & 2020 & & & & \checkmark\tnote{tnote:ca:ot:1} & Surgical task classification & & & & & \checkmark\tnote{tnote:ca:ot:5} \\ 
			\hline
			\dataindex{data:CRT-EPIGGY19} & \href{http://crt-epiggy19.surge.sh/}{CRT-EPIGGY19} & 2019 & & & & \checkmark\tnote{tnote:ca:ot:2} & Heart modeling & \checkmark & & & & \\ 
			\hline
			\dataindex{data:lvquan19} & \href{https://lvquan19.github.io/}{LVQUAN19} & 2019 & & \checkmark & & & Full quantification of cardiac LV & \checkmark & & & & \\ 
			\hline
			\dataindex{data:lvquan18} & \href{https://lvquan18.github.io/}{LVQUAN18} & 2018 & & \checkmark & & & Full quantification of cardiac LV & \checkmark & & & & \\ 
			\hline
			\dataindex{data:EchoNet} & \href{https://echonet.github.io/dynamic/}{EchoNet-Dynamic} & 2017 & & & \checkmark & & Heart tracking & & & \checkmark & & \\ %
			\hline
			\dataindex{data:LUMIC} & \href{https://lumic.grand-challenge.org/}{LUMIC} & 2018 & \checkmark\tnote{tnote:ca:ot:3} & & & & CT registration with phantom images & & \checkmark & & & \\ 
			\hline
			\dataindex{data:hc18} & \href{https://hc18.grand-challenge.org/}{HC 18} \cite{Heuvel2018,VandenHeuvel2018} & 2018 & & & \checkmark & & Fetal head circumference & \checkmark & & & & \\ 
			\hline
			\dataindex{data:slawt} & \href{https://www.doc.ic.ac.uk/~rkarim/la_lv_framework/wall/index.html}{SLAWT} \cite{Karim2018} & 2016 & \checkmark & \checkmark & & & Left atrial wall thickness & \checkmark & & & & \\ 
			\hline
			\dataindex{data:SAACMHS} & \href{http://stacom.cardiacatlas.org/}{Statistical Atlases and Computational Modelling of the Heart - S} & 2016 & \checkmark & \checkmark & & & Left atrium wall thickness & \checkmark & & & & \\
			\hline
			\dataindex{data:MMWHS17} & \href{http://www.sdspeople.fudan.edu.cn/zhuangxiahai/0/mmwhs17/index.html}{MMMWHS17} & 2017 & \checkmark & \checkmark & & & Multi-modalities heart registration & & \checkmark & & & \\ %
			\hline
			\dataindex{data:second-annual-data-science-bowl} & \href{https://www.kaggle.com/c/second-annual-data-science-bowl}{2nd Annual Data Science Bowl} & 2016 & & \checkmark & & & Cardiac ejection fraction & \checkmark & & & & \\ 
			\hline
			\dataindex{data:clust15} & \href{http://clust.ethz.ch/}{CLUST 2015} & 2015 & & & \checkmark & & Liver tracking & & & \checkmark & & \\ 
			\hline
			\dataindex{data:lmdc} & \href{http://www.cardiacatlas.org/challenges/landmark-detection-challenge/}{Landmark Detection Challenge} & 2015 & & \checkmark & & & Landmark location & & & & \checkmark & \\ %
			\hline
			\dataindex{data:clust14} & \href{http://clust14.ethz.ch/}{CLUST 2014} & 2014 & & & \checkmark & & Liver tracking & & & \checkmark & & \\ 
			\hline
			\dataindex{data:MOTIONCORRECTIONCHALLENGE} & \href{http://www.cardiacatlas.org/challenges/moco-perfusion-challenge/}{Motion Correction Challenge} & 2014 & & \checkmark & & & Heart motion correction & & \checkmark & & & \\ %
			\hline
			\dataindex{data:casdqe} & \href{http://coronary.bigr.nl/stenoses/}{Coronary Artery Stenoses Detection and Quantification Evaluation} \cite{Kirisli2013} & 2012 & \checkmark\tnote{tnote:ca:ot:4} & & & & Coronary artery stenoses detection and quantification & \checkmark & & & & \\ 
			\hline
			\dataindex{data:challengeus2012} & \href{http://www.ibme.ox.ac.uk/challengeus2012}{Challenge US: ISBI 2012} \cite{Rueda2014} & 2012 & & & \checkmark & & Fetal biometric measurements & \checkmark & & & & \\ 
			\hline
			\dataindex{data:motion-tracking-challenge} & \href{http://www.cardiacatlas.org/challenges/motion-tracking-challenge/}{Motion Tracking Challenge} & 2011 & & \checkmark & & & Heart motion tracking & & & \checkmark & & \\ %
			\hline
			\dataindex{data:CardiacMotionAnalysis} & \href{http://stacom.cardiacatlas.org/motion-tracking-challenge/}{Cardiac Motion Analysis Challenge 2011} & 2011 & & \checkmark & & & Heart motion tracking & & & \checkmark & & \\ %
			\hline
			\dataindex{data:empire10} & \href{https://empire10.grand-challenge.org/}{RMPIRE10} & 2010 & \checkmark & & & & Lung registration & & & & & \\ %
			\hline
			\dataindex{data:lvmc} & \href{http://stacom.cardiacatlas.org/stacom2014/capwebprd01.its.auckland.ac.nz/web/stacom2014/lv-mechanics-challenge.html}{LV Mechanics Challenge} & 2009 & & \checkmark & & & Modeling & & & & & \\ 
			\hline
			\dataindex{data:RCAAEF} & \href{http://coronary.bigr.nl/}{Rotterdam Coronary Artery Algorithm Evaluation Framework} & 2009 & \checkmark & & & & Reconstruction & & & & & \checkmark\tnote{tnote:ca:ot:6} \\ %
			\bottomrule
		\end{tabular}
		\raggedright
		\begin{tablenotes}{llllll}
			\item[tnote:ca:ot:1] Endoscopy &
			\multicolumn{3}{l}{\item[tnote:ca:ot:2] See official description: \url{http://crt-epiggy19.surge.sh/datasets.html}.} &
			\item[tnote:ca:ot:3] CT Pulmonary Angiography &
			\item[tnote:ca:ot:4] CT Angiography \\
			\item[tnote:ca:ot:5] Classification &
			\item[tnote:ca:ot:6] Generation
		\end{tablenotes}
\end{sidewaystable*}

\subsection{Datasets for other chest and abdomen-related tasks}
\label{sec:chest-abdomen:other}

Besides the classification, detection, and segmentation tasks, there are also several other tasks which are the current focus of research. In the following, we present the datasets and challenges related to these tasks, and report them in Table \ref{tab:ca:ot}.

\paragraph{Chest \& abdomen datasets for regression:}

Similar to attributes classification, regression is another task which aims to compute or measure the target attributes from given images, but the difference is that the outputs of regression are continuous.
A typical example is fetal biometric measurements \dRef{data:hc18,data:challengeus2012}. These challenges provide ultrasound images to help researchers design algorithms to measure such attributes to estimate the gestational age and monitor the fetus's growth. Besides, another example is cardiac measurements \dRef{data:lvquan19,data:lvquan18,data:slawt,data:second-annual-data-science-bowl,data:casdqe,data:lvmc,data:CRT-EPIGGY19}. These datasets and challenges provide MR or ultrasound images to analyze the heart's attributes to detect heart diseases.

\paragraph{Chest \& abdomen datasets for tracking:}

Tracking is a critical task because our body and organs move during imaging. For organs, such as the heart, the characteristics of their motion is informative. Challenges \dRef{data:clust15,data:clust14} provide ultrasound data to track the liver to analyze the following of a surgery and treatments.
Datasets and challenges \dRef{data:EchoNet,data:motion-tracking-challenge,data:CardiacMotionAnalysis} focus on the tracking of the heart. They provide ultrasound images to track and analyze the heart.

\paragraph{Chest \& abdomen datasets for registration:}

Challenge \dRef{data:LUMIC} focuses on the CT registration of lungs and provides CT scans with and without enhanced and contrast agents.
Meanwhile, challenges \dRef{data:MMWHS17,data:MOTIONCORRECTIONCHALLENGE} focus on the registration between different modalities of the heart and provide MR, CT, and other modalities to register images with beating hearts.

\paragraph{Datasets for other chest \& abdomen related tasks:}

Challenges \dRef{data:A-AFMA-localization,data:lmdc} focus on localizing specific landmarks, including the amniotic and the heart, using ultrasound and MR images. Challenge \dRef{data:surgvisdom} focuses on the classification of surgery videos. Dataset \dRef{data:RCAAEF} focuses on the reconstruction of the coronary artery.


\section{Datasets and challenges for pathology and blood}
\label{sec:path-blood}

Though radiography, MR imaging, and other imaging modalities have been used as the basis for diagnosis, pathology images are also used as a gold standard for diagnosis, particularly for tumors and lesions. Digital pathology images are generally obtained by collecting tissue samples, making slices, staining, and imaging. Therefore, pathology images are also one of the mainstream image modalities that are used for diagnosis.

The focus of these datasets and challenges include \textbf{1)} the identity and segmentation of basic elements (e.\,g., cell and nucleus) in pathology images, and \textbf{2)} blood-based diagnosis from images.
In this section, we present datasets and challenges of the pathology images (Subsection \ref{sec:path-blood:pth}), and at the same time, cover the datasets and challenges of blood images in Subsection \ref{sec:path-blood:blood}.

\begin{sidewaystable*}
    \centering
    \caption{Summary of datasets and challenges for pathology-related image analysis. }
    \label{tab:path}
    \small
        \begin{tabular}{cm{22em}cm{14em}ccccm{5em}<{\centering}}
            \toprule 
            \textsc{\textbf{Reference}} & \textsc{\textbf{Dataset/Challenge}} & \textsc{\textbf{Year}} & \textsc{\textbf{Focus}} & \multicolumn{4}{l}{\textsc{\textbf{Tasks}}} & \textsc{\textbf{Stain}} \\ 
            \cline{5-8}
            \textsc{\textbf{Index}} & & & & Classification & Segmentation & Detection & Other & \\
            \midrule
            \dataindex{data:PathologyVQA} & \href{https://pathvqachallenge.grand-challenge.org/}{Pathology VQA} \cite{He2020} & 2020 & Visual question answering & & & & \checkmark\tnote{tnote:path:1} & H\&E \\ 
            \hline
            \dataindex{data:panda} & \href{https://panda.grand-challenge.org/}{PANDA Challenge} & 2020 & Prostate cancer grade assessment & \checkmark & & & & H\&E \\ 
            \hline
            \dataindex{data:paip2020} & \href{https://paip2020.grand-challenge.org/}{PAIP 2020} & 2020 & Colorectal cancer & \checkmark & \checkmark & & & H\&E \\ 
            \hline
            \dataindex{data:radpath-2020} & \href{https://miccai.westus2.cloudapp.azure.com/competitions/1}{MICCAI 2020 CRPCC} & 2020 & Combined radiology and pathology classification & \checkmark & & & & H\&E \\ 
            \hline
            \dataindex{data:herohe} & \href{https://ecdp2020.grand-challenge.org/}{HEROHE} & 2020 & HER2 & \checkmark & & & & H\&E \\ 
            \hline
            \dataindex{data:monusac} & \href{https://monusac-2020.grand-challenge.org/}{MoNuSAC} \cite{Verma2020} & 2020 & Multi-organ nucleus & \checkmark & \checkmark & & & H\&E \\ 
            \hline
            \dataindex{data:tcia:52758117} & \href{https://wiki.cancerimagingarchive.net/pages/viewpage.action?pageId=52758117}{Post-NAT-BRCA} \cite{Peikari2017,Martel2019} & 2019 & Cell & \checkmark & \checkmark & & \checkmark\tnote{tnote:path:2} & H\&E \\ 
            \hline
            \dataindex{data:tcia:52756935} & \href{https://wiki.cancerimagingarchive.net/pages/viewpage.action?pageId=52756935}{Osteosarcoma data for Viable and Necrotic Tumor Assessment} \cite{LeaveyP.2019,Mishra2017,P.2017,Arunachalam2017,Mishra2018} & 2019 & Osteosarcoma pathology image & \checkmark & & & & H\&E \\ 
            \hline
            \dataindex{data:digestpath2019} & \href{https://digestpath2019.grand-challenge.org/}{DigestPath 2019} \cite{Li2019a} & 2019 & Signet ring cell & \checkmark & \checkmark & \checkmark & & H\&E \\ 
            \hline
            \dataindex{data:lyon19} & \href{https://lyon19.grand-challenge.org/}{LYON 19} \cite{Swiderska-Chadaj2019} & 2019 & Lymphocyte & & & \checkmark & & IHC\tnote{tnote:path:5} \\ 
            \hline
            \dataindex{data:anhir} & \href{https://anhir.grand-challenge.org/}{ANHIR} \cite{Borovec2020,Borovec2019} & 2019 & Pathology image registration & & & & \checkmark\tnote{tnote:path:3} & - \\ 
            \hline
            \dataindex{data:Breast+Metastases+to+Axillary+Lymph+Nodes} & \href{https://wiki.cancerimagingarchive.net/display/Public/Breast+Metastases+to+Axillary+Lymph+Nodes}{Breast Metastases to Axillary Lymph Nodes} \cite{Campanella2019,Campanella2019a} & 2019 & Breast cancer metastases to lymph & & & \checkmark & & H\&E \\ 
            \hline
            \dataindex{data:gleason2019} & \href{https://gleason2019.grand-challenge.org/}{Gleason 2019} & 2019 & Gleason grade/score prediction & \checkmark & & & & H\&E \\ 
            \hline
            \dataindex{data:paip2019} & \href{https://paip2019.grand-challenge.org/}{PAIP 2019} & 2019 & Segmentation of liver cancer \& tumor burden & \checkmark & \checkmark & & & H\&E \\ 
            \hline
            \dataindex{data:ACDC-LungHP} & \href{https://acdc-lunghp.grand-challenge.org/}{ACDC@LUNGHP} \cite{Li2018b} & 2019 & Lung cancer & \checkmark & \checkmark & & & H\&E \\ 
            \hline
            \dataindex{data:monuseg} & \href{https://monuseg.grand-challenge.org/}{MoNuSeg} \cite{Kumar2019} & 2018 & Multi-organ nucleus & & \checkmark & & & H\&E \\ 
            \hline
            \dataindex{data:data-science-bowl-2018} & \href{https://www.kaggle.com/c/data-science-bowl-2018}{Data Science Bowl 2018} & 2018 & Cell & & \checkmark & & & H\&E \\ 
            \hline
            \dataindex{data:patchcamelyon} & \href{https://patchcamelyon.grand-challenge.org/}{Patch Camelyon} \cite{Veeling2018,Bejnordi2017} & 2019 & Pathology image patch & \checkmark & & & & H\&E \\ 
            \hline
            \dataindex{data:iciar2018} & \href{https://iciar2018-challenge.grand-challenge.org/}{ICIAR 2018 BACH} \cite{Aresta2019} & 2018 & Patch of breast cancer & \checkmark & & & & H\&E \\ 
            \hline
            \dataindex{data:colorectal-histology-mnist} & \href{https://www.kaggle.com/kmader/colorectal-histology-mnist}{Colorectal Histology MNIST} & 2018 & Colorectal pathology patch & \checkmark & & & & H\&E \\ 
            \hline
            \dataindex{data:tma} & \href{http://www-o.ntust.edu.tw/~cvmi/ISBI2017/}{Challenge for TMA in Thyroid Cancer Diagnosis} & 2017 & Thyroid cancer & \checkmark & & & & H\&E \\ 
            \hline
            \dataindex{data:camelyon17} & \href{https://camelyon17.grand-challenge.org/}{CAMELYON 17} \cite{Litjens2018,Bejnordi2017,Bandi2019} & 2017 & Breast cancer metastases & \checkmark & & \checkmark & & H\&E \\ 
            \hline
            \dataindex{data:camelyon16} & \href{https://camelyon16.grand-challenge.org/}{CAMELYON 16} \cite{Bejnordi2017} & 2016 & Breast cancer metastases & \checkmark & & \checkmark & & H\&E \\ 
            \hline
            \dataindex{data:glas} & \href{https://warwick.ac.uk/fac/sci/dcs/research/tia/glascontest}{GlaS} & 2015 & Colorectal cancer pathology patch & \checkmark & & & & H\&E \\ 
            \hline
            \dataindex{data:soccisc15} & \href{https://cs.adelaide.edu.au/~zhi/isbi15_challenge/}{2nd OCCIS Challenge} & 2015 & Overlapping cervical cell & & \checkmark & & & Papanicolaou \\ 
            \hline
            \dataindex{data:soccisc14} & \href{http://cs.adelaide.edu.au/~carneiro/isbi14_challenge/}{OCCIS} & 2014 & Overlapping cervical cell & & \checkmark & & & Papanicolaou \\ 
            \hline
            \dataindex{data:mitos-atypia-14} & \href{https://mitos-atypia-14.grand-challenge.org/}{MITOS-ATYPIA-14} & 2014 & Mitotic & & & \checkmark & & H\&E \\ 
            \hline
            \dataindex{data:celltrackingchallenge} & \href{http://celltrackingchallenge.net/}{Cell Tracking Challenge} \cite{Ulman2017} & 2020 & Cell & & & & \checkmark\tnote{tnote:path:4} & EM\tnote{tnote:path:6} \\ 
            \hline
            \dataindex{data:particle-tracking} & \href{http://bioimageanalysis.org/track/}{Particle Tracking Challenge} & 2012 & Particle & & & & \checkmark\tnote{tnote:path:4} & EM\tnote{tnote:path:6} \\ 
            \bottomrule
        \end{tabular}
        \raggedright
        \begin{tablenotes}{llll}
            \item[tnote:path:1] Visual question answering &
            \item[tnote:path:2] Cell counting &
            \item[tnote:path:3] Non-linear image registration &
            \item[tnote:path:4] Tracking \\
            \item[tnote:path:5] Immunohistochemistry
            \item[tnote:path:6] Electron microscope
        \end{tablenotes}
\end{sidewaystable*}

\subsection{Datasets \& challenges for pathology}
\label{sec:path-blood:pth}

Pathology images are used as the basis of cancer diagnosis. The pathologists and automatic algorithms analyze images based on specific features, such as cancer cells and cells under mitosis. Many organizations and researchers provide datasets and challenges, which focus on the microcosmic pathology and at the whole slide image (WSI) level. The relevant datasets and challenges are listed in Table \ref{tab:path}.

\paragraph{Imaging datasets \& challenges:}

In most situations, WSI is used in pathology diagnosis. Unlike CT or MR images, the pathology image is an optical image similar to the picture photoed by a camera. However, one major difference is that a pathology image is imaged by transillumination, while the usual photo is imaged by reflection. Another difficulty is in the size of the image. WSI is stored in a multi-resolution pyramid structure. A single multi-resolution WSI is generally achieved by capturing many small high-resolution image patches, and it might contain up to billions of pixels. Thus, WSI is used as a virtual microscope in diagnosis for clinical research, and many challenges use WSI, such as \dRef{data:glas,data:herohe,data:gleason2019,data:paip2019,data:camelyon17,data:camelyon16}. However, in some situations, the WSI is not suitable for analysis tasks, for example, cell segmentation. Therefore, pathology image patches are used in several other challenges, such as \dRef{data:PathologyVQA} for visual question answering, \dRef{data:mitos-atypia-14} for mitosis classification, \dRef{data:monusac,data:monuseg} for multi-organs nucleus detection and segmentation.

\paragraph{Datasets for stain:}

Slides made from human tissues are without color, and required to be stained. The commonly used stains include Hematoxylin, Eosin, and Diaminobenzidine. Usually, two or more stains are used in staining the slide, and the most commonly used combinations include Hematoxylin \& Eosin (H\&E) and Hematoxylin \& Diaminobenzidine (H-DAB).

\paragraph{Pathology datasets according to disease:}

The pathology slides are widely used in the diagnosis of many diseases, especially cancer. The cancer cells and tissues have different shapes compared to their normal counterpart. Thus, the diagnosis via pathology is the gold standard. Many datasets and challenges, such as \dRef{data:celltrackingchallenge,data:monuseg,data:monusac}, do not address any specific disease. At the same time, many datasets and challenges target specific diseases, such as breast cancer \dRef{data:tcia:52758117,data:Breast+Metastases+to+Axillary+Lymph+Nodes}, myeloma \dRef{data:SegPC2021,data:MiMMSBILab}, cancers in the digestive system \dRef{data:digestpath2019}, cervical cancer \dRef{data:soccisc14,data:soccisc15}, lung cancer \dRef{data:ACDC-LungHP}, thyroid cancer \dRef{data:tma}, and osteosarcoma \dRef{data:tcia:52756935}.

\paragraph{Pathology datasets according to task:}

Generally speaking, the tasks used with these datasets and challenges can be classified into two categories: microcosmic task and WSI-level task. The latter targets the diagnosis of diseases, based on a classification task. Expanded from the simple classification tasks, many datasets and research methodologies focus on complex tasks, such as the segmentation of tumor cell areas \dRef{data:monuseg,data:monusac,data:data-science-bowl-2018} and the detection of pathological features \dRef{data:camelyon16,data:digestpath2019}. The microcosmic tasks derive from the clinical analysis to identify cells and detect mitosis to extract key features from pathology images to support further disease diagnosis. The following subsections expand on the microcosmic tasks and WSI-leveling tasks, respectively.

\subsubsection{Microcosmic related datasets}
\label{sec:path-blood:pth:micro}

Microcosmic tasks focus on microcosmic features extraction (e.\,g., nucleus features), for further diagnosis and WSI-level tasks. In this subsection, we introduce the microcosmic task related datasets and challenges.

\paragraph{Data:}

Unlike the WSI-level, the datasets and challenges which focus on microcosmic tasks usually provide small size patch-level images with high-resolution. These patches are suitable for the annotation of microcosmic-level objects and resource-limited algorithms. The size of images varies depending on the image analysis tasks. For the segmentation and detection of cells and nucleus, the size of images is usually a thousand-pixel square to contain the suitable number of cells or nuclei. For individual cell analysis tasks (e.\,g., mitosis determination), the size is usually of a single cell. For other tasks (e.\,g., the patch-level classification), the size varies from dataset to dataset.

\paragraph{Pathology datasets for cell detection \& segmentation:}

Cells are considered to be essential for the pathology image. The analysis of cells is one of the most effective ways to extract pathology image features for diagnosis. The pathologists analyze the size, shape, pattern, and stained color of the cells with their knowledge and expertise to make judgments about these cells and classify them as normal or abnormal. Thus, many datasets and challenges focus on the segmentation and detection of cells. The cells and nucleus can be placed neatly in the slide.
However, during the slide preparation, these cells could overlap or locate randomly on the slide. Aiming at such a problem, challenges \dRef{data:soccisc14,data:soccisc15} focus on the segmentation and detection of overlapping cells and nuclei. The shape and size of cells from different organs might be different and can have different recognition and analysis challenges. Therefore challenges \dRef{data:monusac,data:monuseg} focus on the multi-organ cells or nucleus segmentation.

\paragraph{Pathology datasets for patch-level classification:}

Generally, the size of WSI is too large to be able to analyze every cell and relationships between cells. DL-based methods can easily find essential information from the patch-level image to support the diagnosis based on feature learning.
Many datasets and challenges focus on this problem. The datasets and challenges, which provide patch-level images, mainly focus on the classification, segmentation, or detection tasks. Based on the quality of feature learning, DL has reached the state-of-the-art performance in many areas of computer vision. Therefore, some datasets and challenges focus on the patch itself, and not the cell itself. The tasks can vary from the segmentation, detection, and classification of the cell to the direct classification of the patch. Challenges \dRef{data:patchcamelyon,data:colorectal-histology-mnist,data:lyon19,data:tma} focus on patch-level image classification to determine whether metastatic or a different tissue is present.

\paragraph{Datasets for other pathology tasks:}

Besides the detection and segmentation of cells and the patch-level classification, there are other microcosmic tasks. Challenge \dRef{data:mitos-atypia-14} focuses on the mitotic detection for nuclear atypia scoring. The atypical shape, size, and internal organization of cells are related to the progress of cancer. The more advanced the cancer is, the more atypical the cell looks like. Challenge \dRef{data:celltrackingchallenge} focuses on cell tracking, to know how cells change shapes and move as they interact with their surrounding environment. This is the key to understand cell migration's mechanobiology and its multiple implications in normal tissue development and many respective diseases. Challenge \dRef{data:PathologyVQA} focuses on the visual question answering task of pathology images using AI where the model is trained to pass the examination of the pathologist.

\subsubsection{Datasets for WSI-level tasks}
\label{sec:path-blood:pth:macro}

WSI-level pathology tasks focus on the diagnosis of cancer and pathology image processing. WSI contains all the complete information of a patient to be able to establish an accurate diagnosis. Automatic diagnosis algorithms can quickly analyze the slide. This is useful, especially in developing countries where there is a lack of well-experienced pathologists. However, it is a challenge to directly analyze WSI for both pathologists and algorithms because the size of WSI can be up to $100,000 \times 100,000$ pixels. Thus, such analysis becomes challenging, and to address this, most of the current datasets and challenges focus on the classification and segmentation of biomarkers, cells, and other regions of interest. At the end of this subsection, we introduce other datasets and challenges that are related to the tasks of regression and localization of tumors and biomarkers.

\paragraph{Datasets for classification of WSI:}

The prime goal of the examination of pathological images, especially WSI, is to diagnose cancer. Thus, how to classify WSI with large size and limited computing resources becomes a research challenge. Datasets and challenges \dRef{data:gleason2019,data:herohe,data:radpath-2020,data:panda} focus on predicting cancer or evaluating WSI, such as Gleason grade or HER2 evaluation. At the same time, some datasets and challenges \dRef{data:camelyon16,data:camelyon17,data:Breast+Metastases+to+Axillary+Lymph+Nodes} focus on the classification of metastasized cancer.

\paragraph{Datasets for segmentation and detection of WSI:}

DL-based methods are seen as a black box which process pathology images. The performance of these methods has achieved the state-of-the-art performance, but the interpretability of these methods is still difficult. From the pathologists' point of view, datasets and challenges \dRef{data:camelyon16,data:camelyon17,data:digestpath2019,data:paip2019,data:paip2020} focus on the segmentation and detection tasks to determine the critical elements which led to a particular diagnosis, such as cancer cell area and signet ring cell.

\paragraph{Datasets for other WSI tasks:}

Besides classification and detection, there are a few other tasks based on WSI. This includes the registration of pathology images \dRef{data:anhir} for data pre-processing and the localization of lymphocytes \dRef{data:lyon19}.

\begin{sidewaystable}
    \centering
    \caption{Summary of datasets and challenges of blood-related image analysis tasks. }
    \label{tab:blood}
    \small
        \begin{tabular}{cm{19em}cm{13em}ccccc}
            \toprule 
            \textsc{\textbf{Reference}} & \textsc{\textbf{Dataset/Challenge}} & \textsc{\textbf{Year}} & \textsc{\textbf{Focus}} & \multicolumn{4}{l}{\textsc{\textbf{Tasks}}} & \textsc{\textbf{Stain}} \\ 
            \cline{5-8}
            \textsc{\textbf{Index}} & & & & Classification & Segmentation & Detection & Other & \\
            \midrule
            \dataindex{data:SegPC2021} & \href{https://segpc-2021.grand-challenge.org/}{SegPC 2021}  & 2020 & Myeloma plasma cell & & \checkmark & & & Jenner-Giemsa \\ 
            \hline
            \dataindex{data:MitoEM} & \href{https://mitoem.grand-challenge.org/}{MitoEM Challenge} \cite{Wei2020}  & 2020 & Mitochondria & & \checkmark & & & (electron microscope) \\ 
            \hline
            \dataindex{data:tcia:61080958} & \href{https://wiki.cancerimagingarchive.net/pages/viewpage.action?pageId=61080958}{Single-cell Morphological Dataset of Leukocytes} \cite{Matek2019,Matek2019a} & 2019 & Blast cells in acute myeloid leukaemia & \checkmark & & \checkmark & & - \\ 
            \hline
            \dataindex{data:b-all2019} & \href{https://competitions.codalab.org/competitions/20395}{B-ALL Classification} \cite{Pan2019,Gehlot2020,Goswami2020} & 2019 & Immature leukemic blasts & \checkmark & & \checkmark & & - \\ 
            \hline
            \dataindex{data:malaria-bounding-boxes} & \href{https://www.kaggle.com/kmader/malaria-bounding-boxes}{Malaria Bounding Boxes} & 2019 & Cells in blood & \checkmark & & \checkmark & & Giemsa \\ 
            \hline
            \dataindex{data:lysto} & \href{https://lysto.grand-challenge.org/}{LYSTO}  & 2019 & Lymphocytes & \checkmark & & & & - \\ 
            \hline
            \dataindex{data:MiMMSBILab} & \href{https://wiki.cancerimagingarchive.net/display/Public/MiMM_SBILab+Dataset\%3A+Microscopic+Images+of+Multiple+Myeloma}{ MiMM SBILab Dataset} \cite{Gupta2018,Gupta2017,Gupta2019} & 2019 & Cell & & \checkmark & & & Jenner-Giemsa \\ 
            \hline
            \dataindex{data:snam-stain-norm} & \href{https://wiki.cancerimagingarchive.net/display/Public/SN-AM+Dataset\%3A+White+Blood+cancer+dataset+of+B-ALL+and+MM+for+stain+normalization}{SN-AM Dataset} \cite{Duggal2017,Duggal2016,Gupta2017,Gupta2019a} & 2019 & Stain normalization & & & & \checkmark\tnote{tnote:blood:1} & Jenner-Giemsa \\ 
            \hline
            \dataindex{data:CNMC2019} & \href{https://wiki.cancerimagingarchive.net/display/Public/C_NMC_2019+Dataset\%3A+ALL+Challenge+dataset+of+ISBI+2019}{C NMC 2019 Dataset} \cite{Gehlot2020} & 2019 & Immature leukemic cell classification & \checkmark &  & &  & - \\ 
            \hline
            \dataindex{data:blood-cells} & \href{https://www.kaggle.com/paultimothymooney/blood-cells}{Blood Cell Images}  & 2018 & Cells in blood & \checkmark & & & & Jenner-Giemsa \\ 
            \bottomrule
        \end{tabular}
        \raggedright
        \begin{tablenotes}{l}
            \item[tnote:blood:1] Stain normalization.
        \end{tablenotes}
\end{sidewaystable}

\subsection{Blood-related datasets}
\label{sec:path-blood:blood}

Blood image analysis is the basis of the diagnosis of many diseases. In contrast to the pathology images, blood samples' images mainly contain blood cells, and these datasets and challenges are aimed at blood-related cancer and cell counting. Similar to pathology images, these datasets and challenges also focus on the segmentation, detection, and classification of cells. The relevant datasets and challenges are listed in Table \ref{tab:blood}.

One of the primary tasks of these datasets is the classification of cells, which focuses on identifying the different types of cells. Dataset \dRef{data:blood-cells} focuses on classifying red blood cells, white blood cells, platelets, and other cells. At the same time, dataset \dRef{data:tcia:61080958} focuses on the classification of malignant and non-malignant cells. Other datasets and challenges \dRef{data:MiMMSBILab} (multiple myeloma segmentation), \dRef{data:MitoEM} (mitochondria segmentation), \dRef{data:malaria-bounding-boxes} (malaria detection) focus on the segmentation and detection of blood cells and biomarkers.


\begin{sidewaystable}
    \centering
    \caption{Summary of datasets and challenges of bone-related image analysis tasks.}
    \label{tab:bone}
    \small
        \begin{tabular}{cm{24em}ccccm{16em}ccc} 
            \toprule 
            \textsc{\textbf{Reference}} & \textsc{\textbf{Dataset/Challenge}} & \textsc{\textbf{Year}} & \multicolumn{3}{l}{\textsc{\textbf{Modalities}}} & \textsc{\textbf{Focus}} & \multicolumn{3}{l}{\textsc{\textbf{Tasks}}} \\ 
            \textsc{\textbf{Index}} & & & MR & CT & CR & & Classification & Segmentation & Other \\
            \midrule
            \dataindex{data:KNOAP2020} & \href{https://knoap2020.grand-challenge.org/}{KNOAP2020} & 2020 & \checkmark & & \checkmark & Knee osteoarthritis & \checkmark & & \\ 
            \hline
            \dataindex{data:RibFrac 2020} & \href{https://ribfrac.grand-challenge.org/}{MICCAI 2020 RibFrac Challenge} \cite{Jin2020} & 2020 & & \checkmark & & Rib fracture detection and classification & \checkmark & & \checkmark\tnote{tnote:bone:1} \\ 
            \hline
            \dataindex{data:openneuro:ds002900} & \href{https://openneuro.org/datasets/ds002900}{Spinal Cord MRI Public Database} & 2020 & \checkmark & & & Bone imaging & & & \checkmark\tnote{tnote:bone:2} \\ 
            \hline
            \dataindex{data:verse2020} & \href{https://verse2019.grand-challenge.org/}{VerSe 20} & 2020 & & \checkmark & & Vertebra & & \checkmark & \\ 
            \hline
            \dataindex{data:verse2019} & \href{https://verse2019.grand-challenge.org/}{VerSe 19} \cite{Loffler2020,Sekuboyina2020} & 2019 & & \checkmark & & Vertebra & & \checkmark & \\ 
            \hline
            \dataindex{data:Pelvic+Reference+Data} & \href{https://wiki.cancerimagingarchive.net/display/Public/Pelvic+Reference+Data}{Pelvic Reference Data} \cite{Yorke2019} & 2019 & & \checkmark & & Pelvic images & & & \checkmark\tnote{tnote:bone:3} \\ 
            \hline
            \dataindex{data:aasce19} & \href{https://aasce19.grand-challenge.org/}{AASCE 19} & 2019 & & & \checkmark & Spinal curvature & & & \checkmark\tnote{tnote:bone:4} \\ 
            \hline
            \dataindex{data:mrnetdataset} & \href{https://stanfordmlgroup.github.io/competitions/mrnet/}{MRNet Dataset} \cite{Bien2018} & 2018 & \checkmark & & & Knee & \checkmark & & \\ 
            \hline
            \dataindex{data:mura} & \href{https://stanfordmlgroup.github.io/competitions/mura/}{MURA} \cite{Rajpurkar2017} & 2018 & & & \checkmark & Abnormality in musculoskeletal & \checkmark & & \\ 
            \hline
            \dataindex{data:xVertSeg} & \href{http://lit.fe.uni-lj.si/xVertSeg/}{xVertSeg Challenge} & 2016 & & & \checkmark & Fractured vertebrae & \checkmark & \checkmark & \\ 
            \hline
            \dataindex{data:csi2016} & \href{https://csi2016.wordpress.com/challenge/}{CSI 2016} & 2016 & & & \checkmark & Intervertebral disc and vertebral fracture & \checkmark & \checkmark & \checkmark\tnote{tnote:bone:1} \\ 
            \hline
            \dataindex{data:bone-texture} & \href{https://ieeexplore.ieee.org/document/7163803}{Bone Texture Characterization} \cite{Song2015} & 2014 & & & \checkmark & Bone crisps & \checkmark & & \\ 
            \hline
            \dataindex{data:spineweb} & \href{http://spineweb.digitalimaginggroup.ca/spineweb/}{Spine and Vertebra Segmentation Challenge} \cite{Sekuboyina2020,Loffler2020,Sekuboyina2020a} & 2014 & & & \checkmark & Spine & & \checkmark & \\ 
            \hline
            \dataindex{data:ski10} & \href{http://www.ski10.org/}{SKI 10} & 2010 & \checkmark & & & Cartilage and bone in knee & & \checkmark & \\ 
            \bottomrule
        \end{tabular}
        \raggedright
        \begin{tablenotes}{llll}
            \item[tnote:bone:1] Detection &
            \item[tnote:bone:2] Generation &
            \item[tnote:bone:3] Registration &
            \item[tnote:bone:4] Regression
        \end{tablenotes}
\end{sidewaystable}

\section{Other datasets}
\label{sec:other}

Although we have categorized the datasets and challenges into three parts: ``head and neck'', ``chest and abdomen'', and ``pathology and blood'', several other datasets cannot be categorized under these three areas. In this section, we introduce the datasets and challenges categorized under ``other'' which means that these datasets do not fit under the above categories but they are still relevent to DL methods. The topics of this section include bone (Subsection \ref{sec:other:bone}), skin (Subsection \ref{sec:other:skin}), phantom (Subsection \ref{sec:other:phantom}), and animal (Subsection \ref{sec:other:animal}).

\subsection{Bone-related datasets}
\label{sec:other:bone}

Medical image analysis of bone is currently a major research focus. Radioautography is the most effective way to image bones, because X-Ray is sensitive to calcium that makes up human bones. The segmentation of bone, the detection of abnormalities, and their characterization are meaningful clinical and research tasks. Therefore, the following subsections discuss the datasets and challenges for the classification, segmentation, and other tasks, and Table \ref{tab:bone} reports these datasets and challenges.

\paragraph{Bone datasets for classification:}

The classification tasks for bone related computer-aided diagnosis is the focus for many researchers. Though the classification cannot locate the regions of interest, it can still help orthopedists to judge whether the patient is healthy or not, such as in dataset \dRef{data:bone-texture}. The diagnosis of tears and abnormality is also a research focus, such as meniscal tears \dRef{data:mrnetdataset}, vertebral fracture \dRef{data:csi2016}, and knee abnormality \dRef{data:mrnetdataset}.

\paragraph{Bone datasets for segmentation:}

The segmentation task of bone images plays a vital role in clinical diagnosis and treatment. The computer-aided segmentation algorithms and orthopedist need to segment the different parts of the bone from a given image and make a sound judgment to provide a more adequate treatment. The difficulty with such tasks is the low-resolution of images compared with other image modalities. The focus of these datasets and challenges include the spine \dRef{data:spineweb,data:csi2016}, vertebrae \dRef{data:verse2019,data:verse2020,data:xVertSeg}, and knee cartilage \dRef{data:ski10}.

\paragraph{Other bone related tasks:}

Besides the classification and segmentation tasks, the datasets and challenges of bone also include imaging \dRef{data:openneuro:ds002900}, registration \dRef{data:Pelvic+Reference+Data}, spinal curvature estimation \dRef{data:aasce19}, labeling \dRef{data:verse2019}, and abnormality detection \dRef{data:mura}.

\begin{table*}
    \centering
    \caption{Summary of datasets and challenges for skin, phantom, and animal related image analysis tasks.}
    \label{tab:others}
    \small
        \begin{tabular}{cm{17em}cccccm{18em}}
            \toprule 
            \textsc{\textbf{Reference}} & \textsc{\textbf{Dataset/Challenge}} & \textsc{\textbf{Year}} & \multicolumn{4}{l}{\textsc{\textbf{Modalities}}} & \textsc{\textbf{Focus}} \\ 
            \cline{4-7}
            \textsc{\textbf{Index}} & & & MR & CT & RGB & OT & \\
            \midrule
            \dataindex{data:dfu2020} & \href{https://dfu2020.grand-challenge.org/}{DFU 2020} \cite{Cassidy2020} & 2020 & & & \checkmark & & Diabetic foot ulcer detection \\ 
            \hline
            \dataindex{data:siim-isic} & \href{https://www.kaggle.com/c/siim-isic-melanoma-classification}{SIIM-ISIC Melanoma Classification} \cite{SIIM-ISIC-2020} & 2020 & & & \checkmark & & Melanoma classification \\ 
            \hline
            \dataindex{data:isic2019} & \href{https://challenge2019.isic-archive.com/}{ISIC 2019} \cite{Tschandl2018,Codella2019} & 2019 & & & \checkmark & & Classification of 9 diseases \\ 
            \hline
            \dataindex{data:isic2018} & \href{https://challenge2018.isic-archive.com/}{ISIC 2018} \cite{Tschandl2018,Codella2019} & 2018 & & & \checkmark & & Lesion segmentation and attribution classification of Melanoma and disease classification \\ 
            \hline
            \dataindex{data:isic2017} & \href{https://challenge.isic-archive.com/landing/2017}{ISIC 2017} \cite{Codella2017} & 2017 & & & \checkmark & & Lesion segmentation and attribution classification of Melanoma and disease classification \\ 
            \hline
            \dataindex{data:MATCH} & \href{https://www.aapm.org/GrandChallenge/MATCH/}{MATCH} & 2020 & & \checkmark & & & Tumor tracking in markerless lung \\ 
            \hline
            \dataindex{data:tcia:37224702} & \href{https://wiki.cancerimagingarchive.net/pages/viewpage.action?pageId=37224702}{MRI-DIR} \cite{Ger2018a,Ger2018b} & 2018 & \checkmark & \checkmark & & & Multi-modality registration with phantom \\ 
            \hline
            \dataindex{data:tcia:46334020} & \href{https://wiki.cancerimagingarchive.net/pages/viewpage.action?pageId=46334020}{CC-Radiomics-Phantom-3} \cite{Ger2019,Ger2018} & 2018 & & \checkmark & & & Phantom in different machine \\ 
            \hline
            \dataindex{data:tcia:39879218} & \href{https://wiki.cancerimagingarchive.net/pages/viewpage.action?pageId=39879218}{CC-Radiomics-Phantom-2} \cite{UlHassan2019} & 2018 & & \checkmark & & & Feature variability assessment with phantom \\ 
            \hline
            \dataindex{data:Credence+Cartridge+Radiomics+Phantom+CT+Scans} & \href{https://wiki.cancerimagingarchive.net/display/Public/Credence+Cartridge+Radiomics+Phantom+CT+Scans}{Credence Cartridge Radiomics Phantom CT Scans} \cite{Mackin2017} & 2017 & & \checkmark & & & Phantom research \\ 
            \hline
            \dataindex{data:petseg-challenge} & \href{https://portal.fli-iam.irisa.fr/petseg-challenge/overview}{PET-Seg Challenge} & 2016 & & \checkmark & & \checkmark\tnote{tnote:others:1} & Phantom registration and research \\ 
            \hline
            \dataindex{data:ismrm2015} & \href{http://www.tractometer.org/ismrm_2015_challenge/}{ISMRM 2015} & 2015 & \checkmark\tnote{tnote:others:2} & & & & Fiber imaging from phantom \\ 
            \hline
            \dataindex{data:RIDER+Phantom+PET-CT} & \href{https://wiki.cancerimagingarchive.net/display/Public/RIDER+Phantom+PET-CT}{RIDER Phantom PET-CT} \cite{Muzi2015} & 2015 & & \checkmark & & \checkmark\tnote{tnote:others:1} & Phantom research \& registration \\ 
            \hline
            \dataindex{data:MouseAwakeRest} & \href{https://openneuro.org/datasets/ds002551/versions/1.0.0}{Mouse awake rest} \cite{Takata2020} & 2020 & \checkmark & & & & Mouse brain segmentation \\ 
            \hline
            \dataindex{data:endovissub2019-scared} & \href{https://endovissub2019-scared.grand-challenge.org}{EndoVis 2019 SCARED} & 2019 & & & & \checkmark\tnote{tnote:others:3} & Depth estimation from endoscopic data \\ 
            \hline
            \dataindex{data:MouseLemurAtlasMRIraw} & \href{https://openneuro.org/datasets/ds001945/versions/1.0.0}{MouseLemurAtlas MRIraw} \cite{Nadkarni2019,Nadkarni2019a} & 2019 & \checkmark & & & & Mouse lemur brain image segmentation \\ 
            \hline
            \dataindex{data:CPTAC-GBM} & \href{https://wiki.cancerimagingarchive.net/display/Public/CPTAC-GBM}{CPTAC-GBM} \cite{CPTAC2018} & 2018 & \checkmark & \checkmark & & & Glioblastoma multiforme \\ 
            \hline
            \dataindex{BigNeuron} & \href{https://alleninstitute.org/bigneuron/about/}{BigNeuron} & 2016 & & & & \checkmark\tnote{tnote:others:4} & Animal neuron reconstruction \\ %
            \hline
            \dataindex{data:apples-ct} & \href{https://apples-ct.grand-challenge.org/}{Apples-CT} & 2020 & \checkmark & & & & Apple reconstruction and segmentation \\ %
            \hline
            \dataindex{data:qubiq} & \href{https://qubiq.grand-challenge.org/}{QUBIQ} & 2020 & \checkmark & \checkmark & & & Quantification of uncertainties in biomedical image quantification \\ 
            \hline
            \dataindex{data:AnDi} & \href{https://competitions.codalab.org/competitions/23601}{AnDi} \cite{Munoz-Gil2020} & 2020 & \checkmark\tnote{tnote:others:2} & & & & Anomalous diffusion \\ 
            \hline
            \dataindex{data:OpenKBP} & \href{https://www.aapm.org/GrandChallenge/OpenKBP/}{The Open Knowledge-Based Planning Challenge} & 2020 & & \checkmark & & & Dose distributions predictions \\ 
            \hline
            \dataindex{data:learn2reg} & \href{https://learn2reg.grand-challenge.org/}{Learn2Reg} \cite{Dalca2020} & 2020 & \checkmark & \checkmark & & & Image registration \\ 
            \hline
            \dataindex{data:SNEMI3D} & \href{http://brainiac2.mit.edu/SNEMI3D/}{SNEMI3D} & 2013 & & & & \checkmark\tnote{tnote:others:5} & Segmentation of neurites \\ 
            \hline
            \dataindex{data:SNEMI2D} & \href{http://brainiac2.mit.edu/isbi_challenge/}{SNEMI2D} & 2012 & & & & \checkmark\tnote{tnote:others:5} & Segmentation of neuronal structures \\ 
            \bottomrule
        \end{tabular}
        \raggedright
        \begin{tablenotes}{lllll}
            \item[tnote:others:1] PET &
            \item[tnote:others:2] DWI &
            \item[tnote:others:3] Endoscopy &
            \item[tnote:others:4] General microscopy &
            \item[tnote:others:5] Electron microscope
        \end{tablenotes}
\end{table*}

\subsection{Skin-related datasets}
\label{sec:other:skin}

Skin cancer is one of the most common type of cancer, and melanoma is one of the most lethal types of skin cancer. To diagnose skin cancer, dermoscopy is used to image the skin, and the classification, segmentation, and detection tasks are employed.
The most relevant datasets and challenges are reported in Table \ref{tab:others}.

Aiming at the computer-aided diagnosis of melanoma, ISIC released datasets and a series of challenges for clinical training and for the development of automatic algorithms. The challenges of ISIC include: 2017 \dRef{data:isic2017}, 2018 \dRef{data:isic2018}, 2019 \dRef{data:isic2019}. Challenges \dRef{data:isic2018,data:isic2017} include three sub-challenges: lesion segmentation, lesion attribution detection, and lesion classification with thousands of dermoscopic images. Besides, challenge \dRef{data:isic2019} focuses on the classification of melanoma, melanocytic nevus, basal cell carcinoma, actinic keratosis, benign keratosis, dermatofibroma, vascular lesion, squamous cell carcinoma, and others.
Challenge \dRef{data:siim-isic}, i.\,e., ISIC 2020, focuses on the classification of melanoma to better support dermatological clinical works with 33126 scans of more than 2000 patients.

Moreover, challenge \dRef{data:dfu2020} focuses on Diabetic Foot Ulcers, i.\,e., DFU. The challenges provide more than 2000 images of feet photographed with regular cameras under a consistent light source and annotated by experts for training and testing of automatic detection and classification algorithms.

\subsection{Phantom-related datasets}
\label{sec:other:phantom}

Phantom is an object based on a specific material to mainly evaluate medical imaging equipment. Phantom can be used in the registration of different pieces of equipment and using it as data for the development of automatic algorithms. Registration is essential for clinical diagnosis. For instance, it reduces the difference between different medical devices with or without the same modalities \dRef{data:tcia:46334020,data:tcia:39879218,data:Credence+Cartridge+Radiomics+Phantom+CT+Scans,data:RIDER+Phantom+PET-CT}. When data is scarce, then the phantom becomes useful. Some image analysis tasks or experiments require surgically inserted fiducial markers, which are costly and risky. However, the phantom has low cost and risk, easy to image, and easy to be annotated \dRef{data:MATCH,data:ismrm2015,data:petseg-challenge}. The related datasets and challenges are reported in Table \ref{tab:others}.

\subsection{Animal-related datasets}
\label{sec:other:animal}

Medical image analysis of animal material is relatively a smaller research area. However, it is not as limited by privacy and stricter ethics restrictions as human medical images. The datasets and challenges we found focus on animal brain segmentation \dRef{data:MouseAwakeRest,data:MouseLemurAtlasMRIraw}, depth estimation from endoscopic \dRef{data:endovissub2019-scared}, and multi-modality registration \dRef{data:tcia:37224702}. The relevant datasets and challenges are reported in Table \ref{tab:others}.


\section{Discussions}
\label{sec:dis}

The success of AI algorithms such as DL has led to their widespread use in several fields, including for medical image analysis. Researchers with different knowledge and background tackle image-based clinical tasks using computer vision tools to design automatic algorithms for different applications \cite{Zhang2020a,Liu2019a,Mao2020,Liu2019a,Demircioglu2021,Kim2021,Yim2020,Zhangb}. Though AI algorithms can successfully handle many tasks, several unsolved problems and challenges hinder the development of AI-based medical image analysis.

\subsection{Problems and challenges}

DL-based algorithms learn from input images of real data through gradient descent. Large-scale annotated datasets and a powerful DL model are key to the development of successful DL models. For example, the success of AlexNet \cite{Krizhevsky2017}, GoogleNet \cite{Szegedy2015}, ResNet \cite{He2015} are based on powerful models, which include millions of parameters. At the same time, a large-scale dataset, such as ImageNet \cite{Deng2010}, is also necessary to train the DL model to be able to tune such a large number of parameters.
However, when these methods are applied to medical image analysis, many domain-specific problems and challenges start to appear. This subsection discusses some of these challenges.

\subsubsection{Data scarcity}

The biggest challenge in the development of DL models is data scarcity. Different from other areas, the scale of the medical image datasets is usually smaller due to many limitations, e.\,g., the ethical restrictions.

The commonly used datasets for traditional computer vision are in larger scale compared to medical image datasets. For example, the handwritten digits dataset, MNIST \cite{LeCun1998} includes a training set with 60,000 examples and a testing set with 10,000 examples; the ImageNet dataset \cite{Deng2010} includes three million images for training and testing; Microsoft COCO \cite{Lin2014} includes more than two million images with annotations.
In contrast, many medical image datasets are smaller and only include hundreds or at most thousands of images. For example, the challenge BraTS 2020 \dRef{data:brats20} includes four hundred subjects and different modalities for each subject; the challenge REFUGE \dRef{data:REFUGE} provides about 1200 images of the eye; the challenge LUNA 16 \dRef{data:luna16} provides 888 CT scans; our recently published dataset of pulmonary lesions \cite{Li2021} just provides 694 scans; the challenge CAMELYON 17 \dRef{data:camelyon17} contains only more than 1000 WSI pathology images.

There are multiple reasons for the lack of data.
The main cause is due to the restricted access to medical images by non-medical researchers, i.\,e., barriers between disciplines. The root causes of these barriers relate to the cost and difficulties of annotation and the restricted access due to ethics and privacy.

\paragraph{Access to data:}

As mentioned in the introduction, the direct cause of the data scarcity is that most non-medical researchers are not allowed to access medical data directly. Though many medical data are generated worldwide every day, most non-medical researchers have no authorization to access clinical data.
The easily accessible data are publicly available datasets, but these datasets are not at a large-scale to be able to properly train a DL model.

\paragraph{Ethical reasons:}

Ethics of medical data usage is a major bottleneck and a limitation to researchers, particularly, computer scientists.
Medical data stored in databases always contains sensitive or private information, such as name, age, gender, and ID number.
In some cases, the data records of medical images can be used to identify a patient. For example, if an MR scan includes the face, an intruder can identify them for a possibly evil purpose. In most countries and regions, it is illegal to distribute such data with private information without the patients' permission, and nobody would usually consent to such distribution.
Therefore, for Deep learning researchers, it is impossible to gain authorization to access these datasets.

Before DL researchers are able to gain authorization even to desensitized data, they still need to pass ethical reviews.

\paragraph{Annotation:}

Another root cause is the difficulty to annotate medical images. Unlike other computer vision areas, the annotation of medical images requires specialized professions and knowledge. For example, in autopilot, when annotating objects such as vehicles and pedestrians, there are no specific annotators' requirements because most of us can easily distinguish a car or a human. However, when annotating medical images, domain-specific knowledge is essential. E.\,g., few people if naive would be able to tell the differences between an abnormal and normal tissue. However, it is impossible for a non-specialist to mark the lesion's contour or diagnose a disease.

This difficulty cannot easily be solved even when professionals are employed to annotate data. \textbf{First}, the cost of annotation of medical data is huge. Once the researcher and their organization have obtained some data, they need then to spend more money to employ few medics for its labeling. Such annotation cost is enormous, particularly where medical resources are scarce or where medical costs are high. For example, the challenge PALM \dRef{data:PALM} provides about 1,200 images with annotation, but its organizers involved only two clinical medics.
\textbf{Second}, the physician who annotates the data is required to have a rich clinical and diagnosis experience, thus reducing the number of people who are suitable for this task even further.
\textbf{Third}, to avoid any subjectivity, one image needs to be annotated by two or more physicians. Another problem is what to do if the labels of two annotators are not the same? In many challenges, the organizer employs many junior physicians to annotate and employs a senior physician to decide if the junior physicians' annotations are not the same. For example, in the challenge AGE \dRef{data:AGE}, each data annotation is determined by the mean of four independent ophthalmologists in a group and it is then manually verified by a senior glaucoma expert.

\subsubsection{Limitation of medical data}

The characteristics of medical images themselves pose difficulties for the medical image analysis tasks.

There are many types and modalities of images that are used in medical image analysis.
Similar to computer vision, the modalities include both 2D and 3D. However, the medical images have several other differences.
Though the average scale of a medical image dataset is smaller than computer vision-related field datasets, the size of each sample of data is larger on average than the one of a computer vision-related field.

For 2D images, CR, WSI, and other modalities have large variances in the resolution and color than the other computer vision fields. Some modalities might need more bits to encode a pixel, while some modalities are significantly huge. For example, CAMELYON 17 \dRef{data:camelyon17} only includes about a thousand of pathology images, but the whole dataset is about three terabytes. Such datasets with few large samples pose a challenge for the AI algorithms, and it has become a focus of research to design an algorithm that can learn from limited computational resources (e.\,g., the number of labeled samples) and be useful for clinical diagnosis.

For 3D medical images such as CT and MRI, they are dense 3D data, compared with sparse data, such as point cloud, in autopilot. Like the BraTS serial challenges \dRef{data:brats20,data:brats19,data:brats18,data:brats17,data:brats16,data:brats15,data:brats14,data:brats13,data:brats12}, many researchers face the challenges to design algorithms that can effectively learn from multi-modal dataset.

These characteristics of medical images require well-designed algorithms with a more robust capability to fit the data well and without overfitting. However, that further leads to the need for more data and resources. It is a challenge to learn suitable features from a small sample dataset.

\subsection{No silver bullet}

The ideal scenario is to find or invent a method or an algorithm to simultaneously solve all of these encountered problems.
\textbf{However, there is no silver bullet.}
The problems and challenges related to the data and the adopted methods cannot be entirely resolved, or sometimes, a problem arises as another is solved. Nevertheless, many ideas have been introduced to address the current problems, and they are introduced in this subsection.

With respect to the problems and challenges mentioned above, researchers are working on two research directions: \textbf{1)} a more effective model with less data, and \textbf{2)} a more practical approach to access data.
For the learning methods with small datasets, researchers use approaches such as few-shot learning and transfer learning. In order to access more data, researchers adopt three main approaches, namely federated learning, lifelong learning, and active learning.

\subsubsection{Practical learning from small samples}

Many medical image datasets have a small number of samples. For example, challenge MRBrains13 \dRef{data:MRBrainS13} only includes 20 subjects for training and testing, while challenge KITS 19 \dRef{data:c4kc-kits} has about two hundred subjects. Therefore, many researchers struggle to find a practical approach to learn from small samples.

\paragraph{Few-shot learning and zero-shot learning}

Few-shot learning hits one of the critical spots of DL-based medical image analysis problems, i.\,e., the development of DL models with fewer data. Humans can effectively learn from few samples. Therefore, different from the standard deep learning-based methods, humans learn to diagnose a disease from images, without the need to view tens of thousands of images (i.\,e., from only few-shot). Meta-learning, which is also called learning to learn, is a solution used to solve few-shot learning problem. Meta-learning can learn the meta-features from a small data size. The number of medical images in most datasets and challenges is not as large compared to the regular computer vision-related datasets and challenges.
Mondal et al. \cite{Mondal2018} use few-shot learning and GAN to segment medical images. The GAN is modified for semi-supervised learning with few-shot learning. Similar to few-shot learning, zero-shot learning aims at novel samples. Rezaei et al. \cite{Rezaei2020} cover a review of zero-shot learning from autonomous vehicles to COVID-19 diagnosis. However, zero-shot and few-shot learning have also their disadvantages, such as domain gap, overfitting, and interpretability.

\paragraph{Knowledge transfer:}

Transfer learning is another method, which can recognize and apply knowledge and skills learned from a previous task. For example, both white matter and gray matter segmentation and multi-organs segmentation are segmentation tasks. However, the neural network training is usually independent, which means that almost nobody trains a neural network with two tasks at once. However, it does not mean that these two tasks are unrelated. Besides zero-shot learning and few-shot learning, transfer learning, or say, knowledge transfer, is another method to infer knowledge from a previously learned task. Transfer learning can be applied to two similar tasks and between different domains. The most significant advantage of transfer learning is that they use rich scale datasets to pre-train the neural network and then fine tune and transfer the network to the main task on a few samples dataset.

\subsubsection{Effective access to more samples}

Besides finding a practical approach to learn from small samples, many researchers have been working on active learning and federated learning (which aims to use data without access to sensitive information). This also reduces annotation costs of deep learning algorithms.

\paragraph{Federated learning:}

Federated learning provides another way to access data. As discussed previously, the limitation of accessing data is led by privacy and other problems.
Instead of directly sharing data, federated learning shares the model to protect privacy from being leaked.
With other privacy protection methods, federated learning can effectively use the data from each independent data center or medical center.

However, there are two disadvantages of federated learning: annotation and implementation. The problem of annotation cannot be solved by sharing data but other methods. The main challenges are the implementation, as only a few institutions have attempted federated learning so far. For example, Intel and other institutions have attempted to apply federated learning for brain tumor-related tasks in their research \cite{Sheller2019}.
The main challenges in their implementation include:
\begin{enumerate}
    \item[\textbf{1)}] The implementation and proof of privacy protection,
    \item[\textbf{2)}] The methodology for sharing and updating millions of the model's parameters,
    \item[\textbf{3)}] Preventing attacks on DL algorithms and leaks of data privacy on the Internet or computing nodes.
\end{enumerate}

\paragraph{Natural language processing:}

Natural language processing is also a potential tool to automatically or semi-automatically annotate medical image data.
It is a standard procedure for a medic to provide a diagnostic report of the patient, particularly after the medical image was taken.
Therefore, such large amounts of data (image and text) is useful for medical image analysis after desensitization, and natural language processing can be used for annotation. Several natural language processing-based methods, e.\,g., \cite{Liang2019,Viani2021,Kim2020} have been applied in medical-related research fields.

\paragraph{Active learning:}

Active learning aims to reduce the annotation cost by indirectly using the unlabeled data to select the ``best'' samples to annotate. Generally, data annotation for deep learning requires experts to label data so that the neural network can learn from the data. Active learning does not require too many samples at the beginning of training. In other words, active learning can ``help'' annotators to label their data.
Active learning uses the knowledge learned from the labeled data to select and annotate the unlabeled data. The unlabeled data with annotation from algorithms is used to subsequently train the network over the next number of epochs.
Active learning \cite{Shao2018a,Carse2019} is used in the medical image analysis in a loop of \textbf{1)} algorithm learn from the data annotated by humans, \textbf{2)} human annotate the unlabeled data selected by the algorithm \textbf{3)} algorithm add the newly labeled data to the training set. The advantage of active learning is obvious: annotators do not need to annotate all the data they have, and at the same time, the neural network can learn from data faster from such interactive progress.


\section{Conclusion}
\label{sec:conclusion}

In this work, we have provided a comprehensive survey of the datasets and challenges for medical image analysis, collected between 2013 and 2020. The datasets and challenges were categorized into four themes: head and neck, chest and abdomen, pathology and blood, and others. We provide a summary of all the details about these themes and data.
We also discuss the problems and challenges of medical image analysis and the possible solutions to these problems and challenges.

\section*{Acknowledgments}

We thanks for the projects of National Natural Science Foundation of China (62072358), Zhejiang University special scientific research fund for COVID-19 preverntion
and control, National Key R\&D Program of China under Grant No 2019YFB1311600.

	\bibliographystyle{draft}
	\IfFileExists{./references-1b.bib}{
        \bibliography{references-1b}
    }{
        \bibliography{refs/references-1b}
    }



\end{document}